\input{psfig.sty}
\documentclass[12pt,preprint]{aastex}
\usepackage{graphicx}

\shorttitle{Coronal Densities in Active Stars from 
		{\it Chandra} HETG data}
\shortauthors{Testa et al.}
\usepackage{color} 
\usepackage{amsmath}           

\begin{document}

\title{The Density of Coronal Plasma in Active Stellar Coronae}
\author{Paola Testa\altaffilmark{1,2,3}, Jeremy
J. Drake\altaffilmark{2}, Giovanni Peres\altaffilmark{3}}
\altaffiltext{1}{SAO Predoctoral Fellow}
\altaffiltext{2}{Smithsonian Astrophysical Observatory, MS 
3, 60 Garden Street, Cambridge, MA 02138, USA; 
ptesta@head.cfa.harvard.edu, jdrake@head.cfa.harvard.edu}
\altaffiltext{3}{DSFA, Sezione di Astronomia, Università di 
Palermo Piazza del Parlamento 1, 90134 Palermo, Italy; 
testa@astropa.unipa.it, peres@astropa.unipa.it}
\affil{}
 
\begin{abstract}
We have analyzed high-resolution X-ray spectra of a sample of 22
active stars observed with the High Energy Transmission Grating
Spectrometer on {\em Chandra} in order to investigate their coronal
plasma density.  Densities where investigated using the lines of 
the He-like ions O~VII, Mg~XI, and Si~XIII.  Si~XIII lines in all 
stars of the sample are compatible with the low-density limit (i.e.\ 
$n_{\mathrm{e}} \lesssim 10^{13}$~cm$^{-3}$), casting some doubt on 
results based on lower resolution EUVE spectra finding densities
$n_{\mathrm{e}} > 10^{13}$~cm$^{-3}$.  Mg~XI lines betray the presence
of high plasma densities up to a few $10^{12}$~cm$^{-3}$ for most of
the sources with higher X-ray luminosity ($\gtrsim 10^{30}$~erg/s);
stars with higher $L_{\mathrm{X}}$ and $L_{\mathrm{X}}/L_{\mathrm{bol}}$ 
tend to have higher densities at high temperatures.  Ratios of
O~VII lines yield much lower densities of a few $10^{10}$~cm$^{-3}$,
indicating that the ``hot'' and ``cool'' plasma resides in physically
different structures.
In the cases of EV~Lac, HD~223460, Canopus, $\mu$~Vel, TY~Pyx, and 
IM~Peg, our results represent the first spectroscopic estimates of 
coronal density. No trends in density-sensitive line ratios with 
stellar parameters effective temperature and surface gravity were found, 
indicating that plasma densities are remarkably similar for stars with 
pressure scale heights differing by up to 3 orders of magnitude. 
Our findings imply remarkably compact coronal 
structures, especially for the hotter ($\sim 7$~MK) plasma emitting 
the Mg~XI lines characterized by coronal surface filling factor, 
$f_{\mathrm{MgXI}}$, ranging from $10^{-4}$ to $10^{-1}$, while we 
find $f_{\mathrm{OVII}}$ values from a few $10^{-3}$ up to $\sim 1$ 
for the cooler ($\sim 2$~MK) plasma emitting the O~VII lines.  
We find that $f_{\mathrm{OVII}}$ approaches unity at the same 
stellar surface X-ray flux level as solar active regions,
suggesting that these stars become completely covered by active
regions.  At the same surface flux level, $f_{\mathrm{MgXI}}$ is 
seen to increase more sharply with increasing surface flux.  These 
results appear to support earlier suggestions that hot $10^7$~K plasma 
in active coronae arises from flaring activity, and that this flaring
activity increases markedly once the stellar surface becomes covered
with active regions.  Comparison of our measured line fluxes with
theoretical models suggests that significant residual model 
inaccuracies might be present, and in particular that cascade
contributions to forbidden and intercombination lines resulting 
from dielectronic recombination might be to blame.  
\end{abstract}

\section{INTRODUCTION}
\label{sec:intro}

X-ray and EUV spectra represent the most effective means to 
investigate the physics of the hot magnetically confined
plasma in the outer atmospheres of late-type stars.  
The aim of spectroscopic diagnosis of these astrophysical 
plasmas is to probe the physical conditions of the gas and
to derive relevant parameters such as electron density, 
thermal structure, and relative element abundances:
such information should help us to 
constrain models of coronal heating and structuring.
As an intermediate step toward this goal, by studying the 
emission properties of stellar coronae we can test the 
validity of the solar analogy, i.e.\ whether the basic 
components of stellar coronae are analogous to those 
observed in the solar corona, whose dominant X-ray emitting
structures are defined by the loop-like morphology of 
coronal magnetic fields.  

The new generation X-ray observatories, {\em Chandra} and {\em
XMM-Newton}, with their unprecedented spectral resolution and large
effective areas, are providing us with high-resolution coronal X-ray
spectra in which individual spectral lines can be readily resolved.
The X-ray plasma diagnostics these spectra now allow for stars were
previously accessible only for the Sun.  It is also worth noting that
{\em Chandra} and {\em XMM-Newton} stellar coronal spectra offer
significantly more extensive X-ray wavelength coverage than achieved
with solar instrumentation. In particular, with regard to the paper in
hand, the High-Energy Transmission Grating Spectrometer (HETGS) on
{\em Chandra} provides high resolution ($\lambda/\Delta\lambda\sim 100-1000$) 
spectra in the energy range
0.4-8.0 keV (1.5-30\AA) where we observe simultaneously a large number
of prominent emission lines that, at least in principle, provide both
temperature and density diagnostics.

Of special importance for the purposes of this paper are 
the complexes of He-like ``triplets'' of O, Ne, Mg, and Si,
which include the {\em resonance} line ({\em r} : 
1s$^{2}$ $^{1}$S$_0$ - 1s2p $^{1}$P$_{1}$), the 
{\em intercombination} lines ({\em i} : 
1s$^{2}$ $^{1}$S$_0$ - 1s2p $^{3}$P$_{2,1}$), and the 
{\em forbidden} line ({\em f} : 
1s$^{2}$ $^{1}$S$_0$ - 1s2s $^{3}$S$_{1}$),
which correspond to transitions between the $n=2$ shell
and the ground level, $n=1$.  The utility of these lines 
for plasma diagnostics was pointed out by Gabriel \& 
Jordan\ (1969) (see also Pradhan \& Shull\ 1981,
and Porquet \& Dubau\ 2000 for a recent treatment): 
the ratio $R=f/i$ is mainly density 
sensitive, owing to the metastable 1s2s $^{3}$S$_{1}$ level,
while the ratio $G=(f+i)/r$ is mostly temperature sensitive.
Similar diagnostic techniques have been widely exploited 
for electron density estimates in EUV spectra (see, e.g., 
Mariska\ 1992 for solar applications, and, Laming\ 1998 
for a review of stellar work).  
{\em Chandra} and {\em XMM-Newton} data offer the opportunity 
to apply these techniques to the X-ray range and to explore 
different temperature and density regimes.  In this context, 
analyses of He-like triplets seen in the X-ray spectra of 
a few different stars have appeared in the recent 
literature (e.g., Brinkman et al.\ 2000; Canizares et al.\ 
2000; Audard et al.\ 2001; G{\" u}del et al.\ 2001ab,2002;  
Huenemoerder et al.\ 2001; Ness et al.\ 2001,2002;  
Raassen et al.\ 2002; Stelzer et al.\ 2002, Sanz-Forcada 
et al.\ 2003b).  
The study of Ness et al.\ (2002) is more extensive and
involves several stars observed with the {\it Chandra} Low 
Energy Transmission Grating Spectrograph (LETGS).  

While all the results from these studies cannot be easily 
summarised,  it
is apparent, mostly from O~VII lines, that there is 
a general trend showing that plasma with temperatures of 
up to a few million K has densities similar to that found 
in solar active regions---up to a few $10^{10}$~cm$^{-3}$ 
or so.  
Based on lines of Mg~XI, there is also some evidence for 
higher densities around $\sim 10^{12}$~cm$^{-3}$.  
These results, however, are not without controversy.
Brickhouse\ (2002), for example, notes that different 
density values have resulted from different analyses of the
same stars observed with the different {\it Chandra} and 
{\it XMM-Newton} spectrometers.  
These instruments have different resolving powers, and so 
density-sensitive lines can be blended to greater or lesser 
extents with other lines.  In this regime, the analysis is also
sensitive to the continuum level adopted.  For the case of 
{\it XMM-Newton} spectra, this is further complicated by the 
extended wings of the instrumental profile, which can give
rise to a significant pseudo-continuum.  

The best instrumentation to study the He-like lines of O, 
Ne, Mg and Si is that with the highest resolving power---the 
{\it Chandra} HETGS.  We were therefore motivated to perform 
a systematic study of these lines in a 
large sample of stars observed with the HETGS in order to 
attempt to characterize
plasma physical conditions in a sample of stars with very 
different stellar parameters.  The results can be used to probe 
possible trends of the plasma parameters with more global coronal 
parameters, such as X-ray luminosity or flux, or stellar 
effective temperature, gravity, rotational 
period, etc.  

In this paper, we have analyzed in detail the lines of
Si~XIII (6.7\AA), Mg~XI (9.2\AA), and O~VII (22\AA), in a 
sample of 22 late-type stars, covering a range of spectral 
types and activity levels.  We did not undertake an 
analysis of the Ne~IX triplet lines here, since they are 
heavily blended with iron (predominantly Fe~XIX; see Ness 
et al.\ 2003).  
The Si, Mg and O He-like triplets offer a reasonable 
sampling of different temperature and density regimes, 
with temperatures of formation and density sensitivities 
approximately corresponding to the following values: 10~MK, 
$10^{13}$~cm$^{-3}$ (Si~XIII), 7~MK, $10^{12}$~cm$^{-3}$
(Mg~XI); 2~MK, $10^{10}$~cm$^{-3}$ (O~VII).

We describe the observations and the stellar sample in
\S\ref{sec:observ}, and the detailed line analysis in 
\S\ref{sec:dataanalysis};  we discuss the results 
obtained in the context of coronal activity in section
\S\ref{sec:discussion} and we draw our conclusion in
\S\ref{sec:concl}.

\section{OBSERVATIONS}
\label{sec:observ}

The sample analyzed consists of all the late-type stars 
for which observations were available in the {\it Chandra} 
public archive\footnote{http://asc.harvard.edu/cda}
at the time when this study was undertaken.  The 
sample comprises 22 cool stars of different activity 
levels, and includes single stars (the flare stars AU~Mic, 
Proxima Cen and EV~Lac, the zero-age main-sequence star 
AB~Dor, the T~Tauri star TW~Hya, and the giants HD~223460, 
31~Com, $\beta$~Cet, Canopus and $\mu$~Vel) and multiple 
systems (the close eclipsing binary system Algol; two 
eclipsing binary systems of W~UMa type, namely ER~Vul, and
44~Boo; and the RS~CVn systems TZ~CrB, UX~Ari, $\xi$~UMa,
II~Peg, $\lambda$~And, TY~Pyx, AR~Lac, HR~1099, IM~Peg).

These sources were all observed using the 
{\em Chandra}-HETGS and the Advanced CCD Imaging 
Spectrometer (ACIS-S) detector (Canizares et al.\ 2000).  
The High-Energy Transmission Grating disperses 
X-rays through two different gratings, 
whose spectral coverage partially overlap; they are 
optimized for ``medium'' (MEG) and ``high'' (HEG) energies.  
The spectral resolution obtained with the HETGS is 
in the range $\lambda/\Delta\lambda=100-1000$.  The data 
used here were either recently processed by the standard 
pipeline and obtained directly from the archive, or else, 
in the case of older observations, reprocessed using 
standard CIAO 3.0 tools and analysis threads to take advantage
of recent improvements in the spectrum reduction software 
and ancillary calibration data.  Our analysis of spectral 
lines was based on the summed data from positive and 
negative orders.

Table~\ref{tab1} and Table~\ref{tab2} summarize, 
respectively, the stellar parameters and the characteristics
of the observations.  For each source the listed X-ray 
luminosity was derived by integrating the photon energies 
over all the wavelengths in the spectral ranges of the two
different instruments, HEG and MEG.

As summarized in Table~\ref{tab1}, the single
stars in our sample are four late-type flare stars, namely 
AU~Mic, Proxima Centauri, EV~Lac, and the rapid rotator 
AB~Dor ($P_{\mathrm{rot}} \sim 44$ ksec), the giants, 
HD~223460, 31~Com, $\beta$~Cet, Canopus and 
$\mu$~Vel and the T~Tauri star TW~Hya.
The other sources are binaries and most are 
classified as systems of RS~CVn-type.
In detail, among the RS~CVn systems: II~Peg is a system of 
a K2 V star and an unseen companion; analogously 
$\lambda$~And is a binary system of which we see only one 
component (G8 IV-III); finally, TY~Pyx is an eclipsing 
active close binary system of two G5 IV stars.
As for the remaining sources, $\xi$~UMa is a binary system 
of two optically identical components (G0 V) one of which 
is a rapid rotator, and there are two eclipsing binaries
of W~UMa type: ER~Vul, a short period active binary 
system of solar-like stars (G0 V + G5 V) that are rapid 
rotators (v$\sim 81$, 71 km/s respectively), and 44~Boo, 
a contact binary system of G1V-G2V stars.

It is worth noting that our sample contains sources whose 
X-ray luminosities span a very wide range---from a few 
$10^{26}$~erg/sec of Proxima Cen up to a few 
$10^{31}$~erg/sec  of the giant HD~223460.
As for the ratio of X-ray to total luminosities, 
$L_{\mathrm{X}}/L_{\mathrm{bol}}$, our sample covers the 
range $L_{\mathrm{X}}/L_{\mathrm{bol}} \sim 10^{-6}$-$10^{-3}$.
We are therefore analyzing spectra of stars whose activity
levels cover almost the whole activity range observed for 
late-type stars, although most of them are very active stars. 

\begin{figure*}[!ht]
\centerline{\psfig{figure=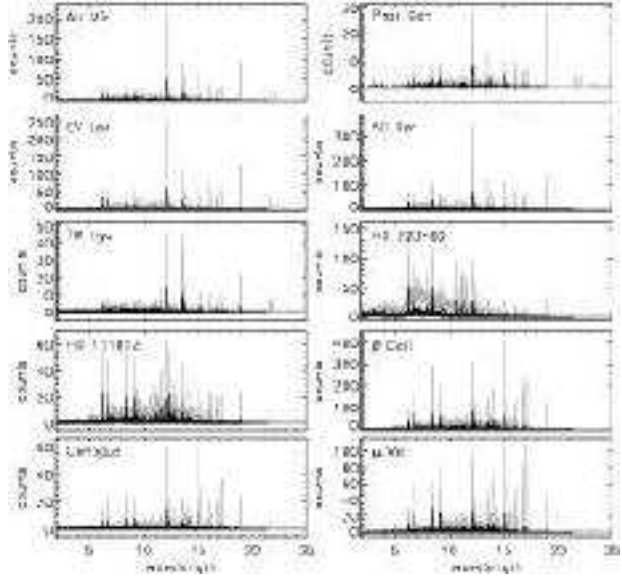}}
\caption{HEG (black) and MEG (grey) spectra of the single stars 
	of the sample. \label{fig1}}
\end{figure*}

\begin{figure*}[!ht]
\centerline{\psfig{figure=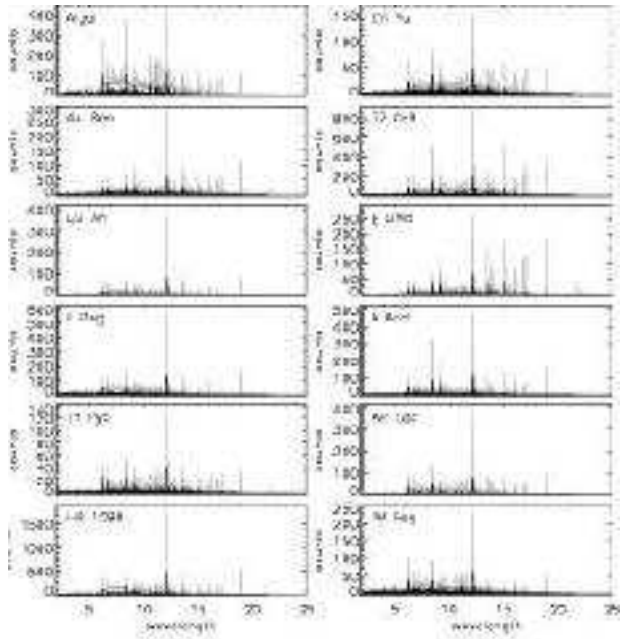}}
\caption{HEG (black) and MEG (grey) spectra of the analyzed
	active binary systems. \label{fig2}}
\end{figure*}

Spectra for each source were extracted from photon event files
obtained from the Chandra Data
Archive\footnote{http://cxc.harvard.edu/cda/} using CIAO 3.0.
Effective areas were calculated using standard CIAO procedures, and we
included the effects of the ACIS-S contamination layer (e.g.\ 
Plucinsky et al.\ 2002) using Version 1 of the effective area
contamination
correction\footnote{http://cxc.harvard.edu/ciao/threads/aciscontam}.
The resulting HETG spectra are shown in Figures~\ref{fig1}
and~\ref{fig2}; spectra of sources of the same type are
grouped together: Fig.~\ref{fig1} shows the spectra of single
stars (both dwarfs and giants); Fig.~\ref{fig2} shows the
spectra of binary systems.  For sources with several different 
observations we show the coadded spectra.

\section{DATA ANALYSIS}
\label{sec:dataanalysis}

\subsection{Lightcurves}

As a first step of the analysis we have extracted the 
lightcurves to check whether the spectra integrated over 
the whole observation time are influenced by significant 
variability or large flare events.
In order to obtain the lightcurves we excluded the 0th 
order events that can be drastically compromised by photon 
pile-up effects, and considered just the dispersed events.
The lightcurves were obtained by summing all the counts of
HEG and MEG spectra for spectral orders $n \geq 1$ in 
temporal bins of 100s.

\begin{figure*}[!ht]
\centerline{\psfig{figure=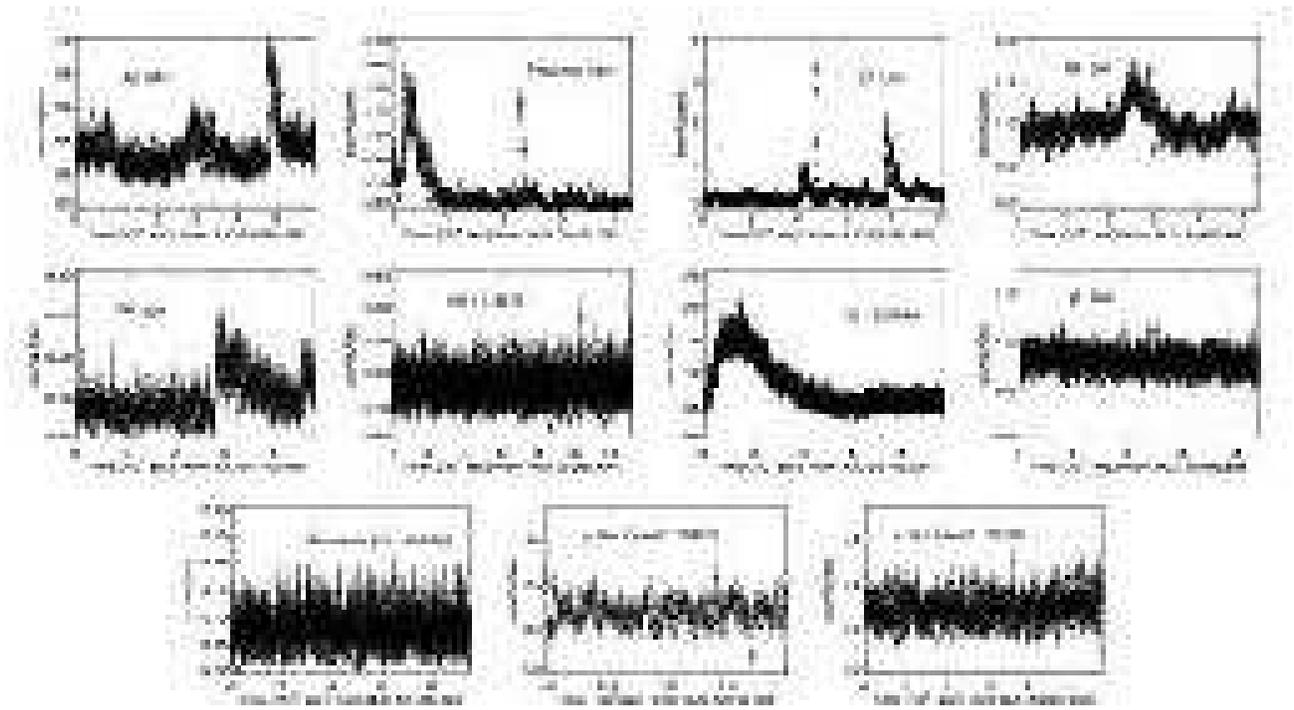,width=17.5cm}}
\caption{Lightcurves, obtained as the sum of total counts of 
	the HEG and MEG	dispersed spectra over temporal bins 
	of 100 sec, for the single stars of the sample. In 
	each plot the initial time in MJD is indicated. 
	\label{fig3}}
\end{figure*}

\begin{figure*}[!ht]
\centerline{\psfig{figure=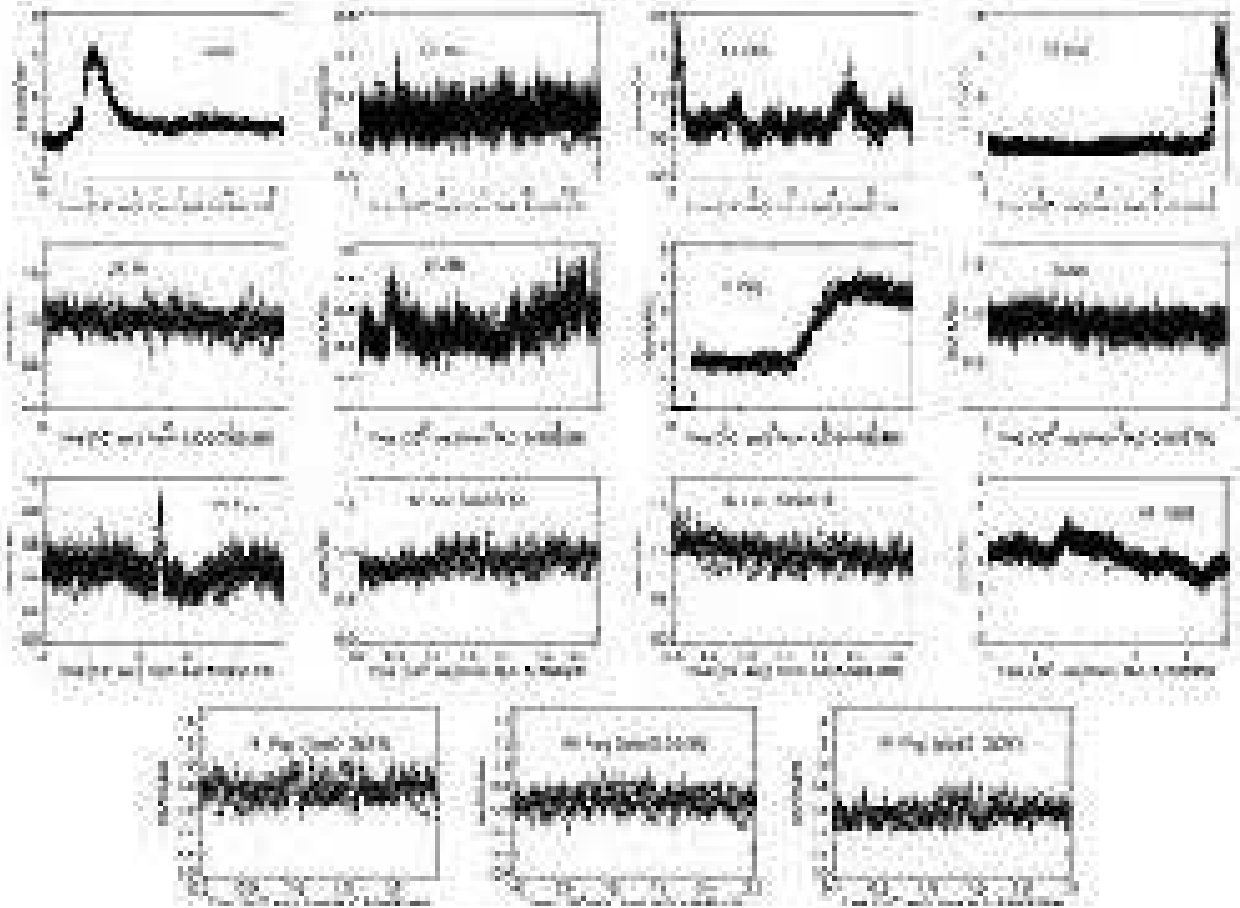,width=17.5cm}}
\caption{Lightcurves, as in Figure~\ref{fig3}, for the
	RS~CVn systems. \label{fig4}}
\end{figure*}

The resulting lightcurves are shown in 
Figures~\ref{fig3} and~\ref{fig4}.  
We see from these figures that there are no remarkable 
variations in X-ray count rate for 31~Com, $\beta$~Ceti, 
Canopus, $\mu$~Vel, ER~Vul, UX~Ari, $\xi$~UMa, 
$\lambda$~And, AR~Lac, and IM~Peg.  In the remaining cases, 
the lightcurves of AU~Mic and EV~Lac show flaring activity, 
and in particular, flares of different intensity and 
temporal extent; the Proxima Centauri lightcurve shows a 
flare at the beginning of the observation and then a decay 
over about 3 ksec and then the emission stabilizes at a 
low level; the AB~Dor emission is almost steady with 
the exception of a moderate flare extending over about 10 
ksec (of the $\sim$ 53 ksec of the total observation); 
HD~223460 shows a large flare over half of the observing 
time---it is worth noting that flares have been thought of as 
fairly unusual on single giants (e.g.\ Ayres et al., 1999,2001);
the lightcurve of Algol shows evidence of a big flare with 
a peak at about 10 ksec after the beginning of the 
observation and then a slow decrease over about 40 ksec to 
approximately the initial emission level; TZ~CrB shows 
a flare with a rapid increase of the emission by a factor 4
over the last $\sim 3.5$ ksec among the $83.7$ ksec of 
total exposure time; II~Peg shows a big flare analyzed in 
detail by Huenemoerder et al.\ (2001); the TY~Pyx total 
emission is quiescent with a slight modulation for almost 
the whole observation except a short flare (lasting $\sim$ 
3 ksec) in the middle of the observation. The lightcurve
of 44~Boo exhibits periodic modulation that was analyzed in detail by 
Brickhouse et al.\ (2001) who exploited, for the first 
time, the possibility of a crude Doppler imaging at X-ray 
wavelengths to derive geometrical properties of coronal 
X-ray emitting structures.

Regarding the influence of these variations on the 
integrated spectra, in most of the cases they do not 
affect substantially the basic properties of line emission, 
whereas the continuum emission changes significantly during 
the flares.  For example, for the II~Peg observation, 
Huenemoerder et al.\ (2001) show that at ``low'' temperatures
(T$\lesssim 10$~MK) the modulation of the line fluxes during 
the flare is low ($\lesssim 15$\%), and in particular, since
the triplet lines do not show substantial variations, 
they use the data integrated over the entire observation for 
the density diagnostics.  In other cases, e.g., AU~Mic, 
TZ~CrB, TY~Pyx, AB~Dor, the flares extend over a very small
portion of the observing time, thus they do not affect 
substantially the total spectra.  Unfortunately, 
in these cases the flaring intervals are also too short to
provide adequate signal for separate analysis.  Finally, 
in the case of Proxima Centauri, the flaring portion of the 
emission is the only detectable signal, and even when 
considering the total spectra, it is not possible to 
measure the fluxes of most of the lines because of the low 
signal.  In the light of the above considerations, our 
analysis here is based on spectra integrated over entire 
observation.

\subsection{Spectra}

\begin{figure*}[!ht]
\centerline{\psfig{figure=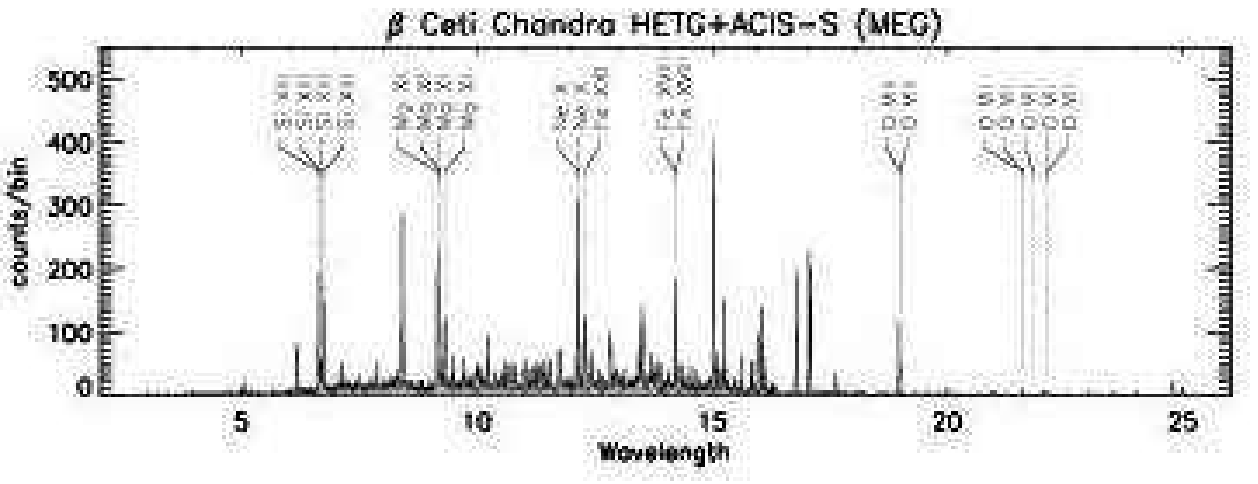,width=18cm}}\vspace{-0.5cm}
\caption{$\beta$~Ceti MEG spectrum. The analyzed lines are marked. 
	\label{fig5}}
\end{figure*}

\begin{figure}[!h]
\centerline{\psfig{figure=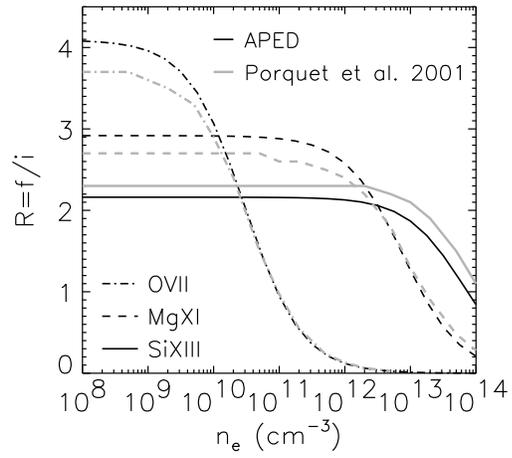,width=8cm}}
\caption{Theoretical R values from APED (Smith et al.\ 2001),
	at the temperature of maximum formation	of each 
	triplet, plotted vs.\ $n_{\mathrm{e}}$ for the 
	three analyzed He-like triplets. The gray curves
	correspond to the theoretical values from Porquet
	et al.\ (2001).  \label{fig6}}
\end{figure}

We illustrate the lines used in this analysis in the MEG spectrum 
of $\beta$~Ceti, illustrated in Figure~\ref{fig5}; the relevant lines 
are labeled by element and ionization stage.  
As already mentioned in Section~\ref{sec:intro}, in the HETGS spectral 
range the lines of He-like ions are among the most intense, though 
our analysis hinges on the measurement of relatively weak components 
of these.  The theoretical wavelengths of these line complexes, as 
they appear in the APED database (Smith et al.\ 2001), are listed 
in Table~\ref{tab3}.
Our plasma density analysis basically consists of a comparison 
between measured and theoretical relative intensities of the 
forbidden and intercombination lines (the R ratio).
For this comparison, we adopt the line emissivities of APED, and in
Figure~\ref{fig6} we show the dependence of the R ratio
on $n_{\mathrm{e}}$ at the temperature of maximum formation of 
each of the He-like ions considered here.  For comparison 
we plot also the values from Porquet et al.\ (2001).

As Figure~6 illustrates, the R ratios at the low density limit differ
by about 10\%\ between the different calculations.  At densities
above the lower sensitivity threshold, the calculations are in much
better agreement for O and Mg, though differences persist for Si at
higher densities.  The Porquet et al.\ (2001) and Smith et al.\ (2001)
emissivities are based on different input atomic data, and so small
discrepancies are to be expected.  Mewe et al.\ (2003) noted the same
differences in predicted R ratios for O~VII, and attributed these to
the different sources used for the direct excitation terms: Smith et
al.\ (2001) used the excitation rate coefficients from Kato \& Nakazaki
(1989) (for n$\leq 5$) and from HULLAC (Liedhal et al.\ 1995; 
for n=6-10), while Porquet et
al.\ (2001) adopted coefficients from Zhang \& Sampson (1987) (for
n$\leq 3$) and from Sampson et al.\ (1983) (for $2 <$ n $< 6$).  We
also note that dielectronic recombination and radiative recombination
rates are from different sources in the two models.  Since the $1s2s\,
^3S_1$ and $1s2p\, ^3P_{1,2}$ levels are primarily populated by
cascades from higher levels that in turn are largely populated by
dielectronic recombination (e.g.\ Smith et al.\ 2001), it seems
plausible that the treatment of these processes could also contribute
significantly to differences in predicted R ratios.

In this paper, we emphasise as much as possible trends in the observed
R ratios, rather than in the exact conversion to density values using
theoretical models; had we adopted the Porquet et al.\ (2001)
theoretical R ratios instead of those of Smith et al.\ (2001), none of
our conclusions would be significantly changed.

Observed spectra were analyzed with the IDL 5.3\footnote{Interactive 
Data Language, Research Systems Inc.} PINTofALE 
1.0\footnote{http://hea-www.harvard.edu/PINTofALE} software 
(Kashyap \& Drake\ 2000).
Line fluxes were determined by fitting with 
a modified Lorentzian function described by the relation
\begin{equation}
F(\lambda)=a/(1+(\frac{\lambda-\lambda_0}{\Gamma})^2)^\beta
\label{e:lorentz}
\end{equation}
where $a$ is the amplitude and $\Gamma$ is a characteristic 
line width, and with a value of 2.5 for the exponent $\beta$.  
This function has been found to be a good match to observed 
line profiles in stellar coronal spectra with counts of a 
few thousand or less (see, e.g., the Chandra Proposers'
Observatory 
Guide\footnote{http://asc.harvard.edu/proposer/POG/html/LETG.html}).

The measured R ratio can be somewhat sensitive to the level assumed
for the underlying continuum, and so we took particular care in
locating this.  For a few spectra of the very active stars the
continuum is quite strong in the vicinity of the Mg and Si triplets
(HD~223460, Algol, TZ~CrB, II~Peg, AR~Lac, HR~1099, and IM~Peg), while
in the 21-22~\AA\ range where the O~VII lines are located the
continuum level was always close to zero and essentially negligible.
The continuum in a coronal plasma at temperatures exceeding a few
million~K is predominantly due to bound-free and free-free emission
from hydrogen.  Over small wavelength intervals this continuum is
quite flat and can be modelled using the instrument effective area
function pinned down in adjacent ``line-free'' regions.  For the very
active stars, we also used a generic continuum model computed for the
RS~CVn-type binary HR~1099 by Drake et al.\ (2001), scaled again
according to spectral regions deemed essentially ``line-free''.  We
also performed sensitivity tests to determine how errors in continuum
location effected our measured fluxes: line fluxes were measured for
continuum levels deemed too high and too low, based on our location
criteria described above.  In the case of the $f$ and $i$ lines, the
differences in measured fluxes were always smaller than the
statistical uncertainties in the fluxes themselves so that uncertainty in
continuum placement was not a significant source of error in the analysis.
 
We measured the spectral line intensities of the Si~XIII, Mg~XI and
O~VII He-like triplets from both HEG and MEG spectra wherever the data
offer useful signal.  The results obtained from the two spectra are in
good agreement, as we will show below.  The O~VII lines were measured
only in the MEG spectra since they are outside the wavelength range of
the HEG.  For all the He-like line complexes we fitted the three lines
simultaneously; the free parameters of the fit were the position and
the flux of each of the lines, while the widths of all lines were
constrained to have the same value.

\subsection{Mg~XI spectral region}
\label{sec:mg}

While the Si~XIII and the O~VII triplet lines are relatively isolated,
the Mg~XI triplet lies in a more crowded region and the lines are more
likely to be affected by blending.  Indeed, the APED database predicts
a number of Fe~XIX-XXII lines in the 9.15-9.35~\AA\ region, though
each of these is considerably weaker than the Mg lines of interest.
Moreover, lines from the Lyman series of H-like Ne with upper levels
$n > 5$ also lie in this region.  These latter lines are listed in
Table~\ref{tab4}.

\begin{figure}[!h]
\centerline{\psfig{figure=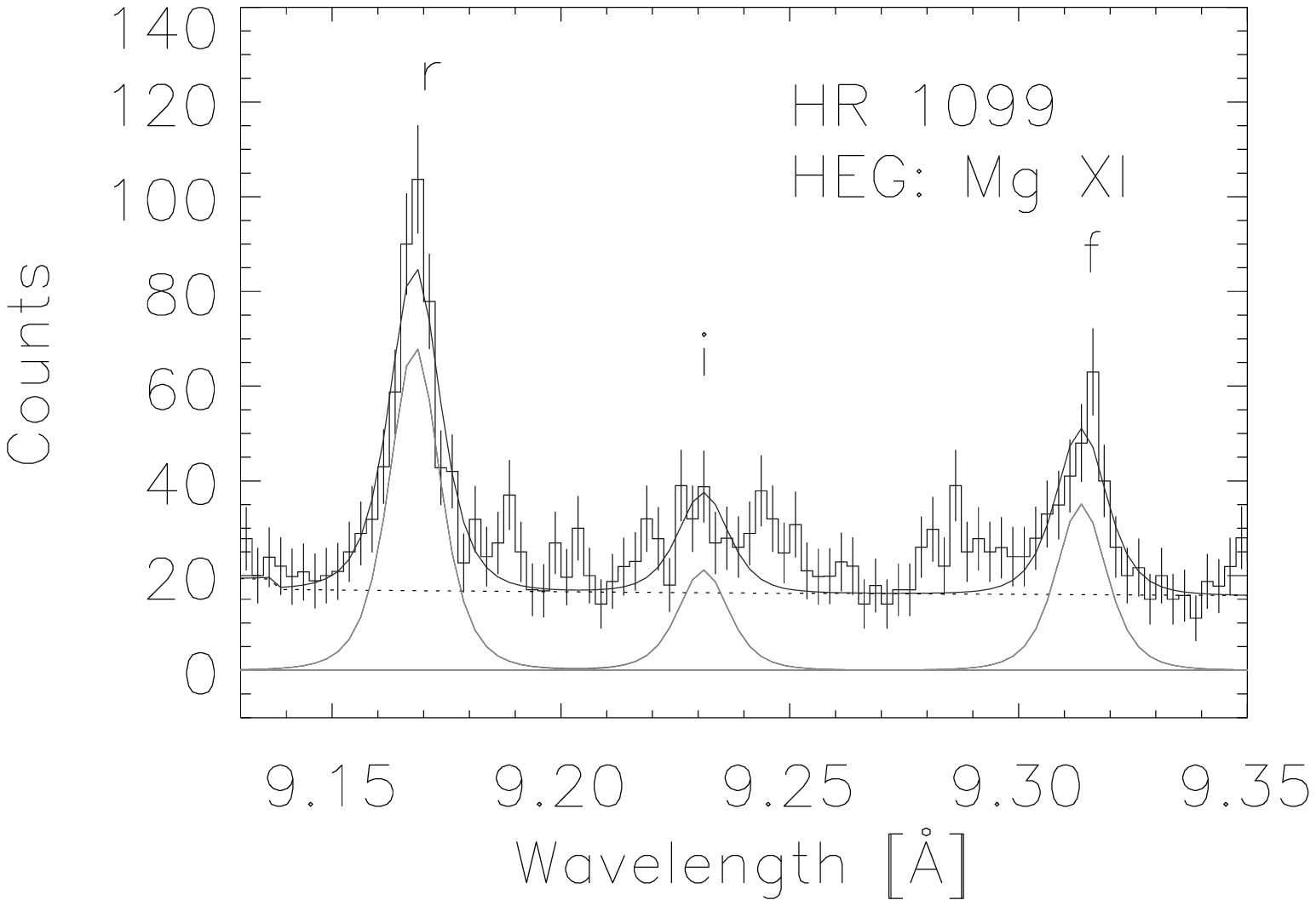,width=8cm}}\vspace{-0.4cm}
\centerline{\psfig{figure=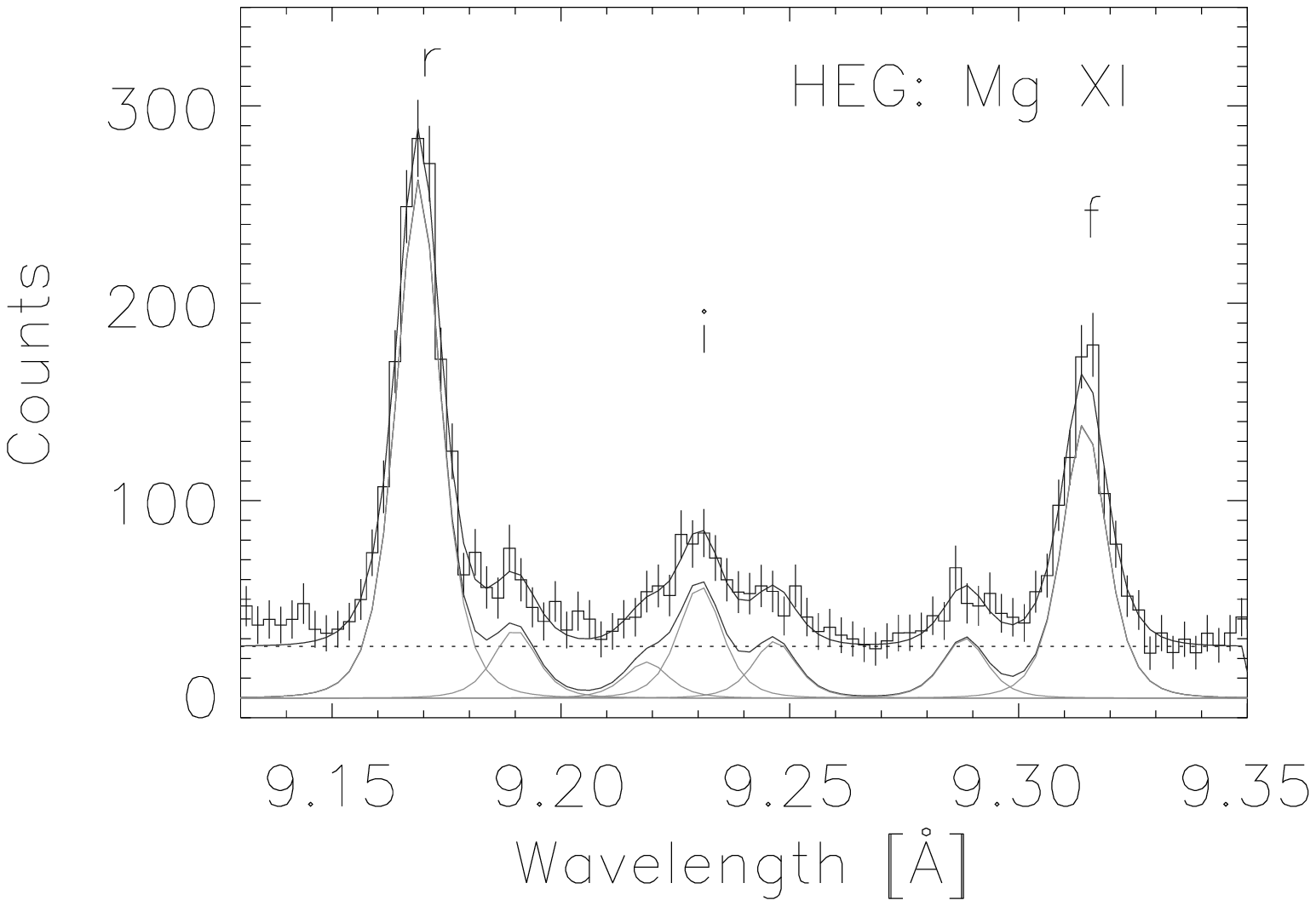,width=8cm}}\vspace{-0.4cm}
\centerline{\psfig{figure=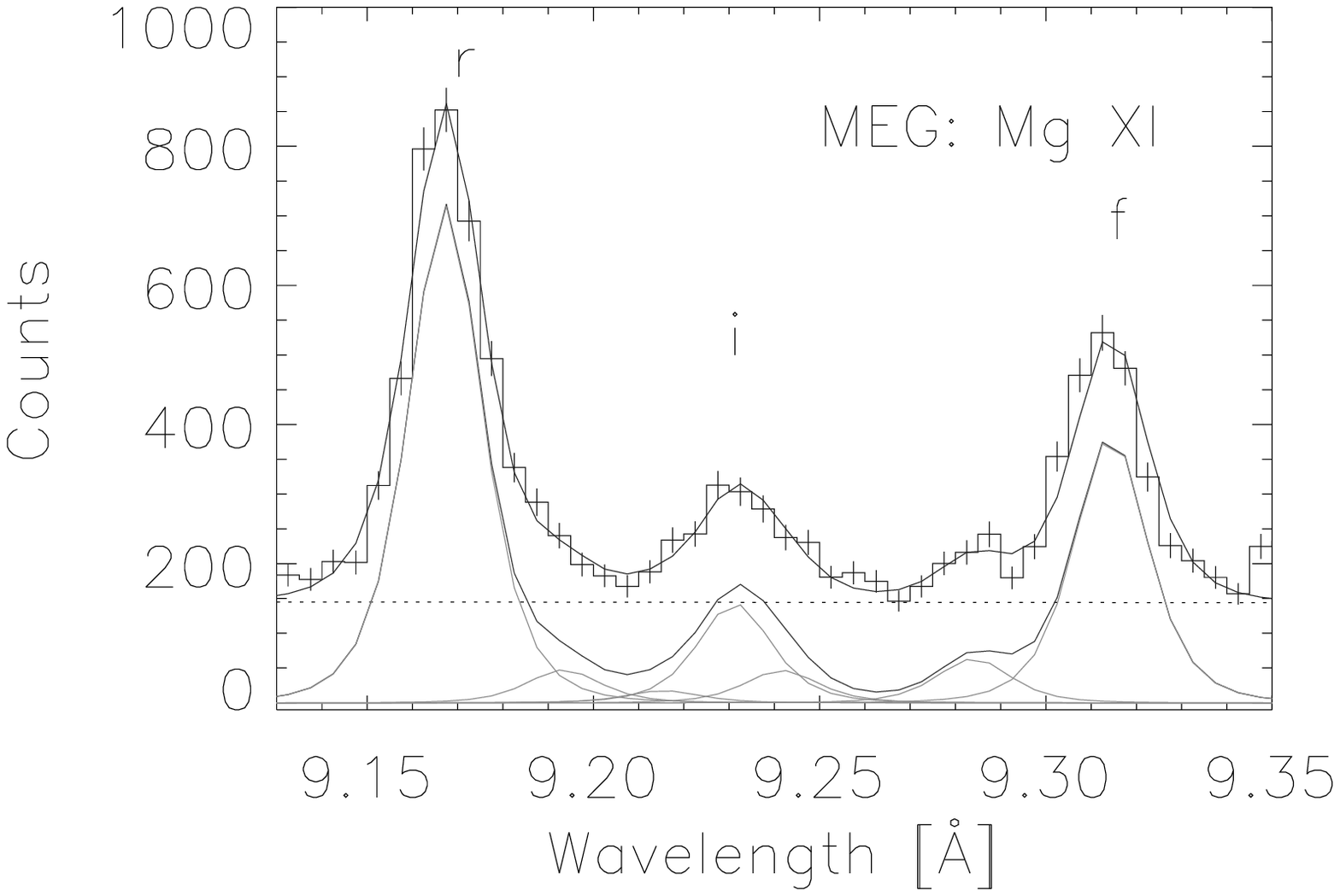,width=8cm}}
\caption{Mg~XI He-like triplet spectral region:
	{\em Top}-- HEG	spectrum of HR~1099: the best fit 
	model, with only the continuum and the He-like 
	triplet components, is superimposed to the data 
	points with their associated error bars.
	{\em Middle}-- coadded HEG spectra of four of the 
	sources with the highest S/N: $\beta$~Ceti, TZ~CrB, 
	AR~Lac and HR~1099.  The best fitting model with the
	blending components is superimposed to the spectrum.
	{\em Bottom}-- Same as in the {\em middle panel} but
	for MEG spectra.   \label{fig7}
	}
\end{figure}

We investigated the possible influence of line blends on our
measurement of the Mg~XI triplet components in some detail.  First, we
carefully examined HEG spectra with both the highest signal-to-noise
ratio and the highest relative intensity of Fe or Ne lines with
respect to Mg lines (e.g., UX~Ari for both Fe and Ne, AB~Dor for Fe,
HR~1099, II~Peg and AU~Mic for Ne) for evidence of spectral features
close to, or blending with, the triplet lines.  In particular, in the
spectra of stars in which Ne has been found to be strongly enhanced
relative to Fe, such as HR~1099 (Drake et al.\ 2001) and II~Peg
(Huenemoerder et al.\ 2001), the Ne Lyman series lines at 9.215
($9\rightarrow 1$), 9.246 ($8\rightarrow 1$) and 9.291 ($7\rightarrow
1$) are readily discernible.  Some lines of highly ionised Fe also
probably blend with the Ne features at some level, though the dominant
components appear fairly well-centred on the expected Ne wavelengths.
A simple fit to the Mg components of the HR~1099 in the HEG spectrum,
ignoring the Ne blends, is illustrated in Figure~\ref{fig7}
({\em top panel}).  It is clear from this figure that the fit to the
intercombination line also encompasses a significant fraction of the
Ne lines: ignoring these blends results in a spuriously large
intercombination line flux.

In order to account for the blends, we constructed an empirical model
of the region that included the major Mg XI lines, together with
blending components.  These components comprised four additional lines
corresponding to the positions of the Ne Lyman series lines with upper
levels $n=7,8,9,10$.  While it is desirable to include explicitly any
significant lines of Fe, this presents difficulties because the
wavelengths of the relevant Fe lines in this region are poorly known.
To our knowledge, the best wavelengths currently available for this
particular region are those predicted by the APED database, and these
are based primarily on theoretical calculations, rather than on
experiment.  The model itself was constrained by comparison with, and
by fitting to, co-added HEG spectra with the highest signal-to-noise
ratio ($\beta$~Ceti, TZ~CrB, AR~Lac and HR~1099).  Since Fe features
might distort and displace the centroids of the Ne features, in the
fitting process we allowed the wavelengths of the four blends to vary.
Line widths were all constrained to the same value, and the wavelength
separations of the Mg lines were also fixed at their accurately known
theoretical values.

In Figure~\ref{fig7} we show the best fit model together with
the individual components superimposed on the coadded HEG spectra.
The wavelength intervals between the blends and the Mg XI resonance
line found from this fit were then fixed into the model, which was
then tested on the coadded MEG spectra of the same four sources.  This
latter fit, which is also illustrated in Figure~\ref{fig7},
provided an excellent match to the MEG spectrum, and yielded relative
line fluxes in agreement with the HEG values to within statistical
uncertainties.  The individual wavelengths of the different line and blend
components in the final model of the Mg XI spectral region determined
in this way from the coadded HEG spectra are listed in
Table~\ref{tab5}, together with the identifications we
attribute to  the features.  

While the blending components include the contribution of all the Ne
lines of the Lyman series present in this spectral range, there are
doubtless contributions at some level from Fe lines, and we list
candidates in Table~\ref{tab5}.  The relative strengths of
the Ne Lyman lines for the higher $n$ transitions should scale in
proportion to the oscillator strengths of the transitions.  This
appears to hold for those with upper levels $n=7$,8 and 9; however,
the $10\rightarrow 1$ transition at 9.194~\AA\ appears too strong and
it is shifted slightly blueward of its predicted location.
Examination of the APED database reveals an Fe~XXI transition in this
vicinity that we attribute to this.  In comparison to this Fe XXI
line, other potential Fe blends predicted in APED appear significantly
weaker by factors of at least 3 and are likely less significant.  The
wavelengths fitted for the other blend components also remained very
close to the values of the Ne lines, and the presence of further
significant blends that we have omitted seem unlikely.

\begin{figure}[!h]
\centerline{\psfig{figure=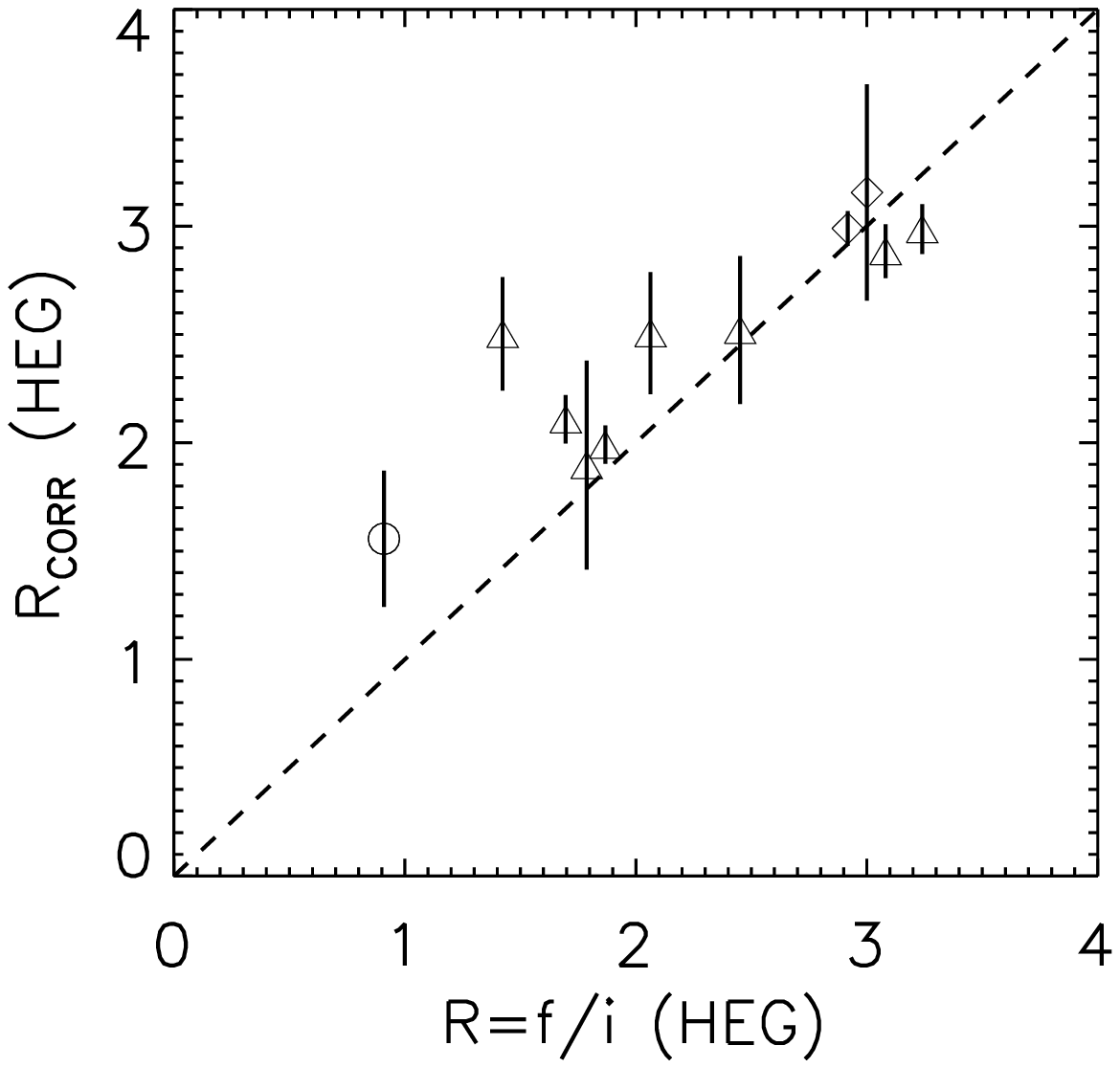,width=5.5cm}\hspace{-1cm}
          \psfig{figure=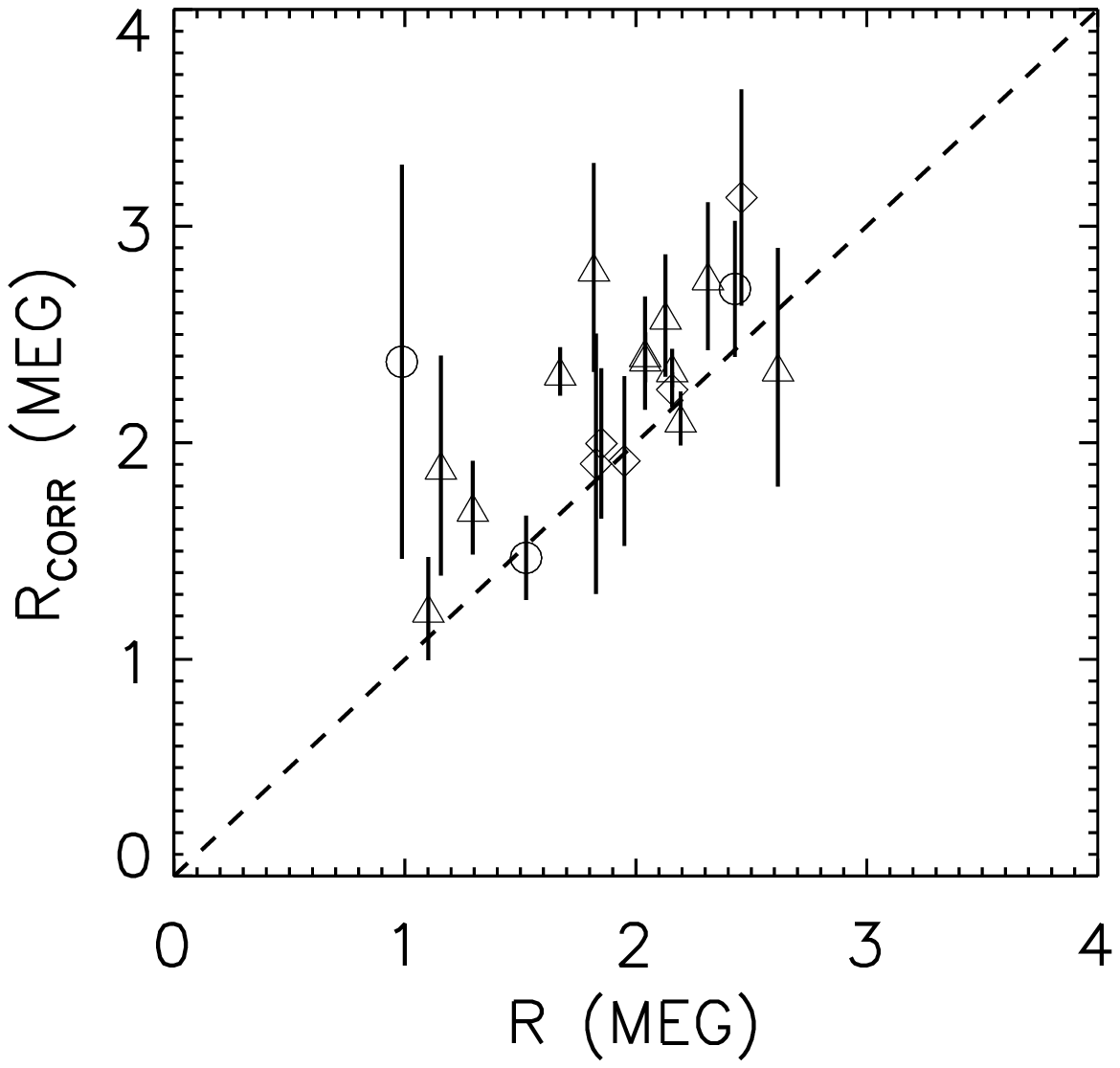,width=5.5cm}}\vspace{-0.6cm}
\centerline{\psfig{figure=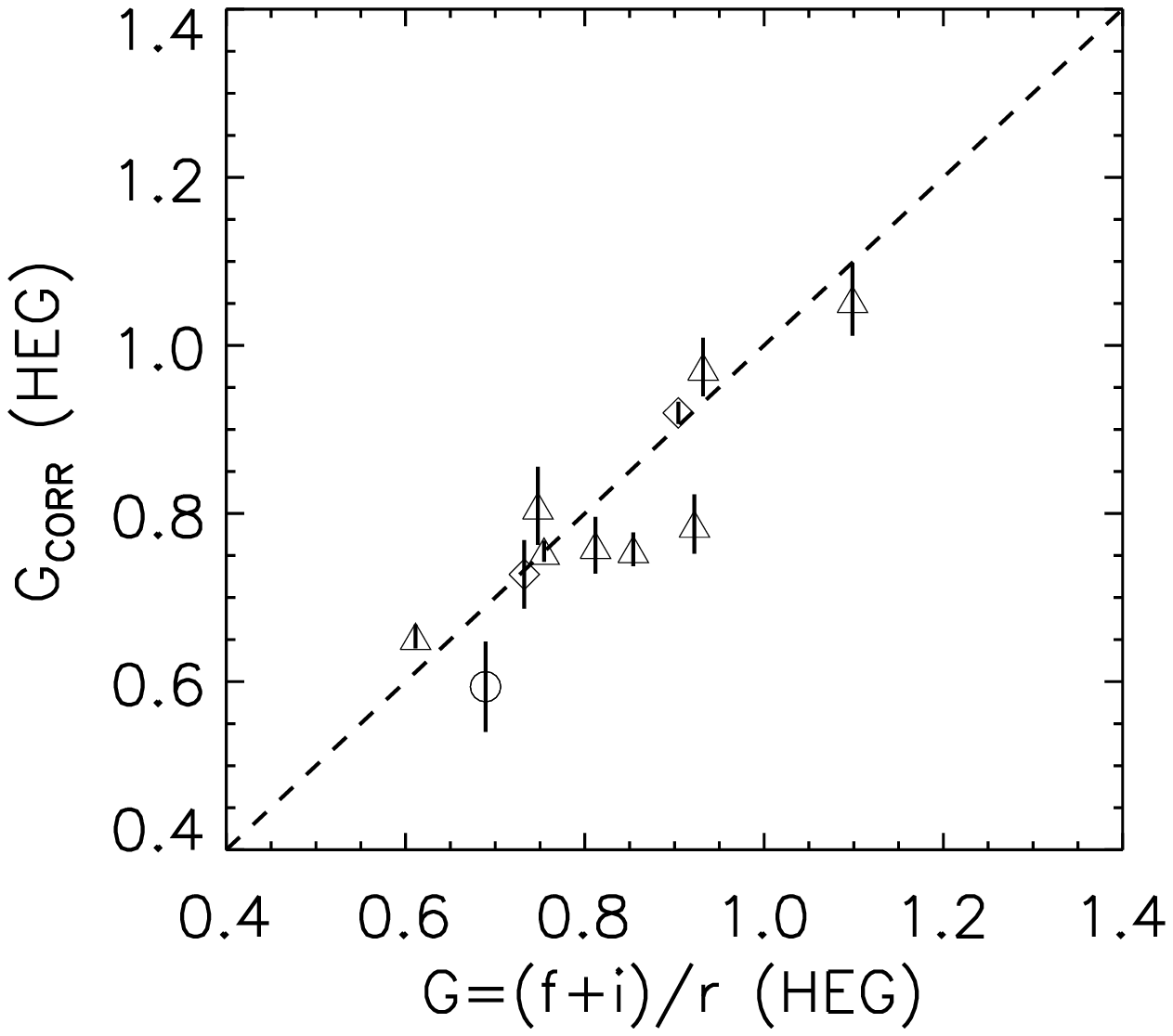,width=5.5cm}\hspace{-1cm}
            \psfig{figure=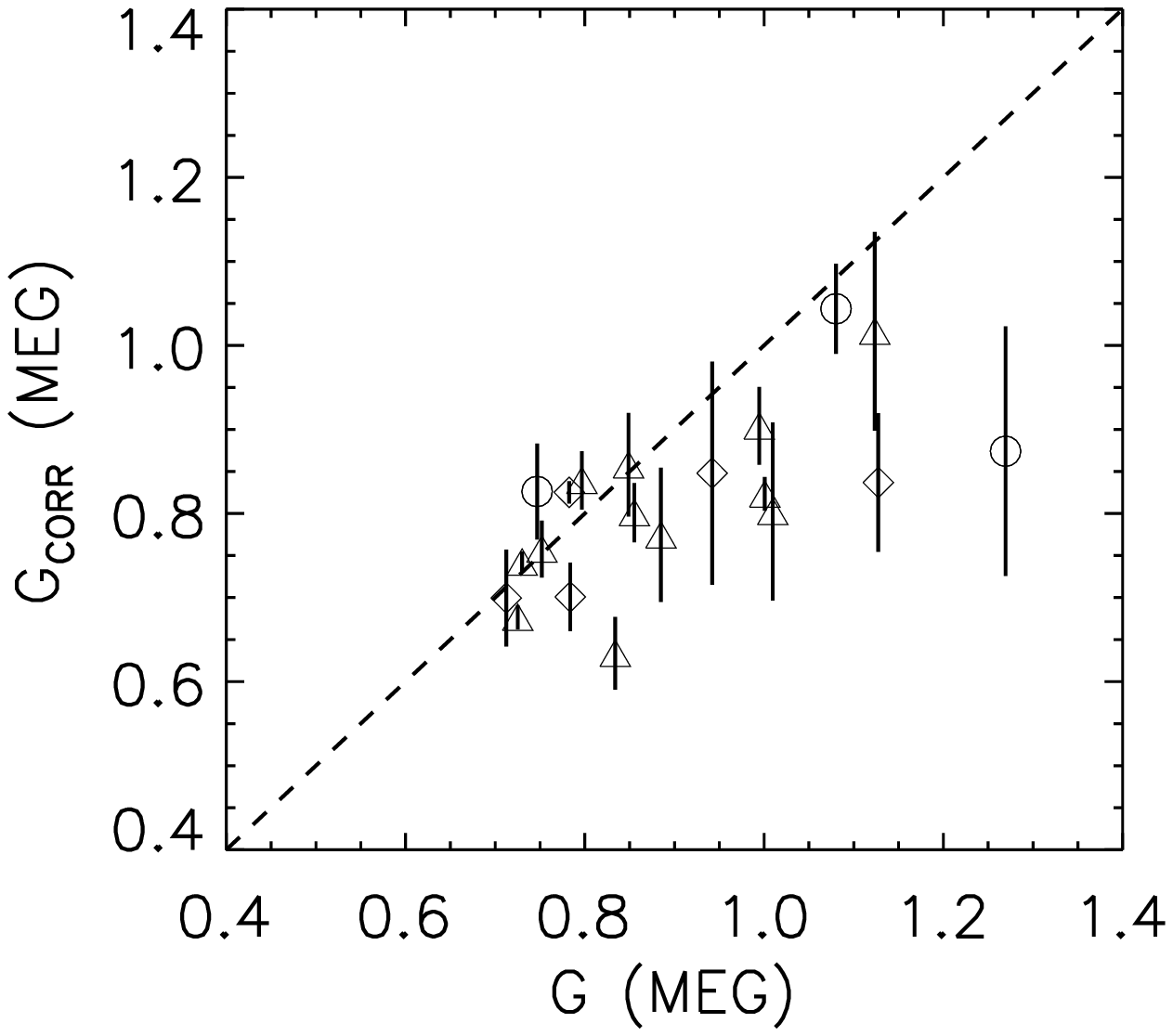,width=5.5cm}}
\caption{Values of R ({\em upper panels}) and G ({\em lower 
	panels}) ratios for Mg~XI, obtained taking into 
	account the blending with the Fe and Ne lines, 
	plotted	vs.\ the values derived not considering the
	blending.  The dashed lines mark the locus of 
	equality.  Different symbols are used for different 
	source types: {\em circles} for single dwarfs,
	{\em diamonds} for giants, {\em triangles} for 
	binary systems. \label{fig8}}
\end{figure}

We compare in Figure~\ref{fig8} the R and G ratios
obtained for the Mg~XI triplet in each star for the case in which only
the continuum and Mg~XI lines themselves were included in the line
fitting with the ratios obtained when we fitted the region using the model
including blends.  As expected, because the intensity of $i$ is
generally diminished compared to that of $f$ in the latter approach,
the effect is of raising the R ratio and slightly decreasing the G ratios
compared to the former approach, i.e.\ including blends implies shifts
toward lower derived densities and slightly higher temperatures.

\begin{figure}[!h]
\centerline{\psfig{figure=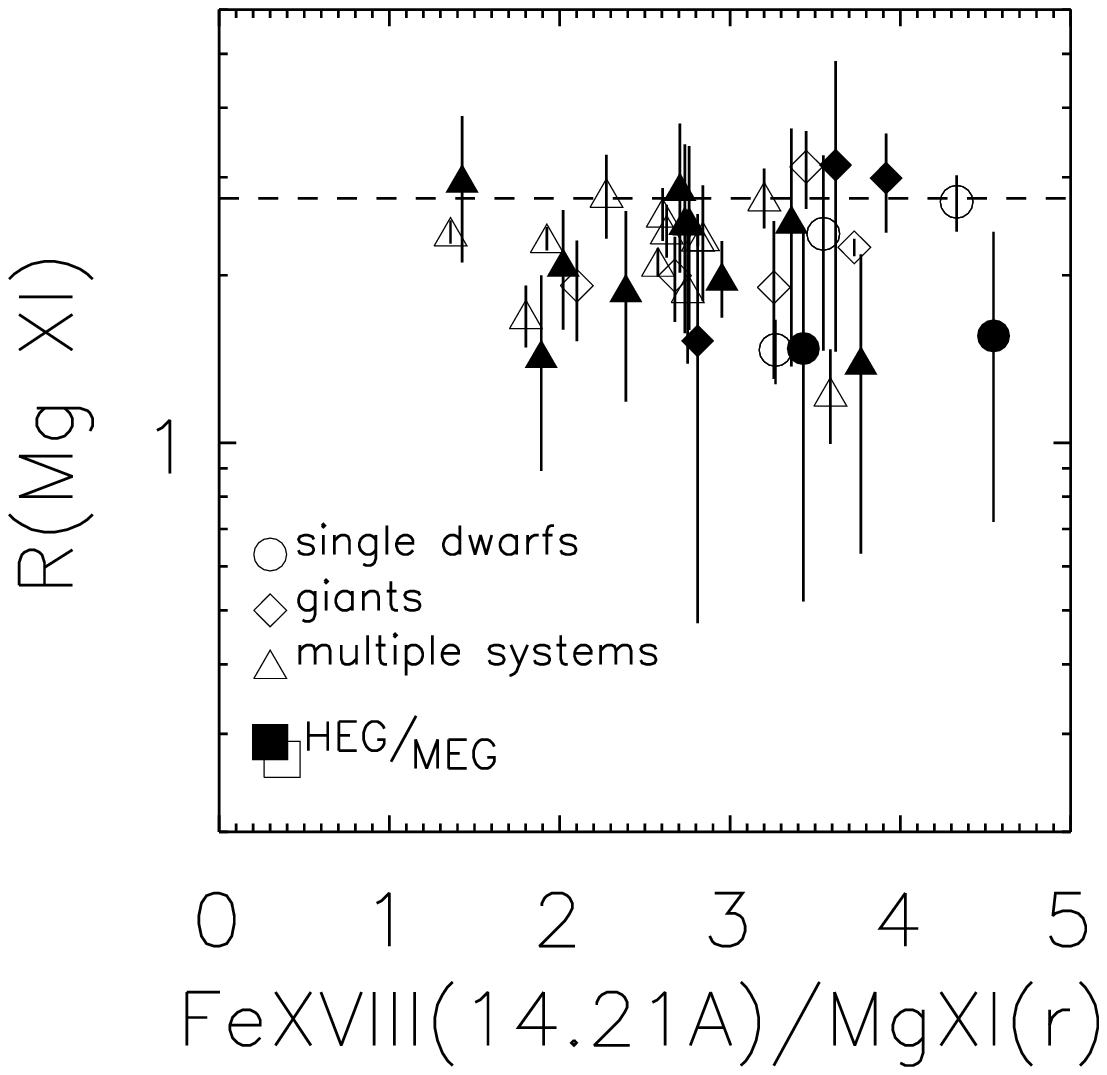,width=6.7cm}\hspace{-1.4cm}
	    \psfig{figure=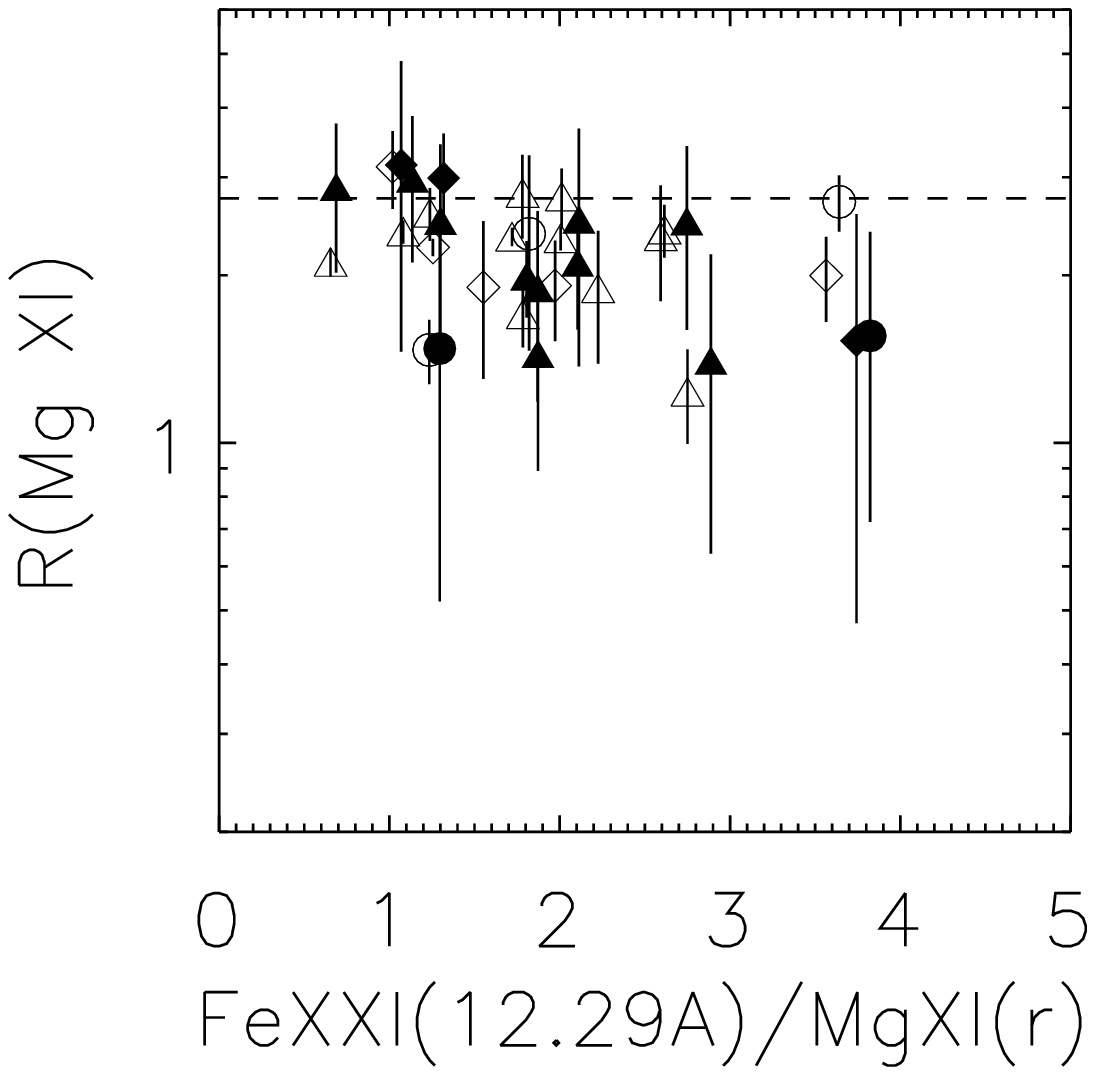,width=6.7cm}\hspace{-1.4cm}
	    \psfig{figure=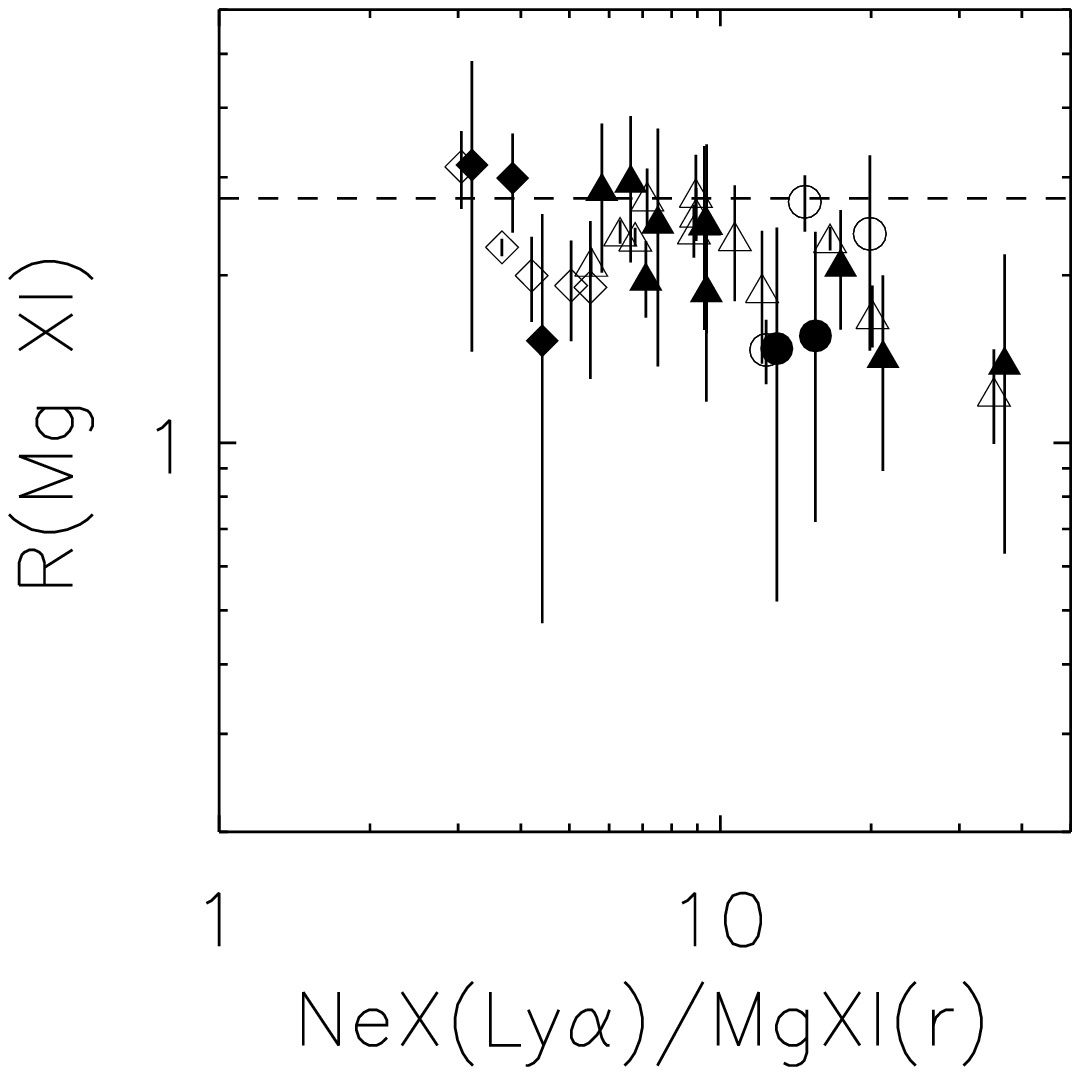,width=6.7cm}}
\caption{Investigation of possible correlations of the R ratio 
	with Fe	or Ne intensity in order to check for blending 
	effects on Mg triplet lines. The dashed line marks the 
	limiting value of the R ratio; lower values of R 
	correspond to higher density (see Fig.~\ref{fig6}).
	Data points from HEG measurements 
	are shifted by +10\% along the x-axis, to distinguish 
	the error bars corresponding to HEG and MEG measurements 
	for the same source, otherwise overlapping. \label{fig9}}
\end{figure}

As an additional test that our final Mg line fluxes, and especially
those of the density-sensitive $i$ and $f$ pair, are not
significantly affected by unseen blends, we have investigated the
observed $f/i$ ratios as a function of both Ne and Fe relative line
strength.  As a measure of relative line strengths of Ne and Fe, we
chose the ratios of the Mg XI resonance line with Fe~XXI~12.29~\AA,
Fe~XVIII~14.21~\AA, and Ne~X~12.13~\AA.
The two lines of iron at different ionization stages provide tests 
at different temperatures, which could be important since the 
thermal structure of our sample stars that cover a fairly wide
range in X-ray activity might be quite varied.  The observed R ratios
are illustrated as a function of our control ratios in Figure~\ref{fig9}.

We note from Figure~\ref{fig9} that the relative
intensities of the chosen Fe and Ne lines with respect to the Mg
resonance line span a wide range of factors of 4-5 (Fe XVIII,XXI) and
$\sim 15$ (Ne X).  Both the thermal structure and the elemental
abundances of the emitting coronae contribute to produce these
differences; the Mg~XI resonance line, for example, is formed at
temperatures intermediate between those of Fe~XVIII and Fe~XXI.  We
might also expect some degree of correlation between coronal density
and both temperature structure and chemical composition.  

Of the three panels in Figure~\ref{fig9}, only the Mg~XI
R ratio variation as a function of Mg $r$/Fe~XVIII line strength
appears completely flat, indicating an R ratio that is independent of
Fe~XVIII line strength.  This is encouraging because Fe~XVIII can be
considered a rough proxy for the coronal Fe abundance, and in the case
where remaining Fe blends were affecting our derived R ratios, we would
expect to see some correlation between the two.  The hidden Fe lines
that we might expect to blend with the Mg lines come from slightly
higher ionisation stages, however.  In the case of Fe~XXI, the R ratio
appears somewhat higher for the very lowest Fe~XXI/Mg~$r$ values, with
a step to lower R ratio for Fe~XXI/Mg~$r > 1.3$.  While the reason for
this is not clear---as we emphasise above, both temperature structure
and relative abundances play a role in the Fe/Mg and Ne/Mg ratios---we
discount blends as the culprit because the trend in R ratio remains
flat while Fe~XXI/Mg~$r$ varies by a factor of 3-4.  The
Fe~XXI~12.29~\AA\ line is also sensitive to density, and decreases in
intensity with densities above several $10^{12}$~cm$^{-3}$.  While
this could contribute to the trend in Figure~\ref{fig9}, we do not
expect this to be a major effect since densities of order
$10^{13}$~cm$^{-3}$ or more would be required; in the analysis
described below we rule out such high
densities based on our observed Si~XIII R ratios. 

Perhaps the most interesting trend is that between R and Ne~X/Mg~$r$:
there does appear to be a weak correlation here, with Ne~X/Mg~$r$
increasing as the R ratio decreases.  At face value this might suggest
residual Ne blending problems; however, the Ne Lyman series lines have
very precise wavelengths and our blending model should accurately
account for them.  We conclude that the observed trend is real: that
there is probably a weak trend of increasing Ne~X line strength with
increasing R ratio.  We return to this in the discussion
(\S\ref{sec:discussion}).

\subsection{Results}
\label{sec:res}
Table~\ref{tab6} and~\ref{tab7} summarize the results for the measured 
line fluxes and $1\sigma$ uncertainties for all the stars.  
The uncertainties were obtained from a full search of parameter 
space for each spectral fit.
HEG and MEG line
fluxes were found to differ beyond their $1~\sigma$ uncertainties to a
small degree.  We first investigated to determine whether or not this
could be the result of the spectral resolution difference between
them, resulting in systematically different estimated continua, but
we were able to rule this out: the required continuum placement errors
are implausibly large.  Instead, we have found that similar flux
differences are obtained by integrating spectral flux in arbitrary
intervals across the bandpass.  We attribute this to effective area
calibration errors, which seem to be at the 10-20\%\ level at this
time.\footnote{http://space.mit.edu/CXC/calib/hetgcal.html} Flux
differences across the bandpass vary only very slowly with wavelength
such that over narrow intervals the effect is grey and does not
influence the measured line {\em ratios}.

\begin{figure}[!h]
\centerline{\psfig{figure=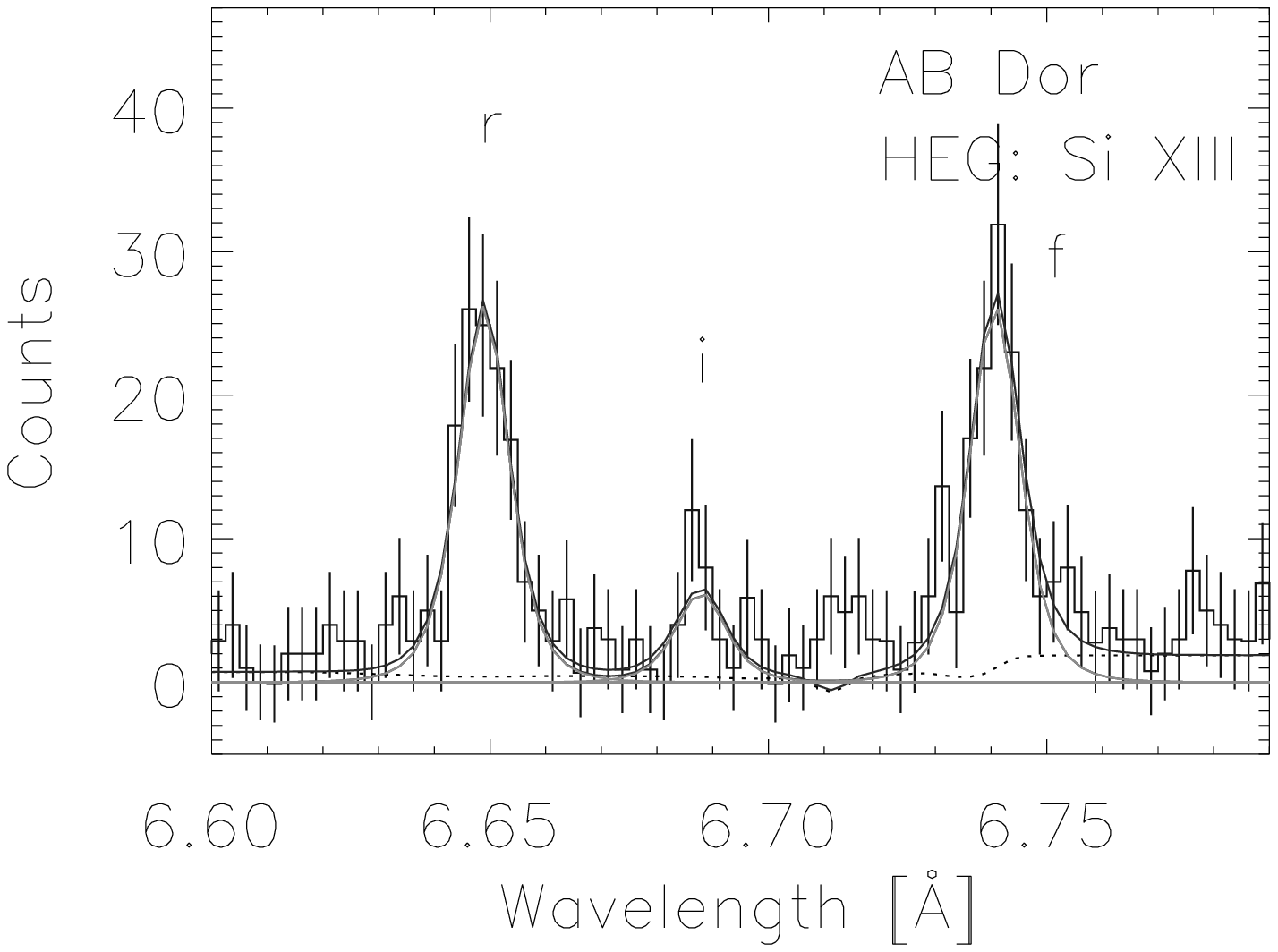,width=6cm}\hspace{-0.8cm}
	    \psfig{figure=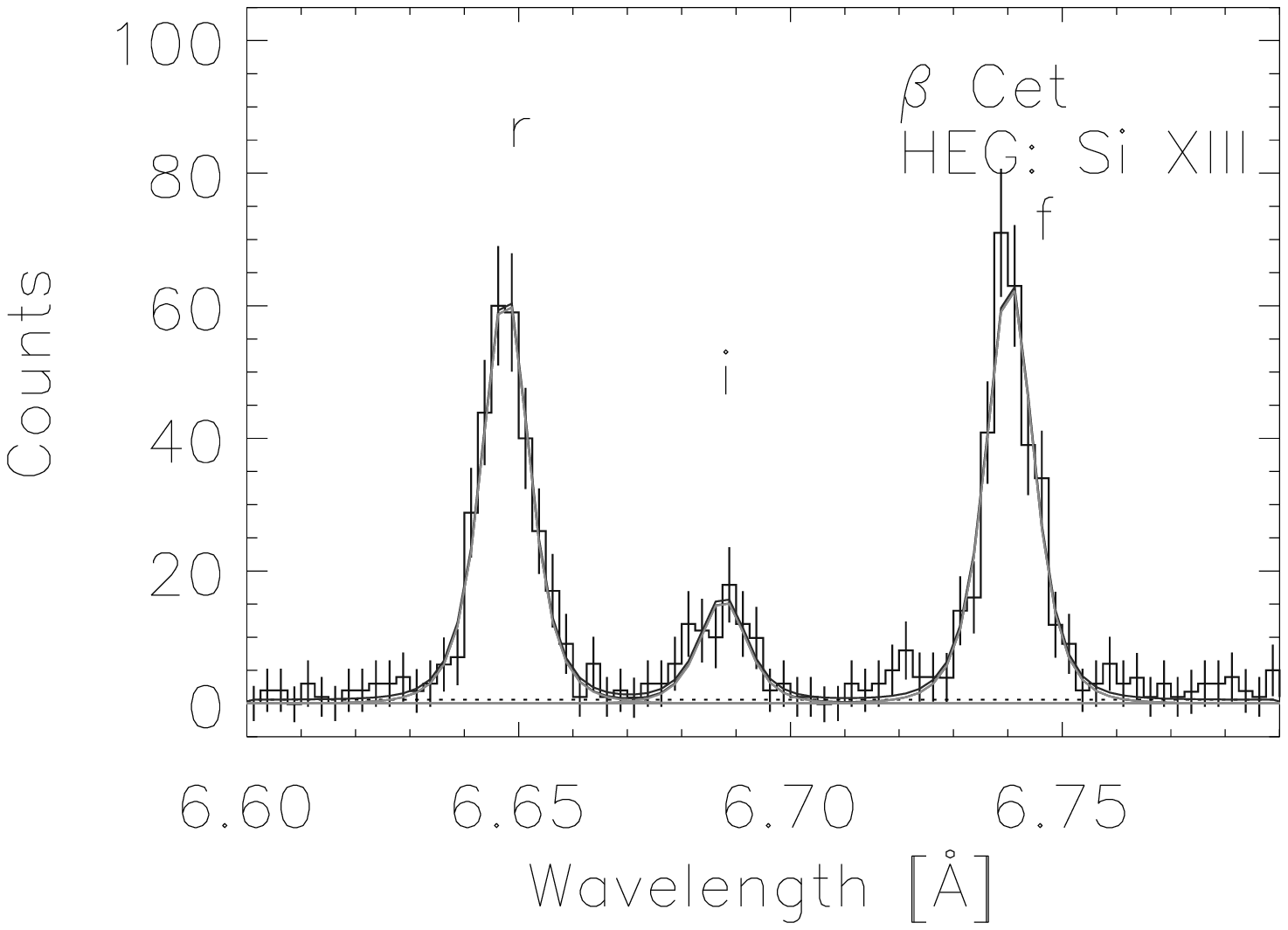,width=6cm}\hspace{-0.8cm}
	    \psfig{figure=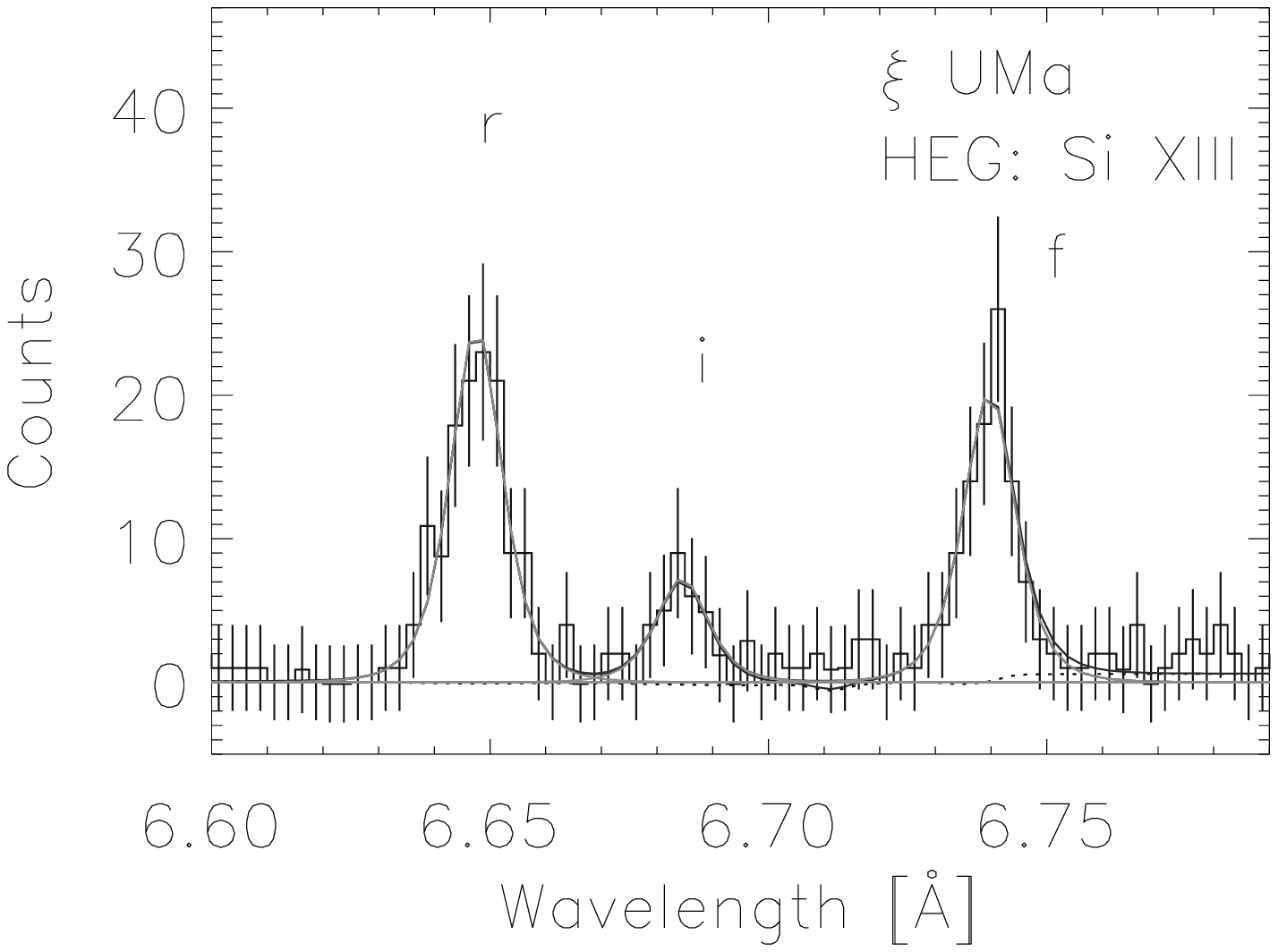,width=6cm}}
\caption{The Si~XIII He-like triplet, as observed with the HEG, and
	examples of spectra and model fits for the three 
	different type of sources: single stars ({\em left}), 
	giants ({\em center}) and binary systems ({\em right}). 
	Data points are shown with their associated error
	bars and a solid line superimposed to the data 
	correspond to the best fitting model; we also 
	show the components of the model, i.e.\ the 
	continuum emission and the triplet emission lines. 
	\label{fig10}}
\end{figure}

Figure~\ref{fig10} shows examples of the HEG spectra in the
region of the Si~XIII He-like triplet for one star of each class:
single stars, giants and multiple systems.  We plot the different
components of the best fitting model (continuum emission and single
spectral line features), together with their sum, superimposed on the
data.

\begin{figure}[!h]
\centerline{\psfig{figure=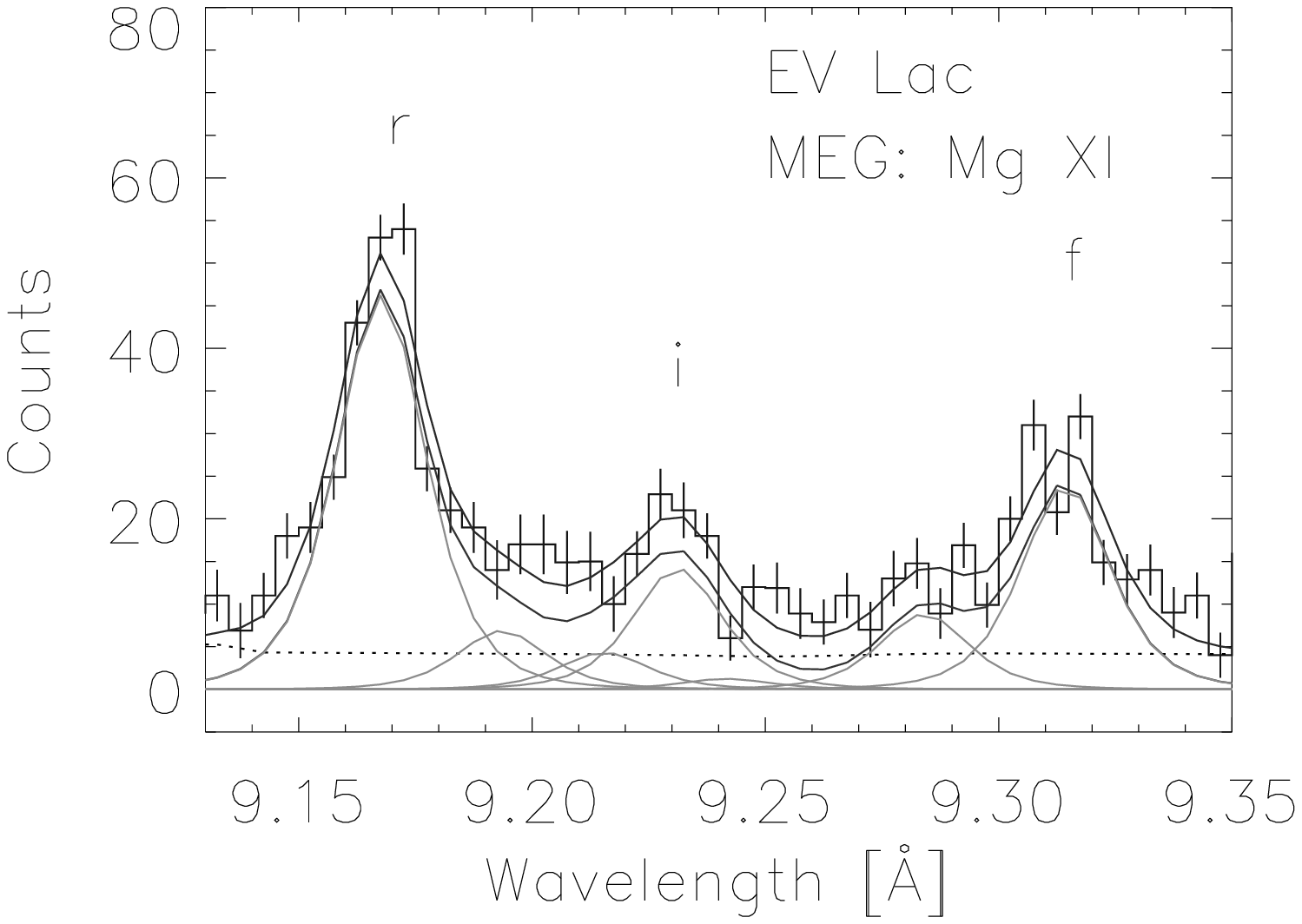,width=6cm}\hspace{-0.8cm}
	    \psfig{figure=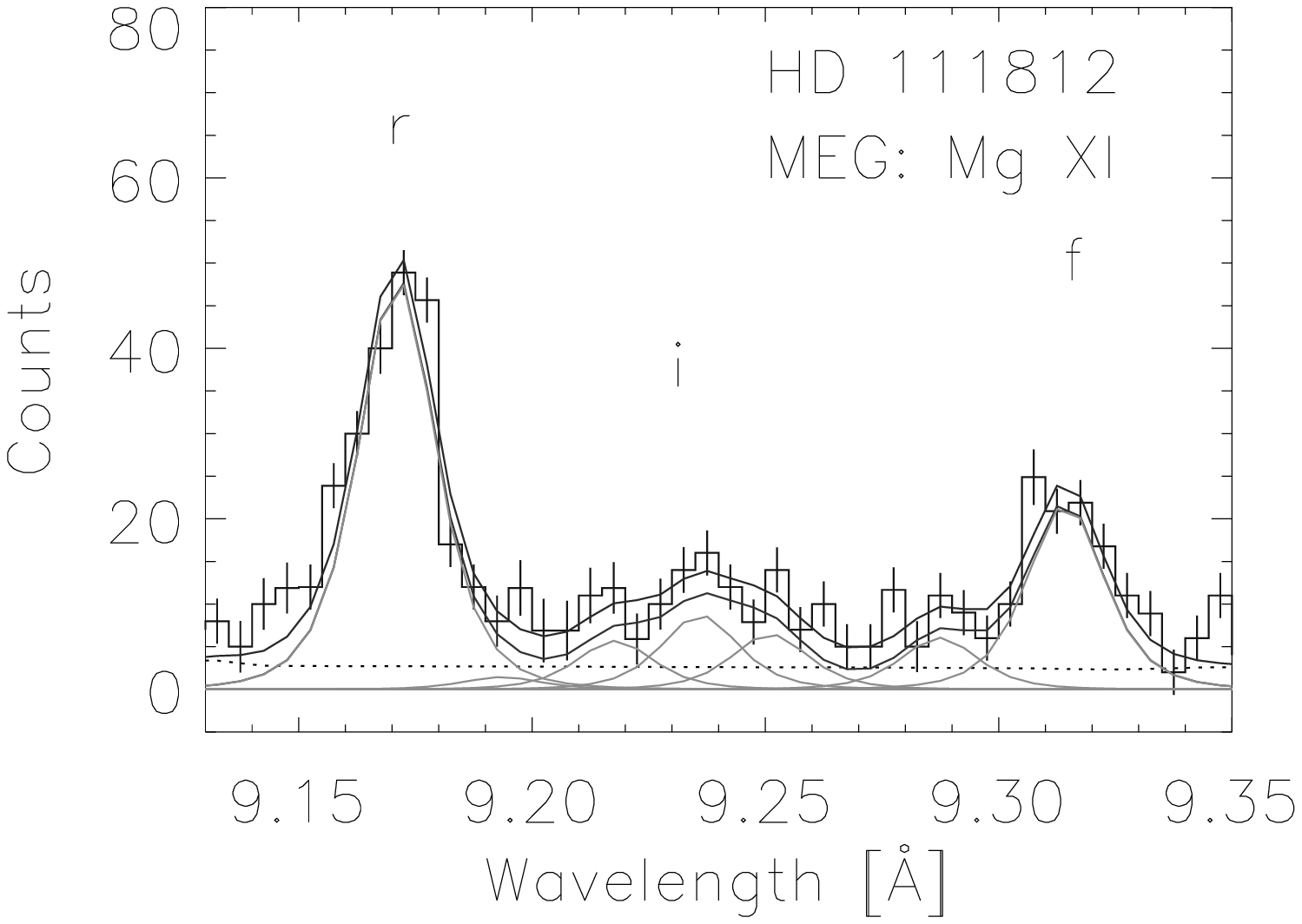,width=6cm}\hspace{-0.8cm}
	    \psfig{figure=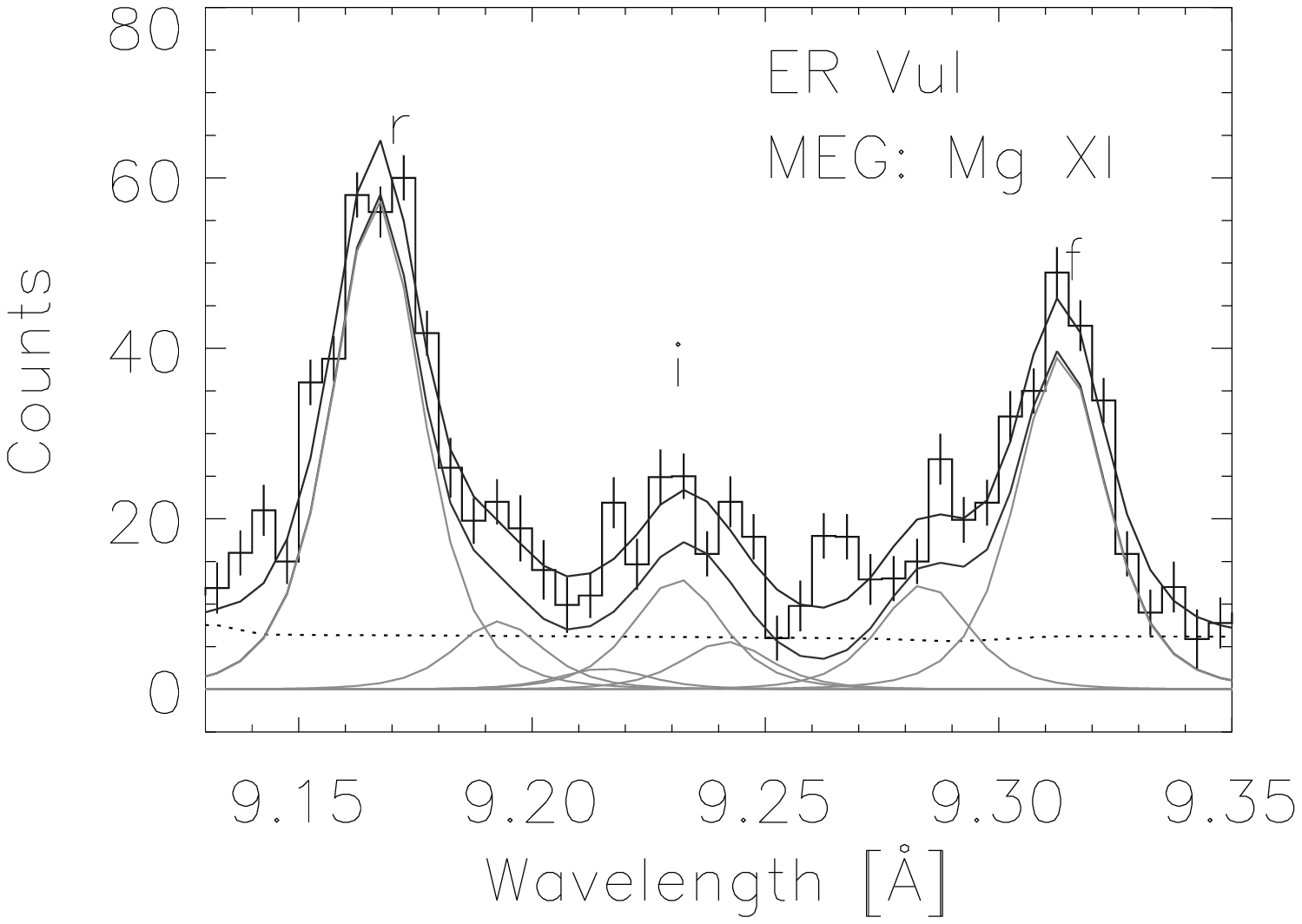,width=6cm}}
\caption{Examples of analyzed spectral regions, as in 
	Figure~\ref{fig10}, for MEG spectra in 
	the Mg~XI triplet region, for single stars 
	({\em left}), giants ({\em center}) and binary 
	systems ({\em right}). 
	\label{fig11}}
\end{figure}

\begin{figure}[!h]
\centerline{\psfig{figure=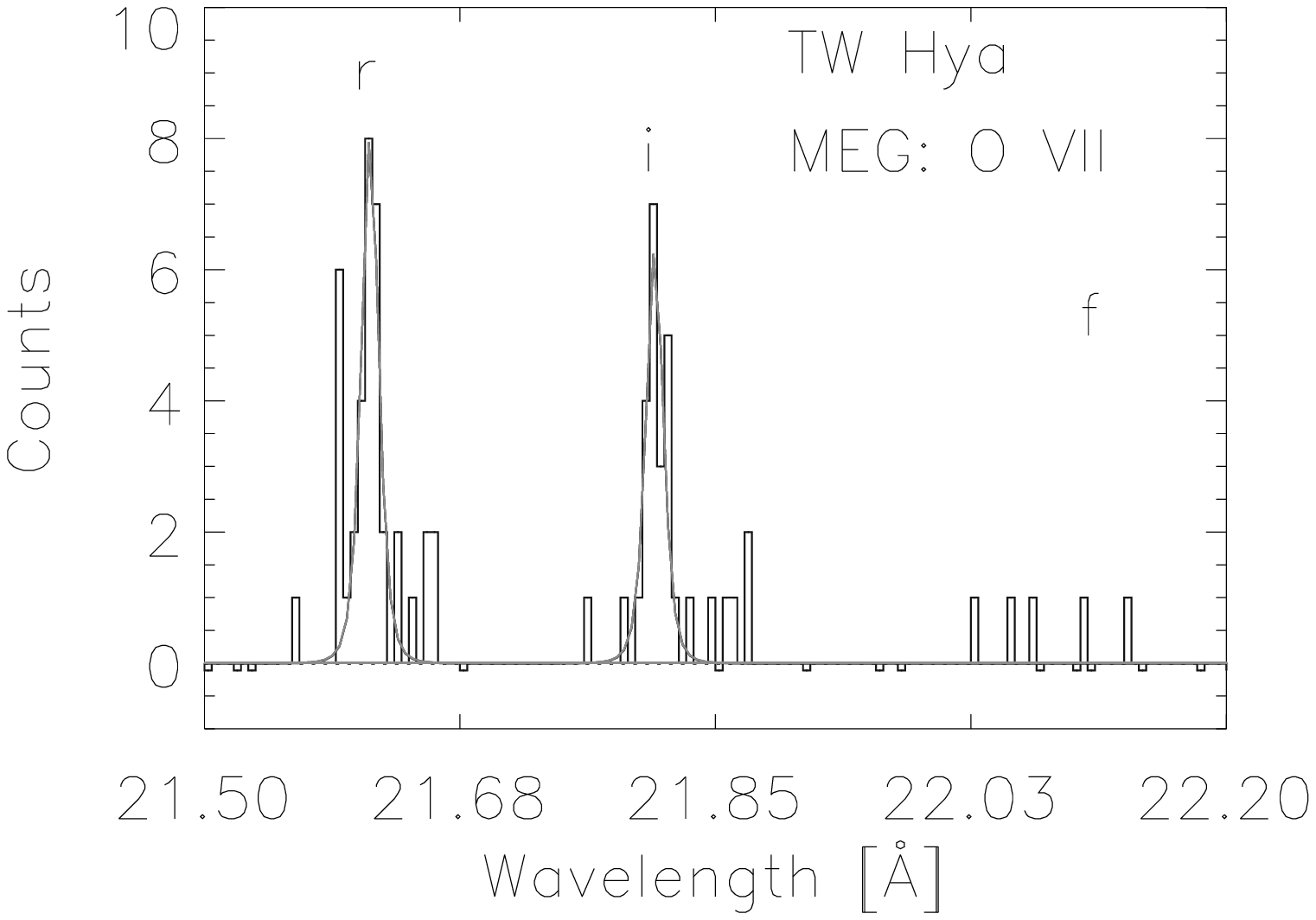,width=6cm}\hspace{-0.8cm}
	    \psfig{figure=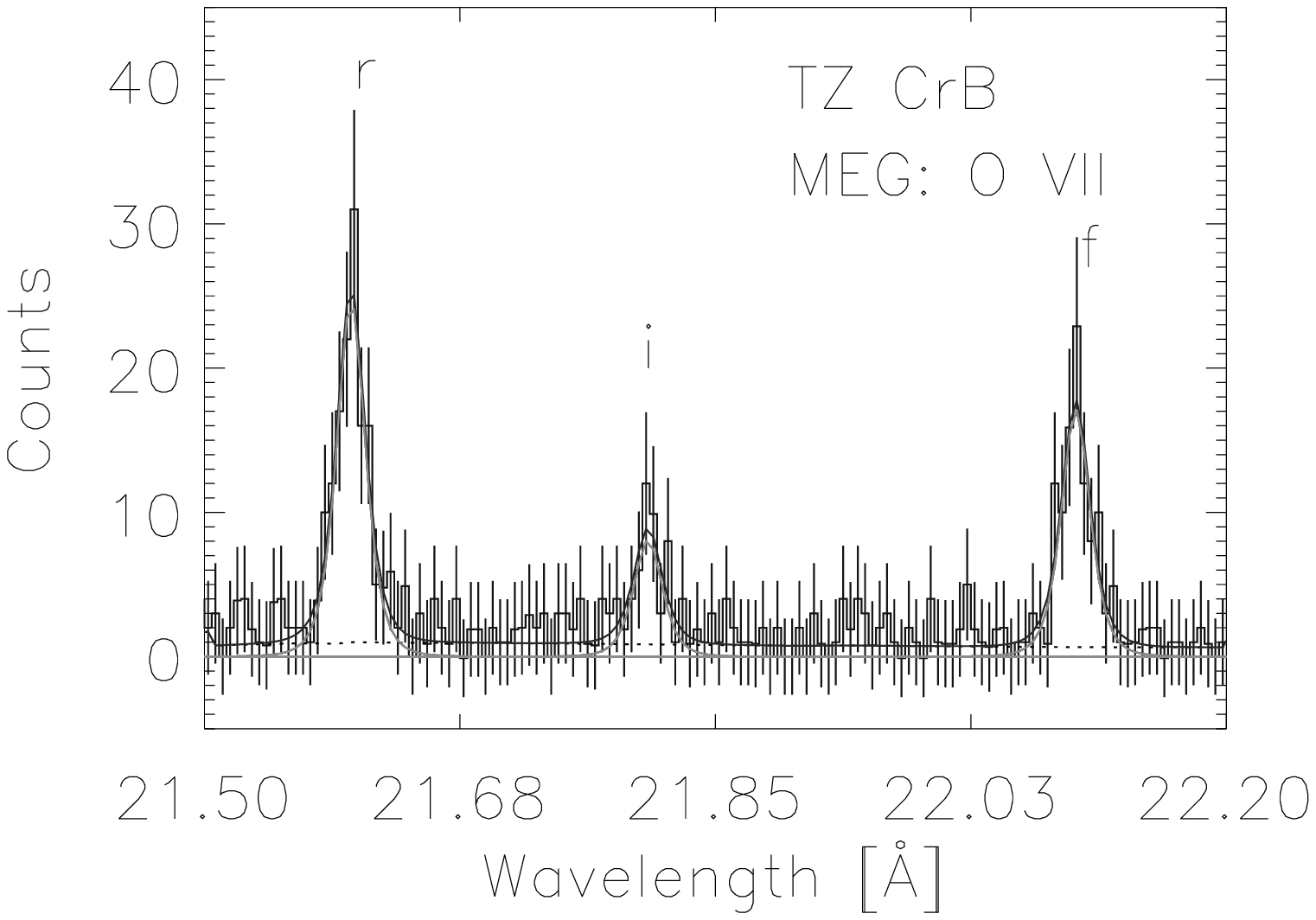,width=6cm}}
\caption{Examples of analyzed spectral regions, as in 
	Figure~\ref{fig10}, for MEG spectra of 
	the O~VII triplet region, for single stars 
	({\em left}), and binary systems ({\em right}). 
	None of the spectra of giant stars yielded 
	significant measurements.  For better readability 
	we omitted the error bars in the TW~Hya spectrum.
	\label{fig12}}
\end{figure}

Analogously, Figure~\ref{fig11} and
Figure~\ref{fig12} show examples of the MEG spectra for the
Mg~XI and O~VII He-like triplet regions, together with spectral fits.
None of the spectra obtained for giant stars yielded significant O~VII
line measurements.  The spectrum of TW~Hya is extremely interesting,
since the forbidden line is essentially not detected, and thus it is
the only case of line fluxes at the high density limit for oxygen (see
also \S\ref{sec:discussion} and Kastner et al.\ 2002).

\begin{figure*}[!ht]
\centerline{\psfig{figure=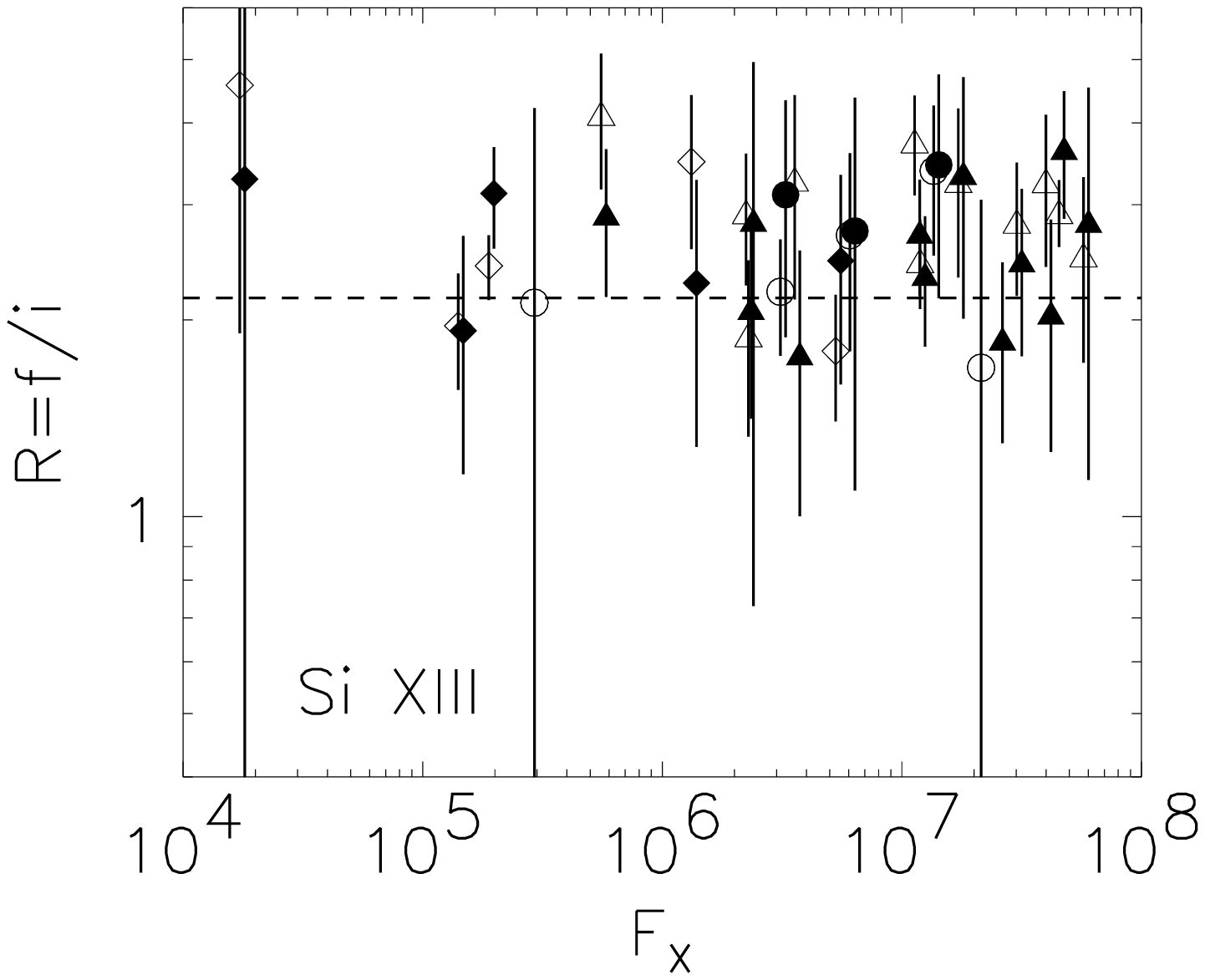,width=6.8cm,height=5.8cm}\hspace{-1.2cm}
            \psfig{figure=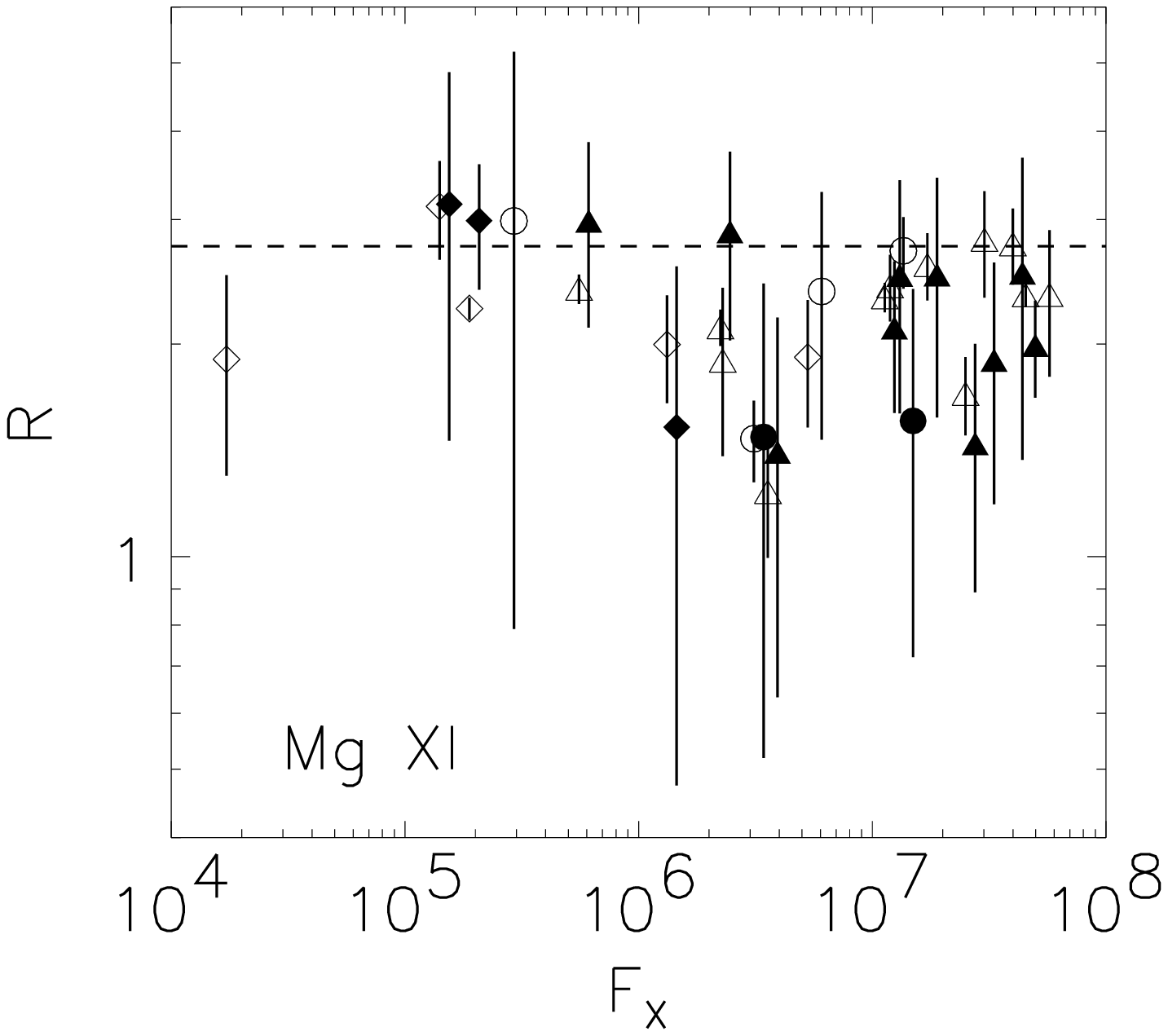,width=6.8cm}\hspace{-1.2cm}
            \psfig{figure=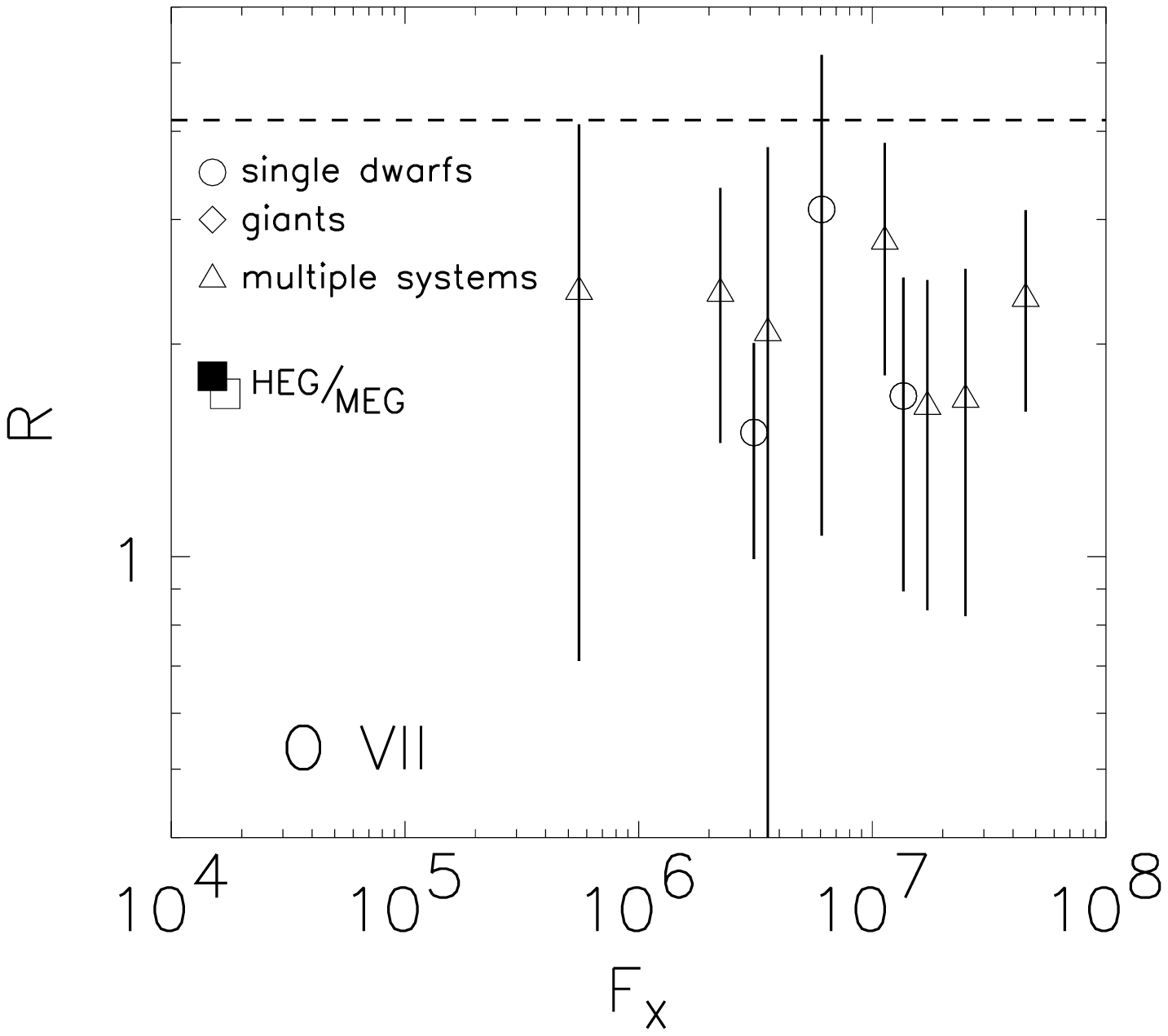,width=6.8cm}}\vspace{-0.7cm}
\centerline{\psfig{figure=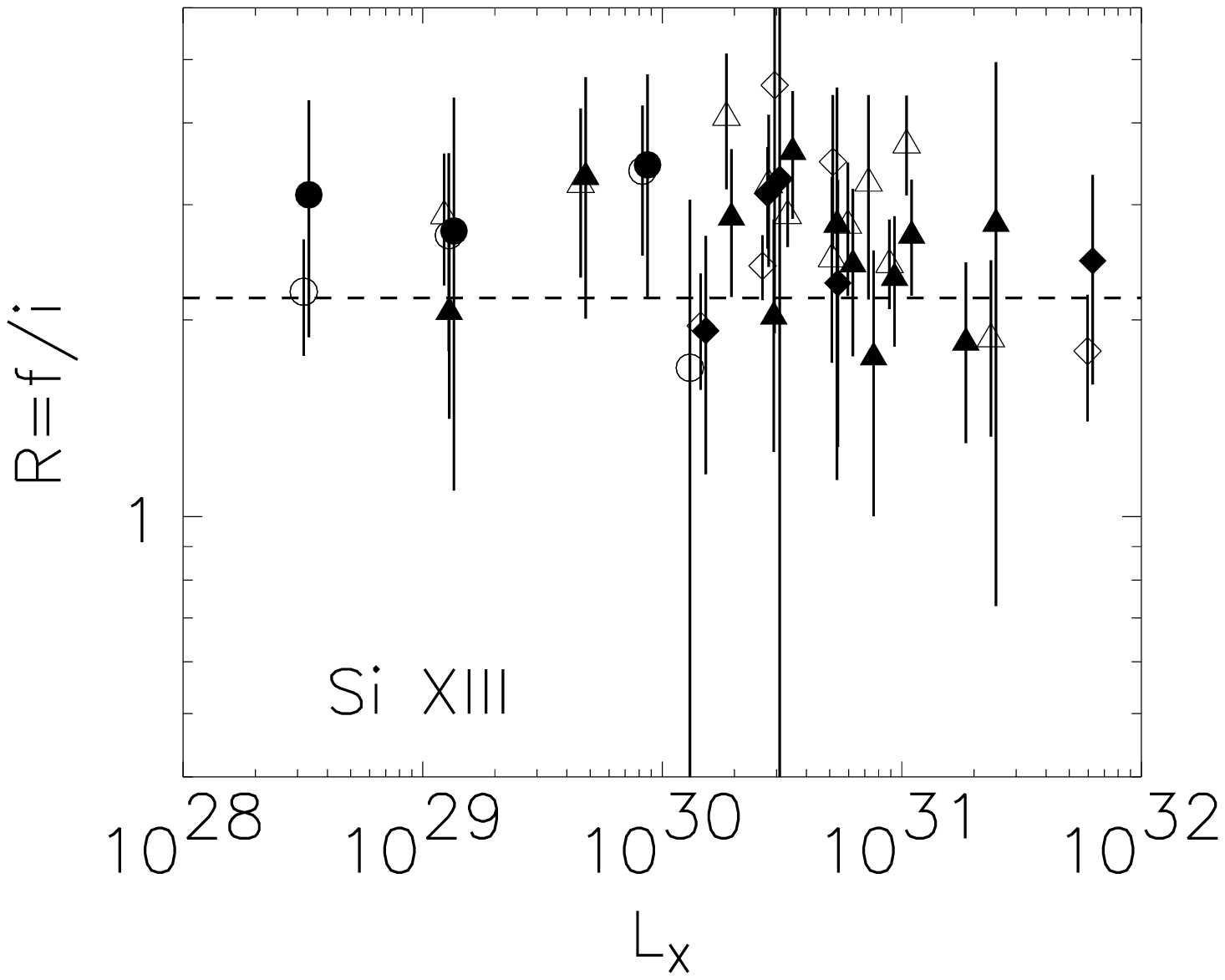,width=6.8cm,height=5.8cm}\hspace{-1.2cm}
            \psfig{figure=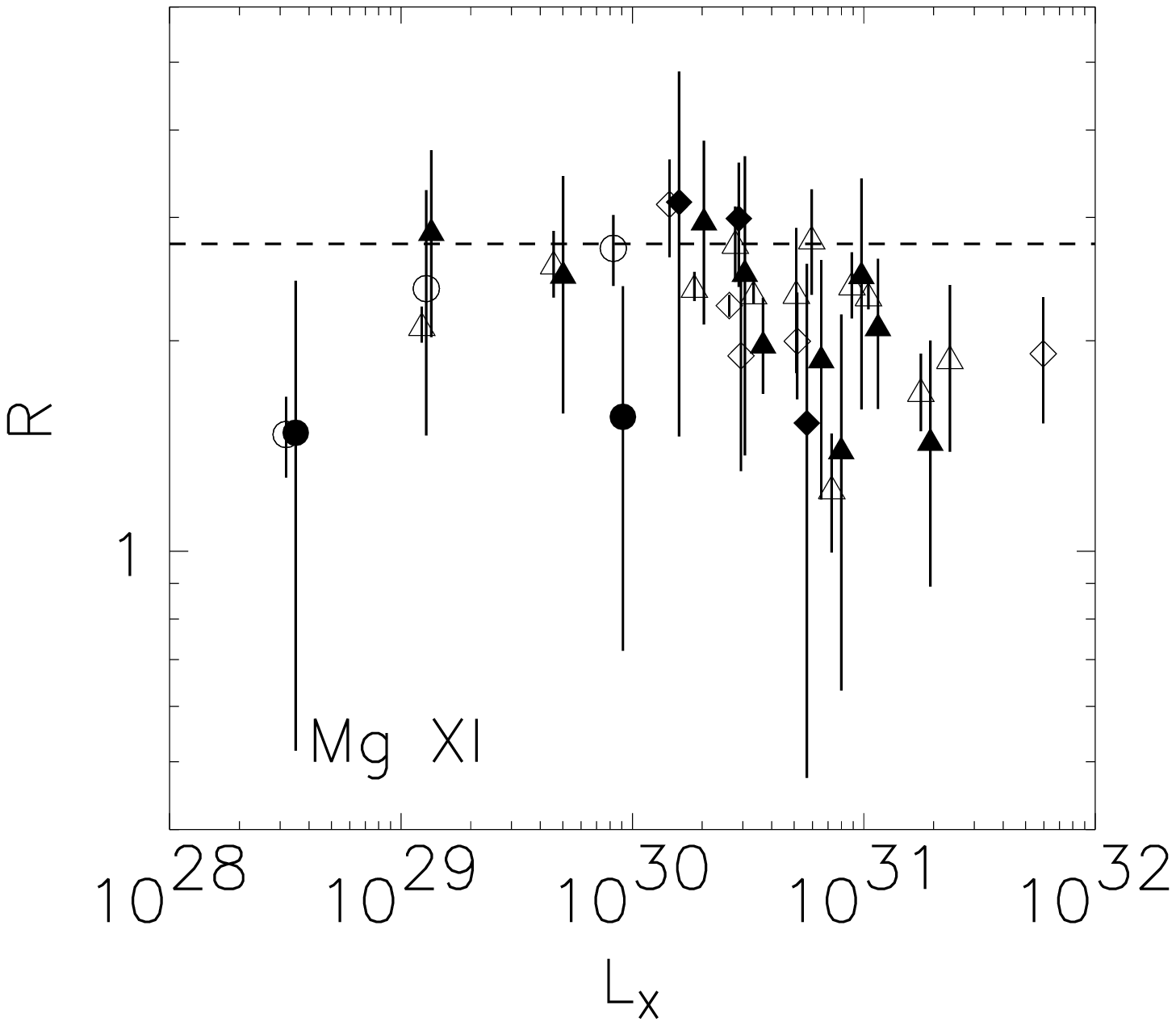,width=6.8cm}\hspace{-1.2cm}
            \psfig{figure=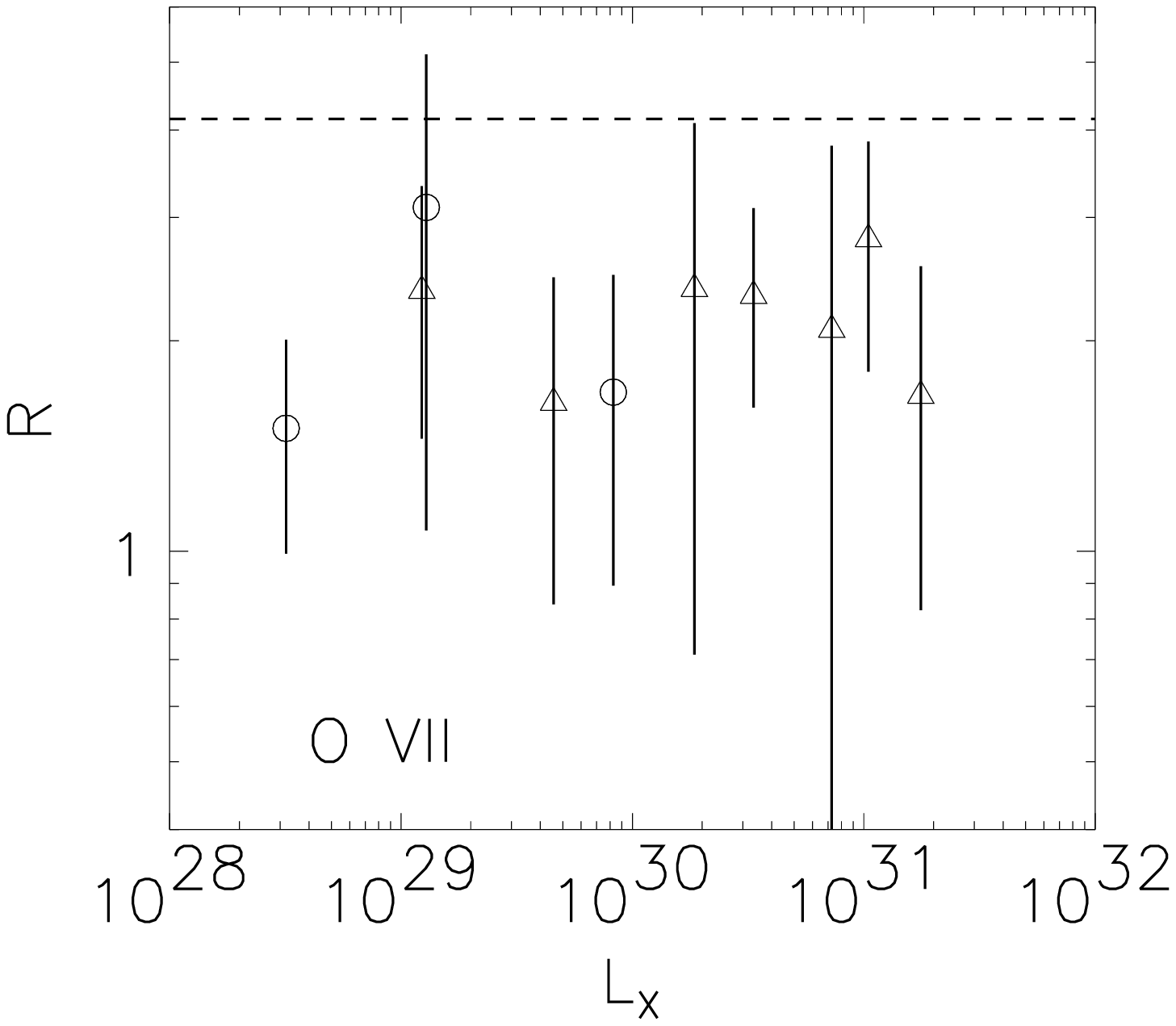,width=6.8cm}}
\caption{Measured R ratios for the He-like emission
	triplets, Si~XIII ({\em left}), Mg~XI ({\em center})
	and O~VII ({\em right}), from both HEG (filled) and
	MEG (empty) data, vs.\ the stellar X-ray surface flux  
	({\em upper panels}) and the X-ray luminosity
	({\em lower panels}) for all the analyzed stars. 
	The dashed line marks the limiting value of the R 
	ratio; lower values of R correspond to higher density 
	(see Fig.~\ref{fig6}). Symbols as in Figure~\ref{fig9}.
	\label{fig13}}
\end{figure*}

Figure~\ref{fig13} summarizes all the results obtained
for the R ratios for all the sources, plotted as a function of surface
X-ray flux and X-ray luminosity.  Some of the ratios, mostly from HEG
spectra, are missing due to the low S/N ratio of the relevant features
(see Table~\ref{tab6}).  We use different symbols for
different classes of sources as indicated in the figures.  It is clear
from these figures that HEG and MEG measurements are generally
compatible within experimental uncertainties.  The dashed lines of
Figure~\ref{fig13} mark the low-density limits of the sensitivity of
the different diagnostics, i.e.\ R
ratios consistent with the dashed lines to within experimental
uncertainty provide only an upper limit to the plasma density, while
those significantly below the limit provide density estimates (see
Fig.~\ref{fig6}).

At face value, all the Si~XIII R ratios are compatible with the 
low-density ratio limit, $n_{\mathrm{e}}\lesssim 10^{13}$~cm$^3$.  
However, we note that a good fraction of the measured ratios lie 
{\em above} this limit.  This suggests that either the theoretical 
low density limit R ratio is too low, or that some fraction of
our observed R ratios are spuriously high.  We return to this in 
\S\ref{sec:discussion}.  

For the Mg~XI triplet lines, formed at lower temperatures with respect 
to the Si~XIII lines, in many cases the R ratio is lower than the 
low-density limit value and yields useful density diagnostics.  
The Mg~XI R ratio vs.\ surface flux (Figure~\ref{fig13}) shows a 
discernible trend that stars yielding high densities tend to have 
higher fluxes.  One exception is the low surface flux supergiant 
Canopus ($F_{\mathrm{X}}= 0.17 \times 10^{5}$~erg/cm$^2$/s), whose 
Mg triplet lines point to high density.  
An analogous trend appears in the figures illustrating the R ratios 
vs.\ X-ray luminosity; in particular for the Mg~XI R ratio this trend 
with luminosity is even more pronounced than with surface flux.  
One exception is the late-type single star EV~Lac, which, owing to 
its late spectral type, has a relatively low X-ray luminosity but 
apparently high plasma density.

Since the He-like triplet features are mildly sensitive to
temperature, in order to derive actual density values from the
measured R ratios we need to determine the plasma temperatures at
which the density-sensitive lines are formed.  The natural choice
would be to use the temperature-sensitive G ratio; however, based on
our measurements the G 
ratio seemed to
yield temperatures systematically lower than what we would
expect. The true temperature of line formation depends on both the
line 
emissivity as a function of temperature (that peaks at $\sim 6.8$MK
in the case of Mg~XI and $\sim 2$MK for O~VII), and 
on the thermal structure of
the corona, i.e.  the emission measure distribution; since most of
these active stars have emission measure distributions peaked around
$10^7$~K or even higher (e.g., Ayres et al.\ 1998 for some giants,
Drake et al.\ 2001 for HR~1099, 
Sanz-Forcada et al.\ 2002 for UX~Ari and AB~Dor, Huenemoerder et al.\ 
2001 for II~Peg, Singh et al.\ 1996b for AR~Lac), we then expect to
get temperatures even higher than the nominal maximum formation
temperature.  Instead, typical derived temperatures from the G ratio
were lower than those of the peak emissivities for both Mg~XI and 
O~VII:  for Mg~XI, typical derived temperatures were in the range
1-$6\times 10^6$~K, while for O~VII temperatures where
only $5\times 10^5$-$2\times 10^6$~K.

\begin{figure}[!h]
\centerline{\psfig{figure=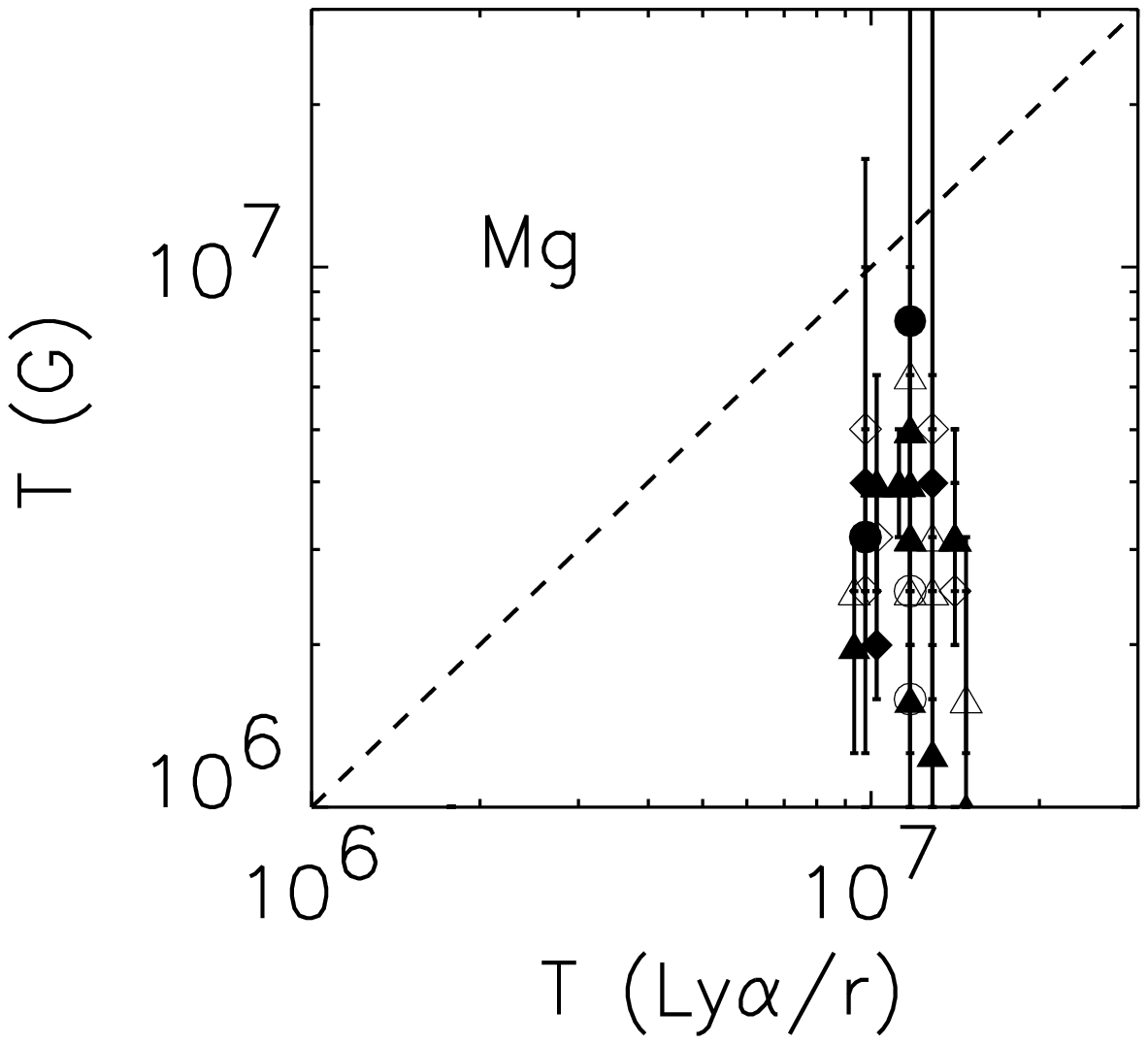,width=8cm}\hspace{-0.8cm}
            \psfig{figure=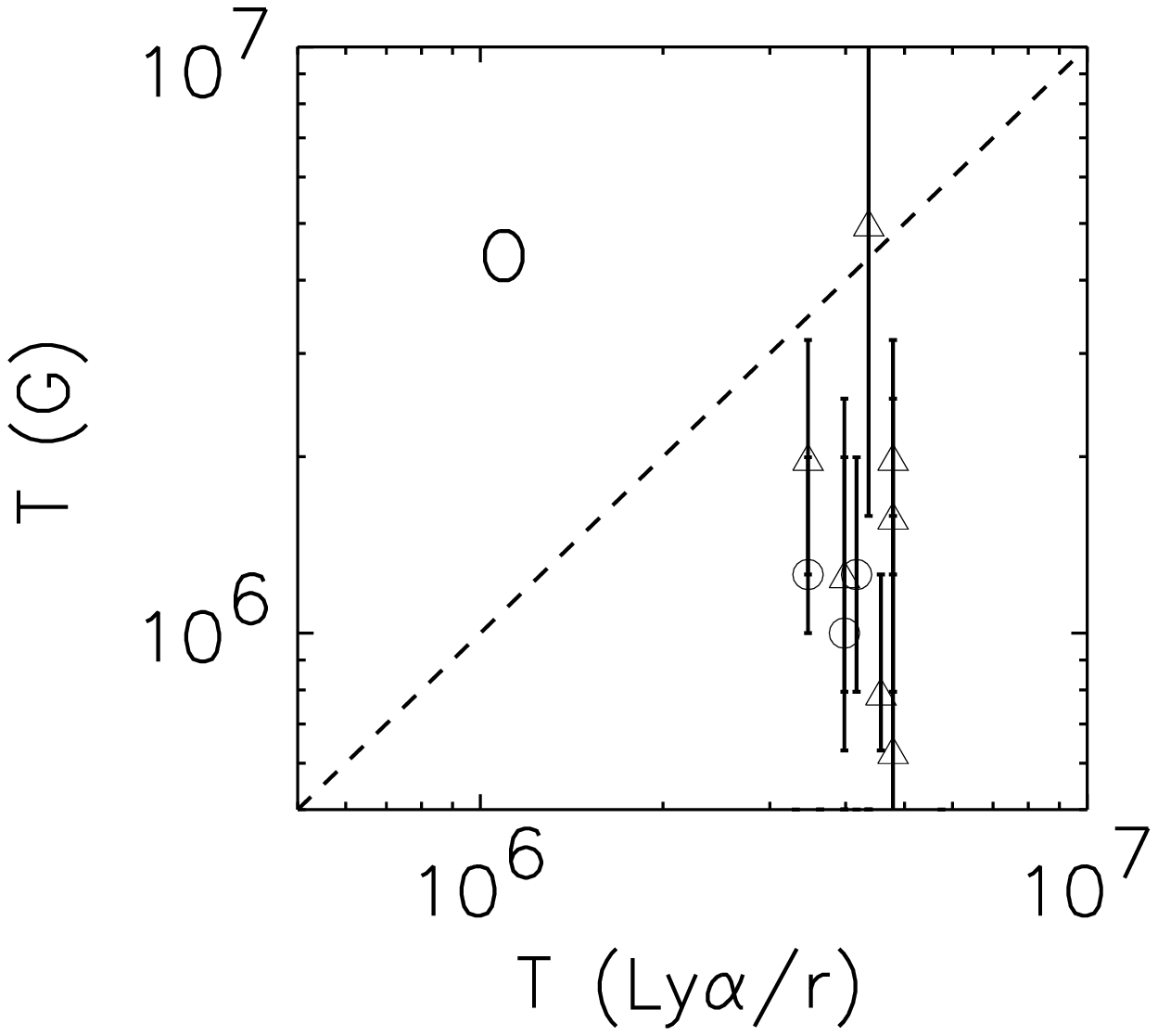,width=8cm}}
\caption{Comparison between the temperatures derived from 
	the G ratio of He-like triplet lines, and the 
	temperatures derived from the ratio of the
	Ly$\alpha$ line of the H-like ion and the resonance 
	line of the He-like ion, for Mg ({\em left}) and O 
	({\em right}). The dashed lines mark the locus of
	equal values.
	Symbols as in Figure~\ref{fig9}.
	\label{fig14}}
\end{figure}

Since the G ratio temperatures seem too low, we compared them to
ionization temperatures derived by comparing predicted and observed
ratios of the He-like to H-like Ly$\alpha$ resonance lines of O and
Ne: the representative ionization temperature is that at which an
isothermal plasma most closely matches the observed ratio.  Line
fluxes for the O and Ne Ly$\alpha$ lines were again measured using the
PINTofALE FITLINES routine, though we only examined the lines in MEG
spectra; fluxes are listed in Table~\ref{tab7}.  The comparison
between G ratio and ionization temperatures is illustrated in 
Figure~\ref{fig14}.  The G ratio yields temperatures
systematically lower than ionization temperatures by factors of
$\sim2$-4.

\begin{figure*}[!ht]
\centerline{\psfig{figure=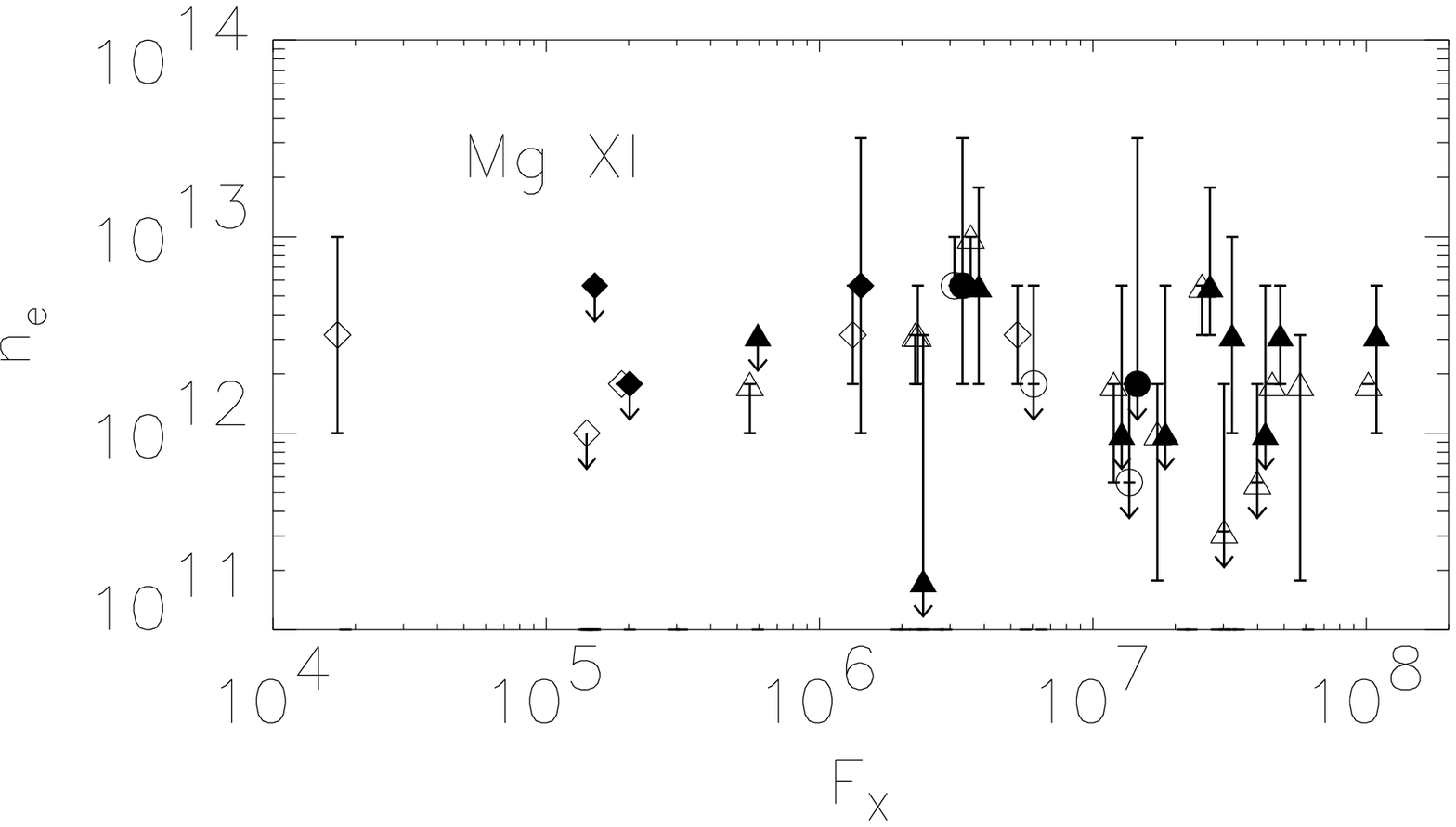,width=9cm}\hspace{-1cm}
            \psfig{figure=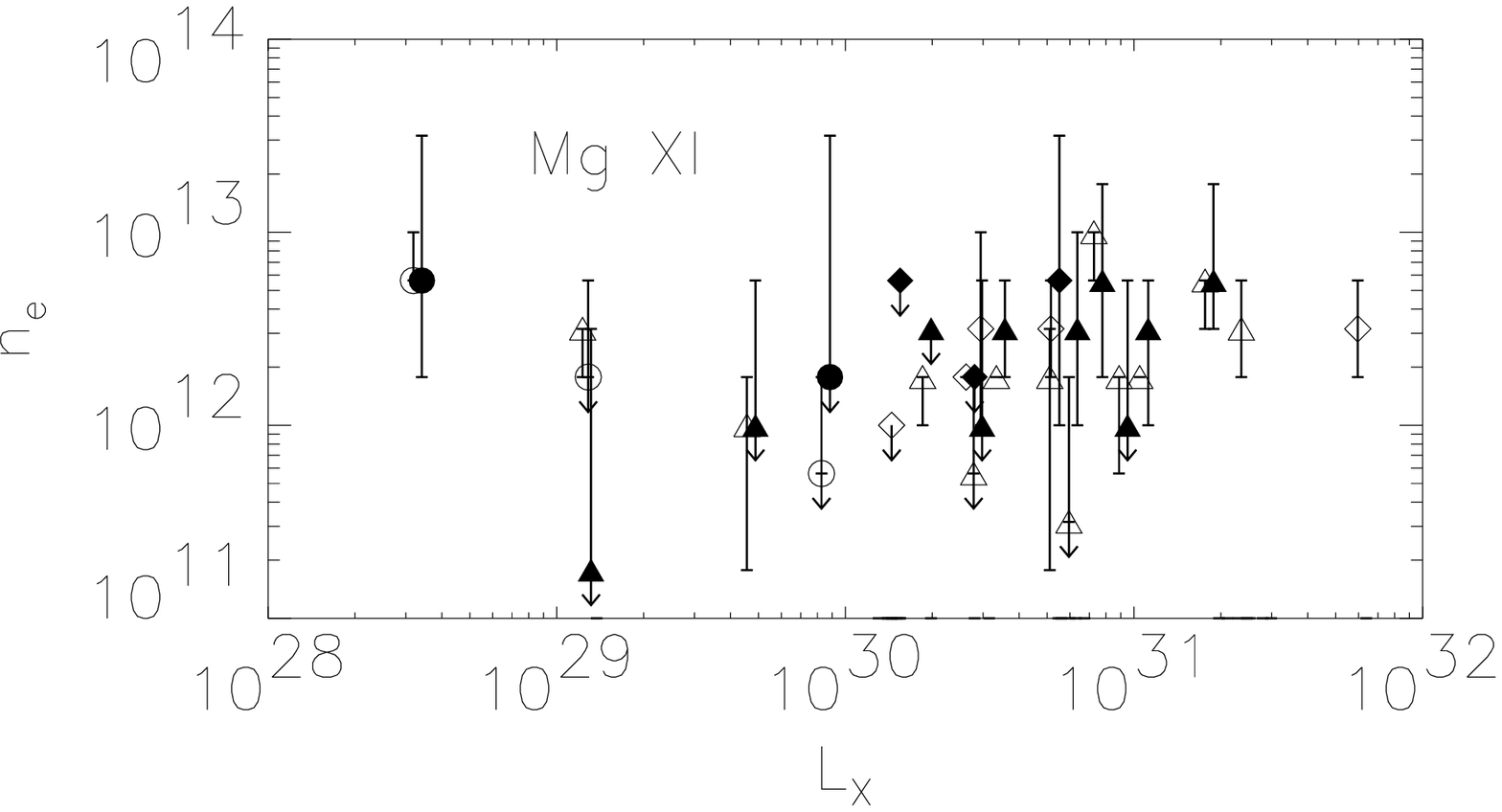,width=9cm}}\vspace{-0.8cm}
\centerline{\psfig{figure=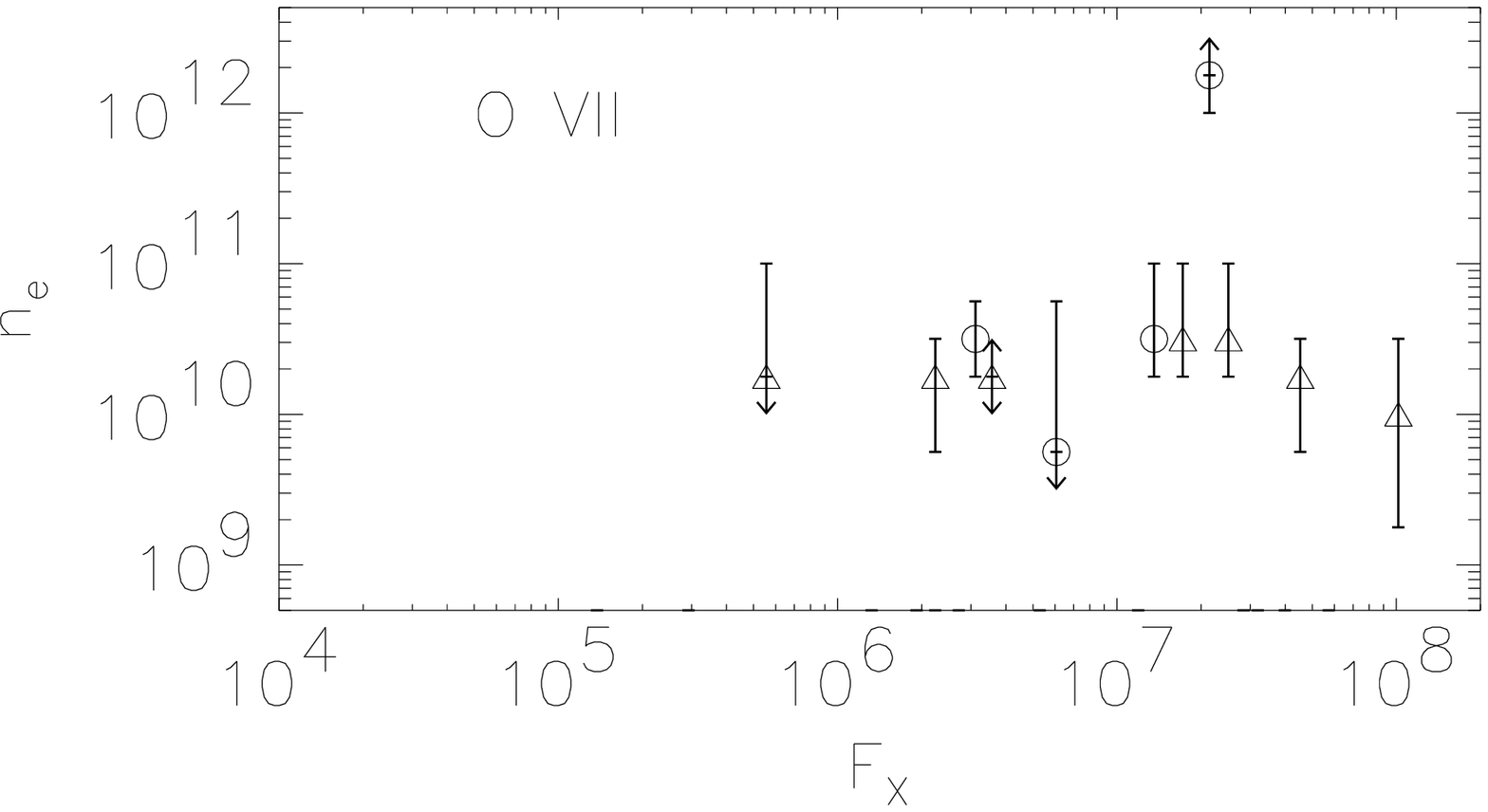,width=9cm}\hspace{-1cm}
            \psfig{figure=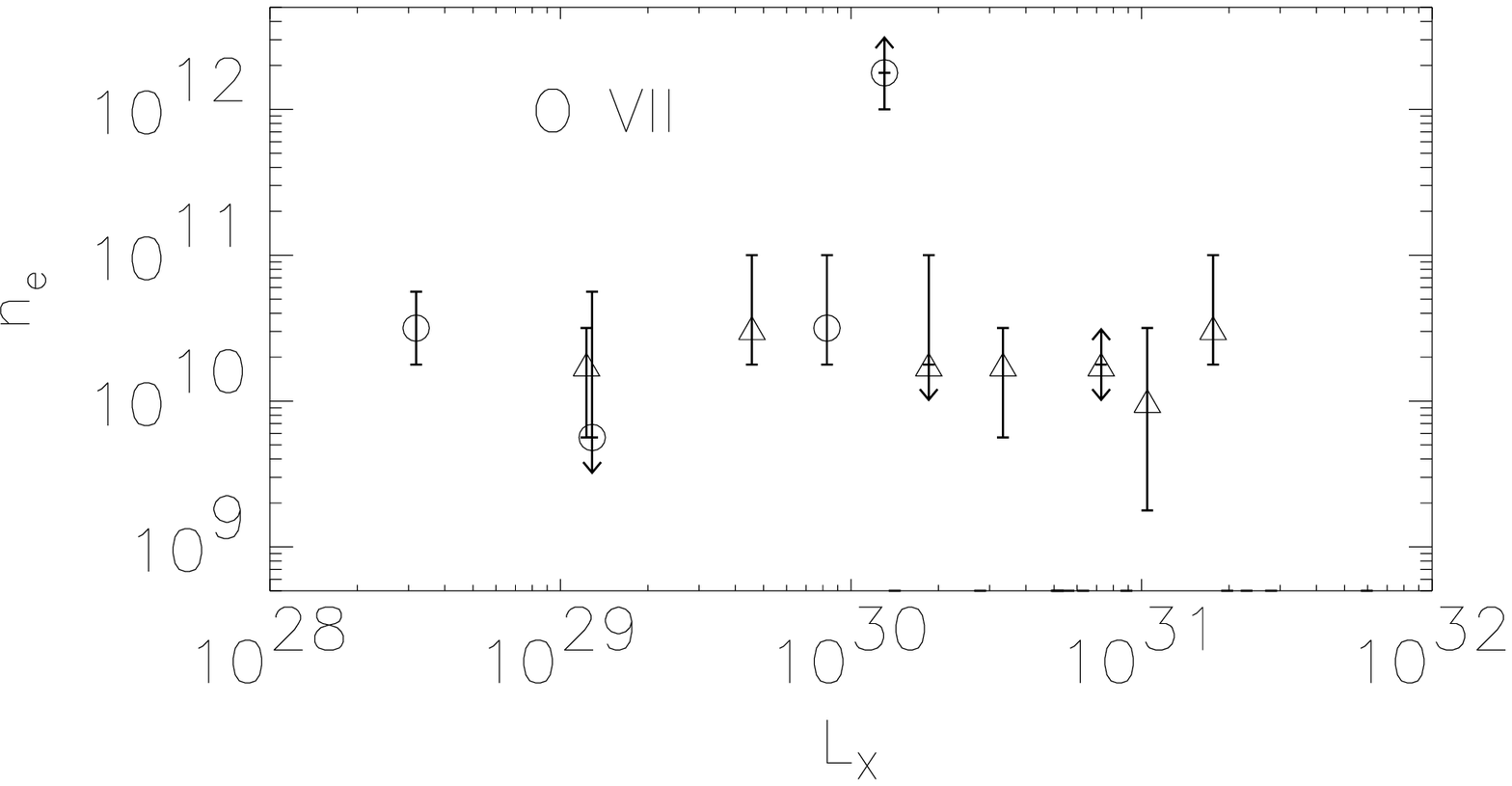,width=9cm}}
\caption{Plasma densities derived from the R ratio of He-like 
	triplets of Mg ({\em upper panels}) 
	and O ({\em lower panels}) as a function of the surface
	X-ray flux ({\em left}) and vs.\ the X-ray luminosity 
	({\em right}).	Error bars compatible with the lower 
	or the upper limit of the density range of 
	sensitivity of the R ratio are represented as down 
	and up arrows, respectively. The upper limit symbols 
	mark densities compatible with the low-density limit.
	Symbols as in Figure~\ref{fig9}.
	\label{fig15}}
\end{figure*}

The scatter in the derived ionization temperatures for the different
stars is very small and amounts to less than a factor of 2.  The
ionization temperature probably overestimates very slightly the true
mean temperature of formation of the He-like lines, since it
represents the temperature intermediate between the H-like and He-like
ions.  Since we discount the G ratio temperatures as spuriously low,
and the ionization temperatures as slightly high, for simplicity in
the derivation of densities we adopted the temperature of peak
emissivity for both O~VII and Mg~XI.  It is worth emphasising here that 
the R ratio is only weakly dependent on temperature, and electron
densities derived using all the different temperatures obtained 
with different methods yield density values that differ by a factor 
at most 2---considerable lower in most cases than the statistical 
uncertainty associated with measurement errors in the R ratio.
The final density values as derived from the R ratios are illustrated
as a function of both X-ray luminosity and surface flux 
in Figure~\ref{fig15} and listed in Table~\ref{tab8}.

\section{Discussion}
\label{sec:discussion}

The survey described here provides information on plasma densities
in the X-ray emitting coronae of a wide sample of stars.  We feel 
in particular that our measured fluxes are robust, and that the
deblending procedure discussed in \S\ref{sec:mg} provides for much
more accurate measurement of the density-sensitive Mg~XI lines than
has been presented in other studies to date.

\subsection{Atomic Data}

The derived plasma densities we will discuss below depend on the
predictions of atomic models, and it is natural to ask what 
systematic errors might be associated with converting the measured R
ratios into electron densities.  In particular, our study highlights
two potential problems with recent calculations of 
He-like emissivities for Si, Mg and O.  

In the case of Si, the APED predicted low density limit value of the R
ratio is about 50\%\ lower than what we observe, assuming the source
plasmas are indeed at the low density limit.  The Porquet et al.\ 
(2001) predicted ratio is slightly higher (Figure~\ref{fig6}), though
still discrepant with regard to our measured ratio.  Instead, for Mg
the low density limit R ratio from both APED and Porquet et al.\ 
(2001) agrees very well with the upper envelope of the observed ratios, as
we would expect if some of the coronae were at the low density limit
and some comprised higher density plasma.  The case of O~VII is more 
difficult to evaluate, since it appears that all the measurements
imply densities above the theoretical low density limits, though some
of these are of course not
statistically significant.

The temperatures derived from the APED G ratios of both Mg and O are
much lower than we would expect for these coronal plasmas.  In the
case of O~VII, the observed G ratios should have reached a limiting
value of $\sim 0.6$ for most of the stars in our study, according to
the APED emissivity ratio.  The situation is improved for the case of
the Porquet et al.\ (2001) ratios, though G ratio temperatures still
seem too low.

While direct excitation dominates other mechanisms in populating the
$1s2p\; ^1P_1$ upper level of the resonance line, in the case of the
forbidden line cascades from higher $n$ levels are much more
important.  These cascade contributions predominantly come from
dielectronic recombination.  The G ratio is therefore quite sensitive
to the balance between dielectronic recombination and direct
excitation processes.  Since our G ratio temperatures are too low,
this suggests that either the model populations of the $f$ and $i$
upper levels are underestimated, or the resonance line, $r$, $^1P_1$
upper level population is overestimated.  Considering that the
predicted Si R ratio low density limit seems too low to explain the
observations, our results suggest that the culprit is likely in the
cascade contributions and the equipartition between cascade rates
into $^3P$ and $^3S$.  
The size of the discrepancy we see suggests errors in the {\em
relative} level populations of about 20\%\ or so.  As we noted
earlier in connection with the comparison of Smith et al. (2001) and
Porquet et al. (2001) predicted R ratios, uncertainties of at least
10\%\ or so are clearly present in these calculations.

In this study, we emphasize trends with the observed R ratio itself,
in addition to the values of the derived plasma densities.  While it
appears that only the Si R ratio is significantly affected by the
model discrepancies described above, it should be kept in mind in the
following discussion that there is an inherent {\em systematic} uncertainty
in converting the observed ratios to densities, in addition to photon
counting statistics.

\subsection{{\em Densities}}

We characterise the general R ratio and plasma density findings
here as follows:
\begin{enumerate}
\item R ratios and, consequently, plasma densities at temperatures of
$\sim 10^6$~K are all very similar, with typical densities of
$n_{\mathrm{e}}\sim 2\times 10^{10}$~cm$^{-3}$ and a scatter between
different stars of approximately a factor of 2.

\item We find a trend of higher plasma density in sources with higher
coronal temperatures, and with higher coronal X-ray luminosities. 
This trend is less obvious when cast in terms of surface flux, and
more obvious when viewed as a function of the ratio of X-ray to
bolometric luminosities.

\item In all the stars studied, the observed Si~XIII R ratios  are
above or similar to the predicted low density limit.  At face value
this indicates that no star has a coronal plasma density
$\gtrsim 10^{13}$~cm$^{-3}$ at temperatures of $\sim 10^7$~K or higher.
\end{enumerate}

The general finding of higher plasma densities at higher temperatures
in active stars has already been suggested by the growing bulk of work
from other EUV and X-ray observations of active stars (e.g., Bowyer,
Drake \& Vennes 2000 and Drake 2001 for summaries of EUVE studies;
Brickhouse\ 2002, Sanz-Forcada et al.\ 2002,2003ab, Huenemoerder et al.\ 
2001, Argiroffi et al.\ 2003).  Our survey here confirms the findings
of the earlier work; the EUV studies were often not unambiguous, owing
to the possible influence of lines blending with the density-sensitive
diagnostics (e.g.\ Drake 2001).  Indeed, the upper limit of
$n_{\mathrm{e}} \lesssim 10^{13}$~cm$^{-3}$ imposed by Si~XIII for
{\em all stars of our sample} casts doubt on findings from some Fe EUV
line ratios of densities $n_{\mathrm{e}} \sim 10^{13}$~cm$^{-3}$, as
reported, e.g., by Sanz-Forcada et al.\ (2003a), though we have to
keep in mind possible inaccuracies in the theoretical Si~XIII R ratio.

The results for different groups of stars are summarized below.
Among our results, coronal density estimates are presented here
for the first time for several sources of our sample: the single
dwarf EV~Lac, the giants HD~223460, Canopus and $\mu$~Vel,
and the RS~CVns TY~Pyx and IM~Peg. 
For the remaining sources, the results we obtain here are compared with
findings of previous works in Tables~\ref{tab9} and \ref{tab10}. 

\subsubsection{Single Dwarfs}

In single dwarfs, excluding the peculiar case of TW~Hya that
we will shortly discuss below, high density seems to be associated
to the presence of evident flaring events as observed for AU~Mic
(Monsignori Fossi et al.\ 1996; Magee et al.\ 2003), Proxima~Cen
(G{\" u}del et al.\ 2002), and EV~Lac\footnote{EV~Lac X-ray
observations prior to the {\em Chandra} and {\em XMM-Newton} era
comprise only low resolution studies, including those of BeppoSAX
and ROSAT (e.g., Sciortino et al.\ 1999), and ASCA (e.g., Favata
et al.\ 2000), in which direct density diagnostics are unavailable.}
(this work).

The results obtained for single dwarfs are generally compatible
with the results present in the literature (see Table~\ref{tab9}),
at lower temperatures (2-3~MK), while some disagreement is present
for higher temperatures (6-10~MK). For example for the rapidly
rotating zero age main sequence K0 dwarf AB~Dor, our density
estimates are in good agreement with the finding of
Sanz-Forcada, Maggio \& Micela\ (2003b), based on both
{\em XMM-Newton} and {\em Chandra} spectra, though their estimate
from Mg~XI is slightly higher than our estimate, possibly because
of the blending due to Ne and Fe that we have taken into account here.
However, our results for Mg are somewhat in conflict with the
analysis of Fe~XIX-XXII lines in EUVE, {\it Chandra} and
{\em XMM-Newton} spectra by Sanz-Forcada et al.\ (2002,2003a),
who find somewhat higher densities of
$n_{\mathrm{e}} \sim 10^{12}-10^{13}$~cm$^{-3}$.  In particular,
the densities estimated from X-ray Fe lines by Sanz-Forcada et
al.\ (2003a) span an order of magnitude.  While it is possible
that the slightly higher temperatures probed by these Fe lines
could be characterized by significantly higher densities than
the plasma responsible for the Mg~XI lines, the limits placed
on the density by the Mg lines tend to rule out the presence
of significant amounts of plasma at densities close to
$10^{13}$~cm$^{-3}$.

\paragraph{TW Hya}

The same {\em Chandra}-HETG spectra of the pre-main sequence T~Tauri
star TW~Hya have been analyzed by Kastner et al.\ (2002), who reported
similar results to those obtained here: the Mg~XI lines have not been
detected, and the O~VII lines yield $n_{\mathrm{e}} \gtrsim
10^{12}$~cm$^{-3}$. The O~VII spectral region of TW~Hya is strikingly
different from the spectra of all the other sources in our sample that
yield $n_{\mathrm{e}}$ invariably close to $10^{10}$~cm$^{-3}$.
Stelzer \& Schmitt\ (2004) analyzed {\em XMM-Newton} observations of
TW~Hya at both high (with the Reflection Grating Spectrometer) and
intermediate (with the European Photon Imaging Camera) spectral
resolution, finding densities based on O~VII in good agreement with
our results and those of Kastner et al.\ (2002).  Since, during the
{\em XMM-Newton} observations no clear flare is detected, Stelzer \&
Schmitt\ (2004) conclude that the unusually high density found in
TW~Hya is not due to any flaring activity.  Echoing Kastner et
al.\ (2002) and Stelzer \& Schmitt (2004), the X-ray emission from
TW~Hya seems radically different from that typical of stellar coronae,
supporting conjectures that it originates from a fundamentally
different plasma, such as one heated by a shock at the bottom of
an accretion column.

\subsubsection{Single Giants}

None of the spectra of the single giants analysed here yielded useful
measurements of the O~VII lines.  In the case of Mg~XI, however, we
find clear indication of high densities of a few $10^{12}$~cm$^{-3}$
for all the giants except $\mu$~Vel, for which both HEG and MEG
measurements suggest that $n_{\mathrm{e}}$ is compatible with the
low-density limit ($n_{\mathrm{e}} \lesssim 10^{12}$~cm$^{-3}$).

One remarkable result among the giants is that the coronal density in
the case of the supergiant Canopus appears to be high---of the order of
$3\times 10^{12}$~cm$^{-3}$ indicated by Mg~XI at temperatures $\sim
10^7$~K.  Consequently, the surface filling factor is tiny---a few
millionths of the stellar surface.  We discuss filling factors for the
sample as a whole in more
detail below.

\subsubsection{Active Binaries}

Conspicuous among the active binaries is the clear indication of high
densities---$n_{\mathrm{e}} \gtrsim 10^{12}$~cm$^{-3}$---from the Mg
line analysis for most systems.  Most of the upper limits found
correspond to HEG measurements that are generally affected by
larger statistical uncertainties.

Several previous analyses of spectra of the active stars of our sample
exist, as summarized in Table~\ref{tab10}.  EUVE spectra of active
binaries were analyzed by Sanz-Forcada et al.\ (2001, 2002), who
estimated plasma densities using Fe~XIX-XXII lines (T$\sim$10~MK), and
invariably found $n_{\mathrm{e}} \sim 10^{12}$-$10^{13}$~cm$^{-3}$.
Other earlier EUVE results are summarized by Drake (2001) and Bowyer,
Drake \& Vennes (2000); Ness et al.\ (2002a, 2002b) presented an
analysis of X-ray spectra obtained with the {\em Chandra} Low Energy
Transmission Grating Spectrometer (LETGS) that provides a slightly
higher sensitivity to the O~VII lines than the HETG+ACIS-S
combination.

Some inconsistencies exist between our and earlier results for high
temperature plasma ($\sim 10^7$~K). For example, Brickhouse \& Dupree
(1998) estimated $n_{\mathrm{e}} > 10^{12}$~cm$^{-3}$ up to $\sim
10^{14}$~cm$^{-3}$, diagnosed from Fe~XIX-XXII ($T\sim 10$MK) lines in
EUVE spectra of 44~Boo.  Again, our Si~XIII, and possibly Mg~XI,
densities seem incompatible with the EUV estimates based on Fe lines,
suggesting that the latter might be spuriously high. As previously
mentioned, the effect of lines blending with the density-sensitive
diagnostics (e.g.\ Drake 2001), or an inaccurate theoretical Si~XIII R
ratio, might be responsible for this disagreement.

\subsubsection{General Trends}

Taking advantage of the large star sample analyzed, we investigated the 
presence of any specific trend related to the physical characteristics 
of the observed stars. We mainly concentrated on the results obtained 
from the Mg~XI measurements, since this is the most complete sample.

We first investigated trends in R ratio with stellar fundamental
parameters such as effective temperature and surface gravity.  
A correlation with surface gravity might be expected, for example, 
because surface gravity dictates the density and pressure that 
would characterise the plasma under hydrostatic equilibrium.  
The underlying stellar effective temperature might be expected 
to have a more subtle influence on coronal gas properties.  
In fact, we did not find any evident trend of the measured R 
ratios or derived densities with surface gravity or effective 
temperature.  The most graphic illustration of this is when 
contrasting the giants, including the supergiant Canopus, and 
the dwarfs.  R ratios are indistinguishable between these groups.

Perhaps more surprising, we also do not see obvious trends of the
measured R ratio with rotation period or Rossby number.
Instead, the only obvious trend in the R ratios is with stellar X-ray
luminosity, and in the ratio of X-ray luminosity to bolometric
luminosity, L$_{\mathrm{X}}$/L$_{\mathrm{bol}}$.  The former trend is
shown in Figure~\ref{fig13}, where we show the O~VII,
Mg~XI and Si~XIII R ratios as a function of both $L_{\mathrm{X}}$ and 
the surface flux $F_{\mathrm{X}}$.  
While there is arguably no discernible trend in any of
the ratios with $F_{\mathrm{X}}$, the Mg~XI ratio shows a decrease 
in R ratio for $L_{\mathrm{X}} > 10^{30}$~erg~s$^{-1}$ or so.  
It is tempting to see the same trend in the Si~XIII R ratio for
$L_{\mathrm{X}} > 10^{31}$~erg~s$^{-1}$, but there are simply too 
few observations at the highest X-ray luminosities.  
The O~VII R ratio is essentially the same for the
whole range of surface fluxes and X-ray luminosities.

\begin{figure*}[!ht]
\centerline{\psfig{figure=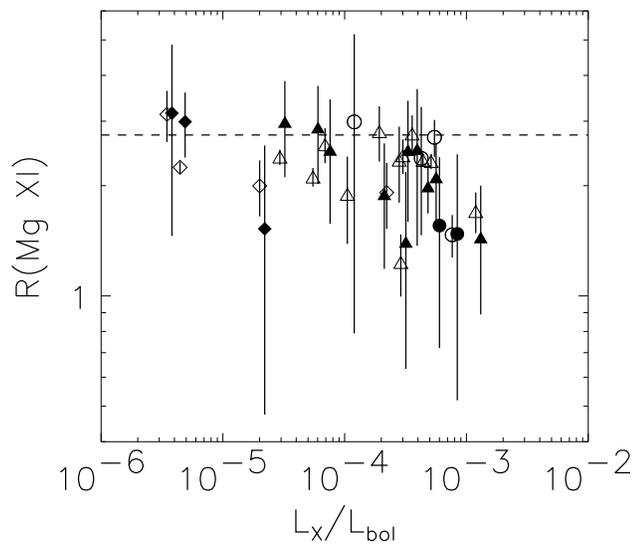,width=10cm}}
\caption{R ratios from the Mg~XI triplet plotted vs.\ 
	L$_{\mathrm{X}}$/L$_{\mathrm{bol}}$. The dashed line 
	marks the limiting value of the R ratio; lower values 
	of R correspond to higher density (see Fig.~\ref{fig6}). 
	Symbols as in Figure~\ref{fig9}. 
	\label{fig16}}
\end{figure*}

The trend of declining Mg~XI R ratio in more active stars becomes more
clear when shown as a function of L$_{\mathrm{X}}$/L$_{\mathrm{bol}}$,
as shown in Figure~\ref{fig16}.  We consider the derived densities for
O~VII and Mg~XI as a function of $L_{\mathrm{X}}$ and $F_{\mathrm{X}}$ 
in Figure~\ref{fig15}:
a significant fraction of the measurements here formally yield upper
limits to the plasma density and these tend to obscure the results
that are more obvious in Figure~\ref{fig13}---it is only when the
measured R ratio sample is taken together than the trends clearly
emerge.  Nevertheless, the trend of increasing density derived from
Mg~XI is still apparent at $L_{\mathrm{X}} > 10^{30}$~erg~s$^{-1}$ 
or so; below
this value the results are characterised more by upper limits.

We emphasise that this trend of Mg~XI R ratio and derived density with
L$_{\mathrm{X}}$ is not determined by a possible selection effect
related to the signal-to-noise ratio of the HETG spectra; in fact the
relative error on the Mg~XI R measurement, $\delta$(R)/R, has no
specific trend with L$_{\mathrm{X}}$.

\subsubsection{Density correlated with Ne/Mg?}

In \S\ref{sec:mg}, we pointed out an apparent trend of measured Mg~XI
R ratio with Ne~X/Mg ~XI line strength seen in Figure~\ref{fig9}.  We
do not believe this trend is a result of residual blending; instead we
suggest that we are seeing a trend of coronal abundance variation in
which the Ne/Mg ratio is higher in stars characterised by higher Mg~XI
R ratios---higher density at temperatures of $\sim 10^7$~K.  Such a
result might be expected based on the emerging trend of coronal
abundance anomalies in active stars.  The coronae of RS~CVn-type
binaries, in particular, exhibit high Ne/Fe abundances compared to
their expected photospheric compositions (see e.g.\ reviews by Drake
et al.\ 2003, Audard et al.\ 2003).  Since we see in this study good
evidence for a trend of decreasing Mg~XI R ratio with increasing
$L_{\mathrm{X}}/L_{\mathrm{bol}}$, the correlation of Ne~X/Mg~XI 
line strength with Mg~XI R ratio would be expected if the Ne/Mg 
abundance ratio increased with $L_{\mathrm{X}}/L_{\mathrm{bol}}$.
While the Ne/Mg abundance ratio remains to be investigated thoroughly 
in the stars of our sample, the evidence from analyses to date 
summarised by Drake (2003) and Audard (2003) does suggest that the 
relative Ne enhancement increases with stellar activity.

\subsection{{\em Filling factors}}

One of the most important aspects of density measurements is the
associated insight into the structuring of the plasma and into the
emitting volume.  In the stellar application, we can also use the
density information to investigate coronal filling factors.  For
coronal line emission we can write $I^{\mathrm{obs}}_{k} = \int_{V}
n^2_{\mathrm{e}} G_k(T,n_{\mathrm{e}}) dV$, for the intensity of a
line $k$ emitted by optically thin plasma.  In the isothermal
approximation we have $I^{\mathrm{obs}}_{k} \equiv
G_k(T,n_{\mathrm{e}}) EM$ where $EM \equiv n^2_{\mathrm{e}} V$ 
is the volume emission measure and $G_k(T,n_{\mathrm{e}})$ is
the {\em contribution function} of the emission line $k$.  If we
further assume that the plasma is at the temperature of maximum
formation of the line $k$, we end up with a lower limit for the EM of
the plasma emitting that line. In turn, from EM we obtain an estimate
of the surface coronal filling factor defined as
\[
f = \frac{V}{\mathcal{L}} \cdot \frac{1}{A_{\star}}
  = \frac{EM}{n^2_{\mathrm{e}} \mathcal{L}} \cdot \frac{1}{A_{\star}} =
  \frac{I^{\mathrm{obs}}_{k}}{n^2_{\mathrm{e}} G_k(T,n_{\mathrm{e}})}
    \cdot \frac{1}{\mathcal{L}} \cdot \frac{1}{A_{\star}}
\]
where $\mathcal{L}$ is the scale height of the emitting plasma 
and $A_{\star}$ is the surface area of the star.  A possible choice 
of $\mathcal{L}$ is given by the hypothesis that the emitting coronal
plasma is confined in hydrostatic loops smaller than, or comparable
to, the pressure scale height.  Under this assumption we can use 
the scaling laws of an hydrostatic loop model, such as that of Rosner, 
Tucker \& Vaiana (1978; RTV) for which 
$\mathcal{L} \sim [T/(1.4 \times 10^3)]^3/p$.
In this way, from the measured densities and assuming a temperature
equal to the temperature of maximum formation of the line we
can derive an estimate of the filling factors.

\begin{figure*}[!ht]
\centerline{\psfig{figure=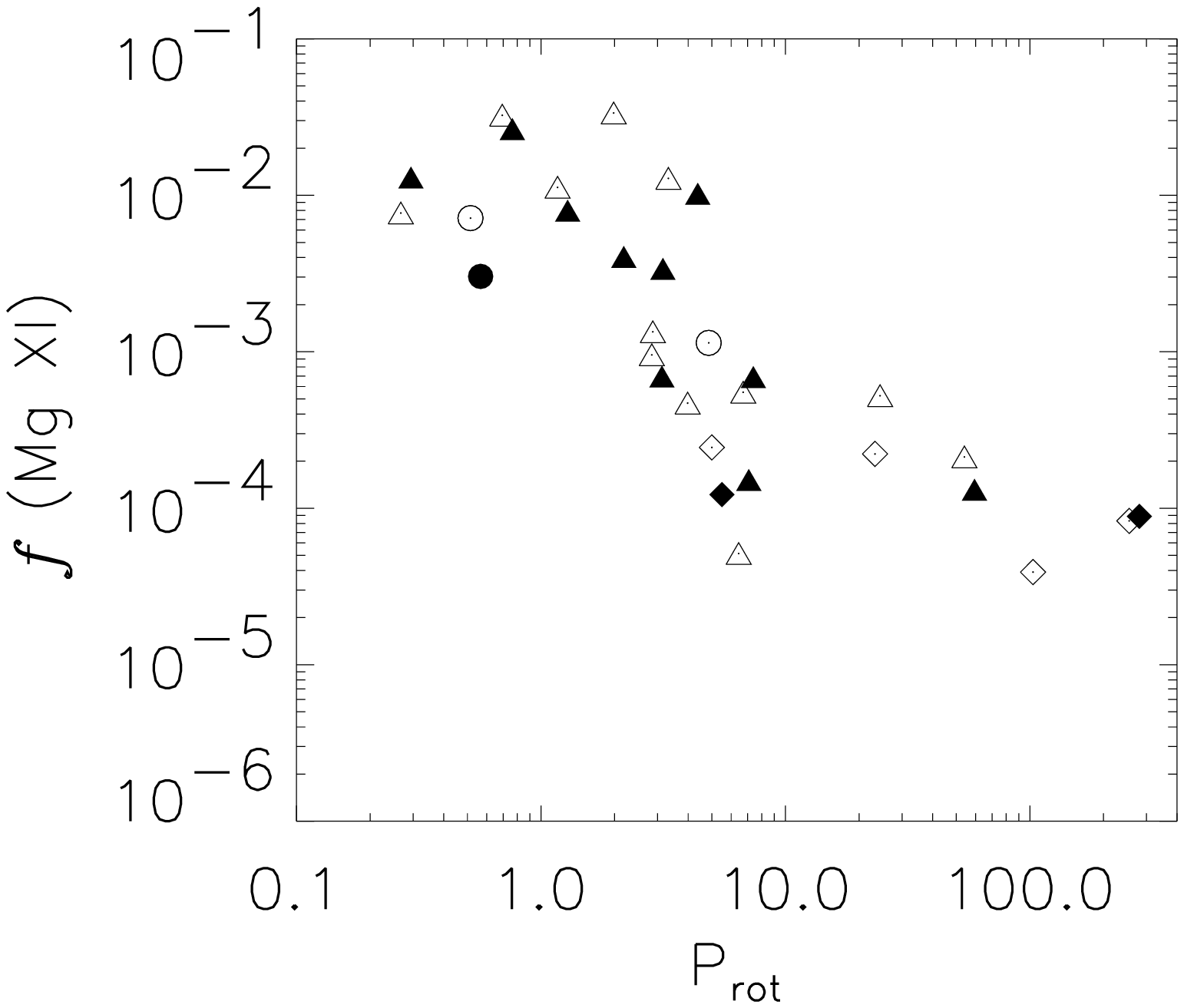,width=9.5cm}\hspace{-1.cm}
            \psfig{figure=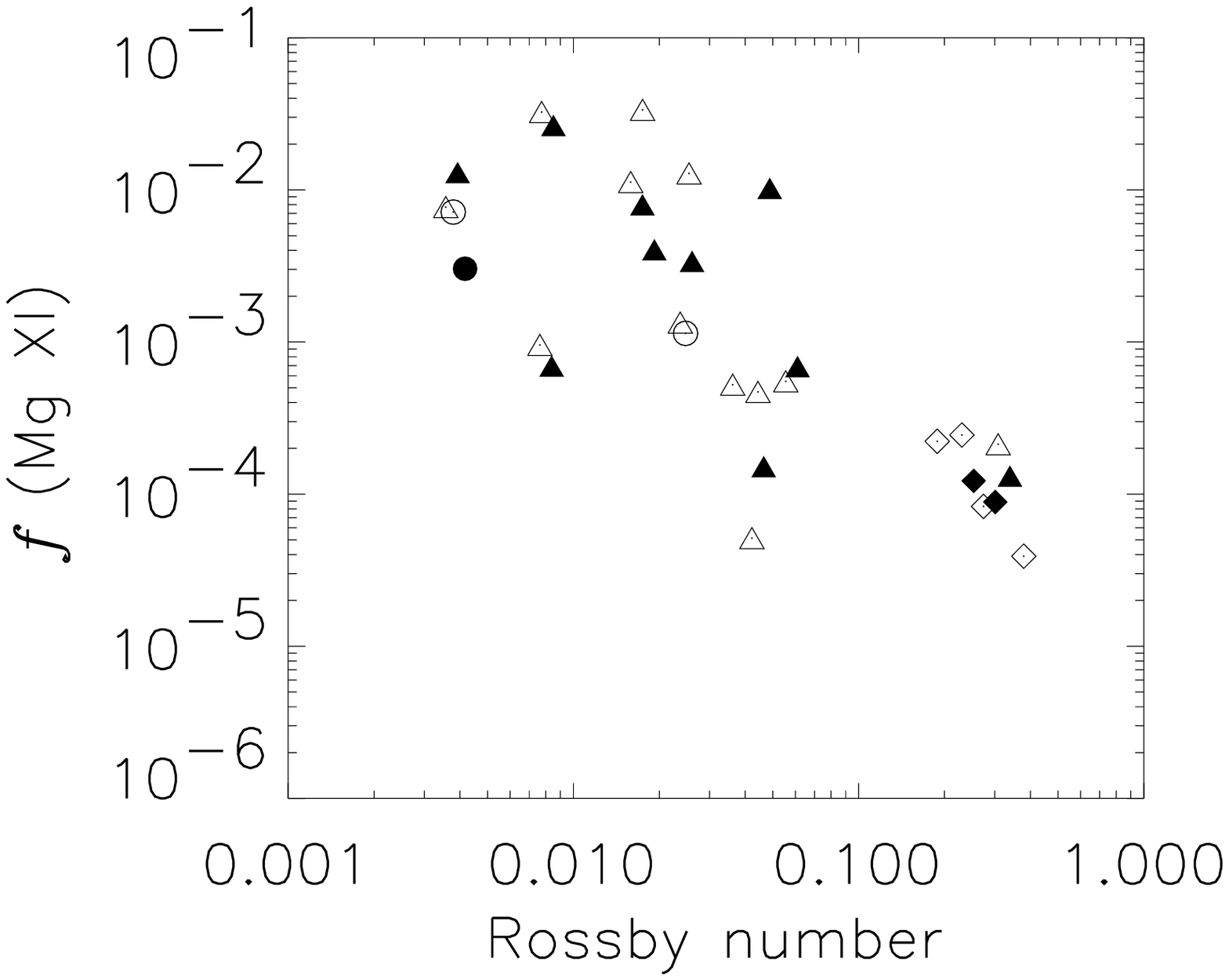,width=9.5cm}}
\caption{Surface filling factors estimated from the 
	measured densities, vs.\ the rotation period
	({\em left}) and vs.\ the Rossby number ({\em right})
	of the sources. Symbols as in Figure~\ref{fig9}.
	\label{fig17}}
\end{figure*}

In Fig.~\ref{fig17} we illustrate the surface filling factors, $f$, 
derived from Mg~XI as a function of the rotation period and of 
the Rossby number.  
There is a sharp increase in filling factor at rotation periods of 3-4
days, below which the filling factor appears to saturate at values of
several percent of the stellar surface, and above which there is an
approximately linear trend between logarithmic rotation period and
filling factor.  A rotation period of 3-4 days also marks the general
location of coronal ``saturation'', and the saturation of filling
factor here is not a surprise.  We note that Canopus, whose rotation
period is not accurately known (e.g.\ Decin et al.\ 2003), is missing
from Fig.~\ref{fig17}.  Nevertheless, if its rotation period is of
order several hundred days it would not be out of place in the
relation between filling factor and period for $P_{rot} > 4$d.

\begin{figure*}[!ht]
\centerline{\psfig{figure=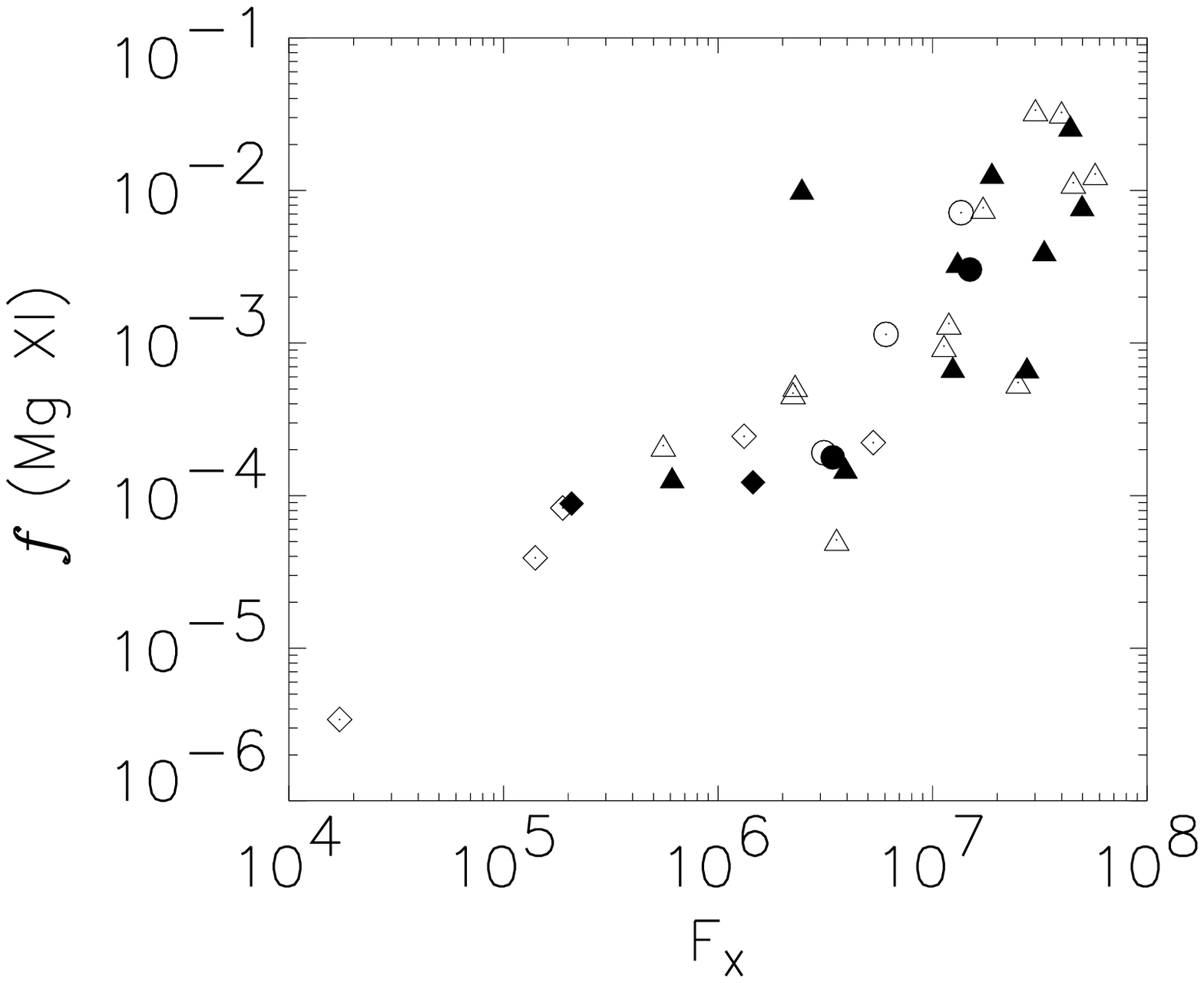,width=9cm}\hspace{-0.7cm}
            \psfig{figure=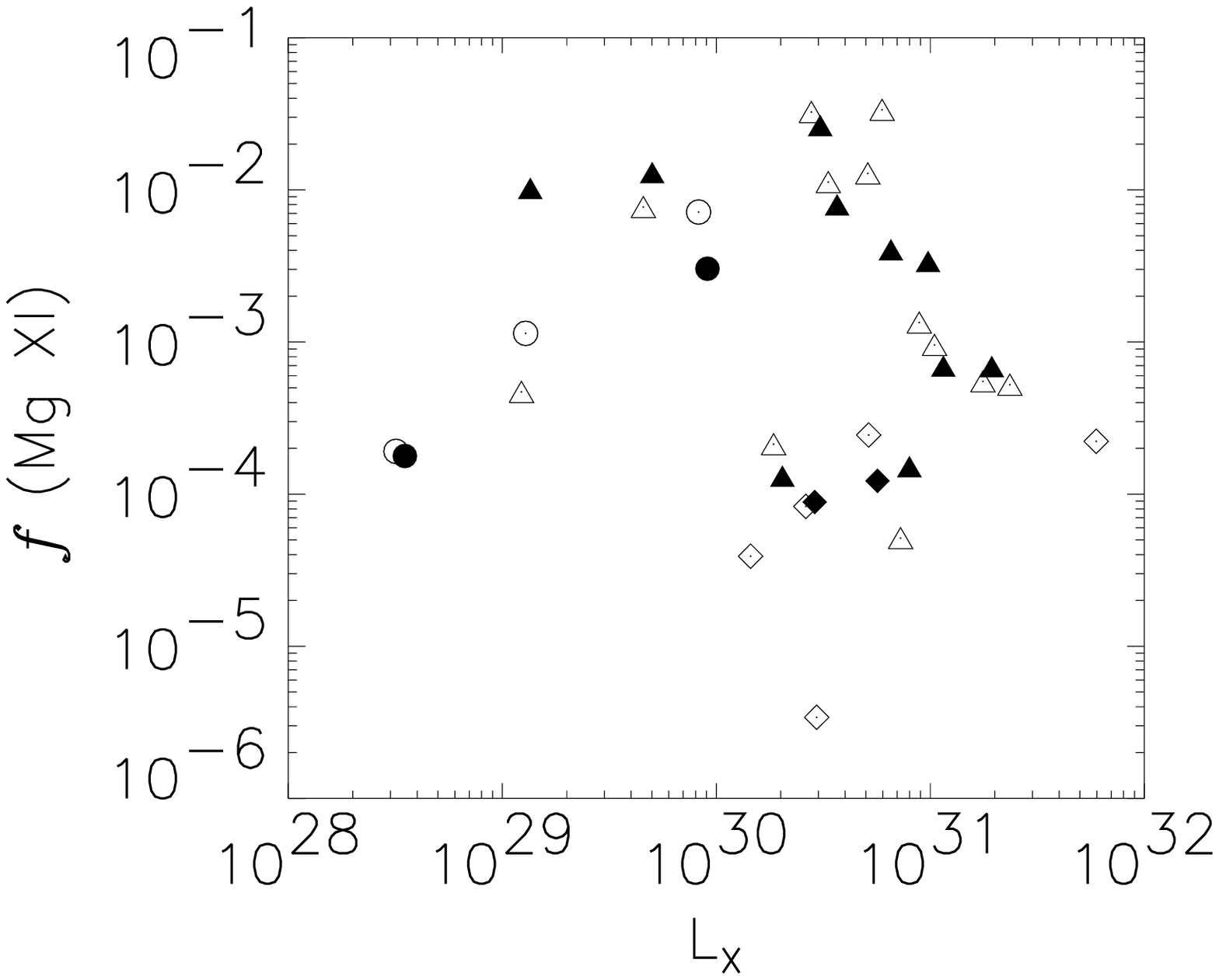,width=9cm}}\vspace{-0.8cm}
\centerline{\psfig{figure=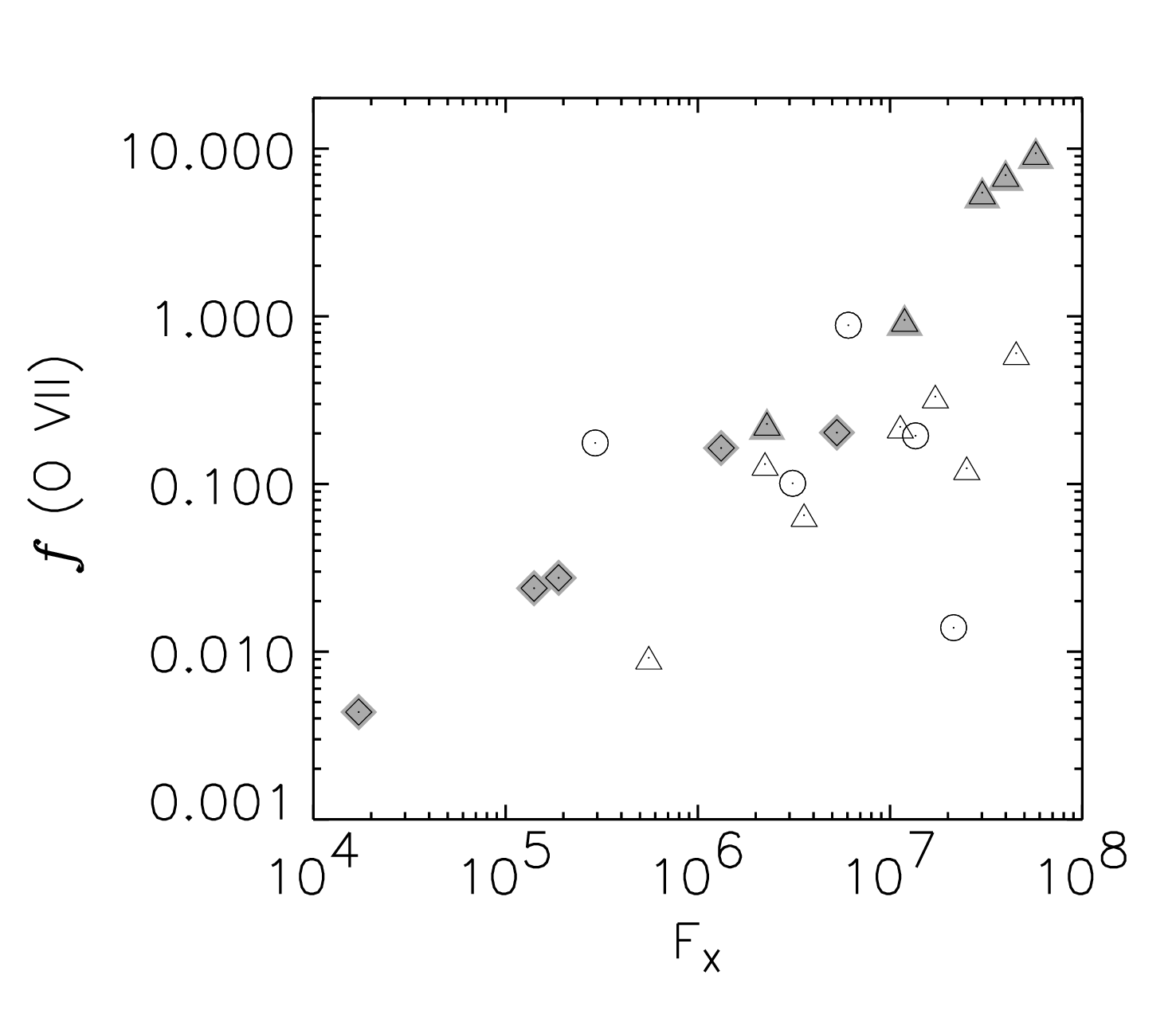,width=9cm}\hspace{-0.7cm}
            \psfig{figure=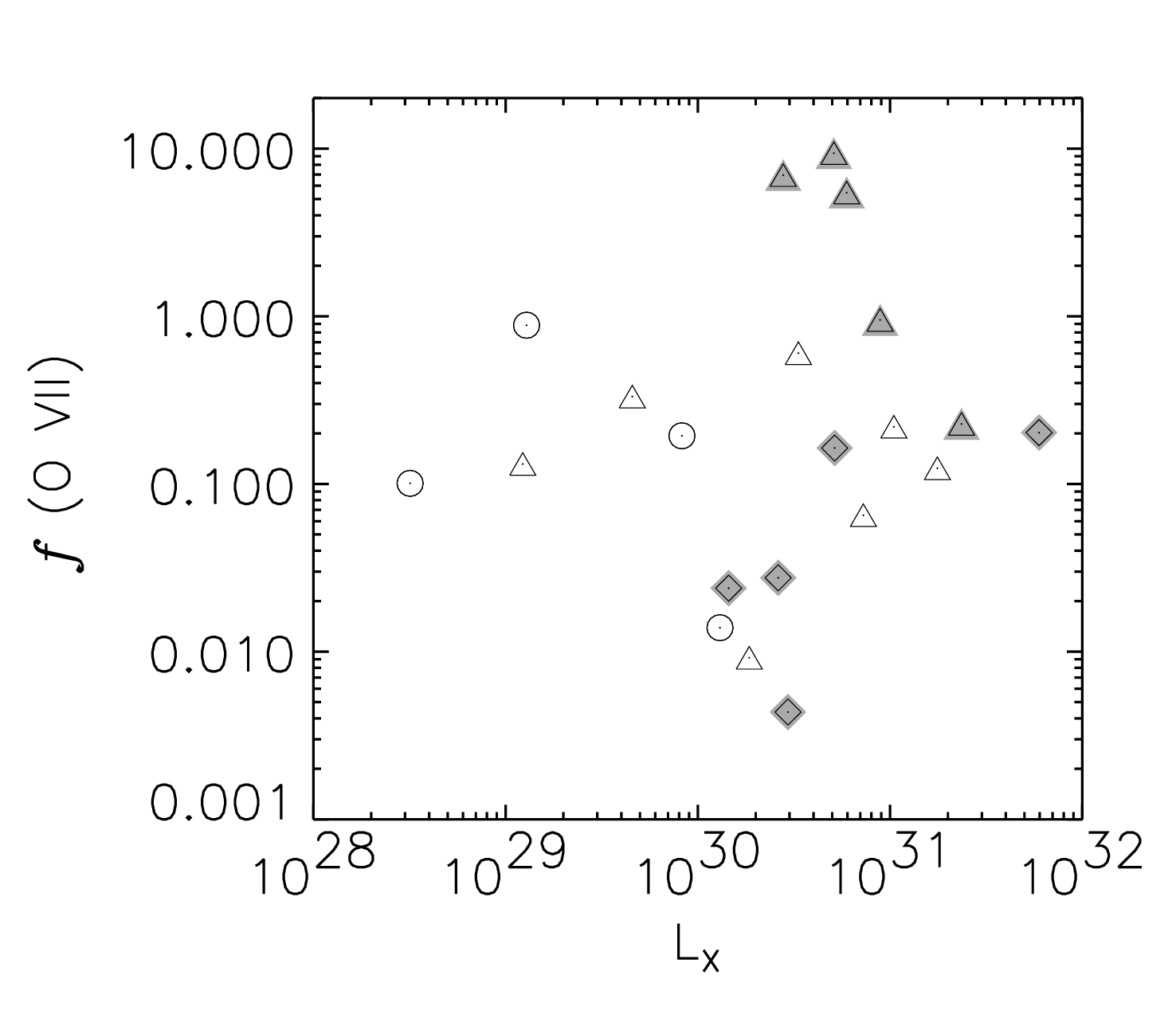,width=9cm}}
\caption{Surface filling factors derived from Mg~XI 
	({\em top}) and O~VII ({\em bottom}) lines vs.\ 
	X-ray surface flux ({\em left}) and the X-ray 
	luminosity ({\em right}) of the sources. 
	Symbols as in Figure~\ref{fig9}. The gray 
	symbols in the bottom panels mark the sources 
	for which we assumed 
	$n_{\mathrm{e}}=2\times 10^{10}$~cm$^{-3}$.
	\label{fig18}}
\end{figure*}

Figure~\ref{fig18} illustrates the filling factors derived from both 
Mg~XI and O~VII lines vs.\ X-ray luminosity and surface flux. In order 
to have a more complete sample and investigate the relation with X-ray 
emission, we assumed an electron density of $2\times 10^{10}$cm$^{-3}$ 
for the stars whose O~VII lines were not measurable.  This is not 
unreasonable given that almost all the spectra of our active stars 
(with rare exceptions like the case of TW~Hya) yield $n_{\mathrm{e}}$ 
close to this value; this estimate is also supported by analyses of 
{\it Chandra} LETGS and {\it XMM-Newton} observations of some of the 
sources of our sample for which we did not obtain O~VII measurements, 
like 31~Com (Scelsi et al.\ 2004), Algol (Ness et al.\ 2002b), UX~Ari 
(Ness et al.\ 2002a) and Prox~Cen (G{\" u}del et al.\ 2002).  
Nevertheless, it is notable that we find filling factors $>1$---a 
nonsensical value---for three of the active binaries (ER~Vul, TY~Pyx, 
and AR~Lac) for which we assume the common density value, indicating 
that the true density at a few $10^6$~K in these stars must be higher
than $2\times 10^{10}$cm$^{-3}$, or that the assumed scale height is too low.

With some degree of scatter, the O~VII filling factor, $f_{\mathrm{OVII}}$, 
is directly proportional to the X-ray surface flux.  It is interesting
that this filling factor begins to saturate---i.e.\ reach values between
0.1 and 1---at a mean surface flux of $\sim 10^7$~erg~cm$^{-2}$~s$^{-1}$.  
While our data are sparse here, this saturation of $f_{\mathrm{OVII}}$ 
seems to occur at surface flux values approximately one order of magnitude 
lower than the surface fluxes of the most active stars in the sample.  
The same value of surface flux, $10^7$~erg~s$^{-1}$~cm$^{-2}$, was found 
for the areas of the solar surface covered with active regions by 
Withbroe \& Noyes (1977).  This suggests then, that at plasma temperatures 
of up to a few $10^6$~K these ``O~VII saturated'' stars are essentially 
covered in active regions possibly similar in nature to those of the Sun.

The behaviour of the Mg~XI filling factor is different.  Firstly,
there is a clear break in the $f_{\mathrm{MgXI}}$-$F_{\mathrm{X}}$
relation {\em at the same surface flux level, 
$F_{\mathrm{X}}\sim 10^7$~erg~s$^{-1}$~cm$^{-2}$, as we
see $f_{\mathrm{OVII}}$ saturate}.  At lower $F_{\mathrm{X}}$ levels, 
the $f_{\mathrm{MgXI}}$-$F_{\mathrm{X}}$ relation appears to have a 
shallower slope than that for $f_{\mathrm{OVII}}$ vs.\ $F_{\mathrm{X}}$ 
though the level of scatter precludes a definitive statement.

\begin{figure}[!ht]
\centerline{\psfig{figure=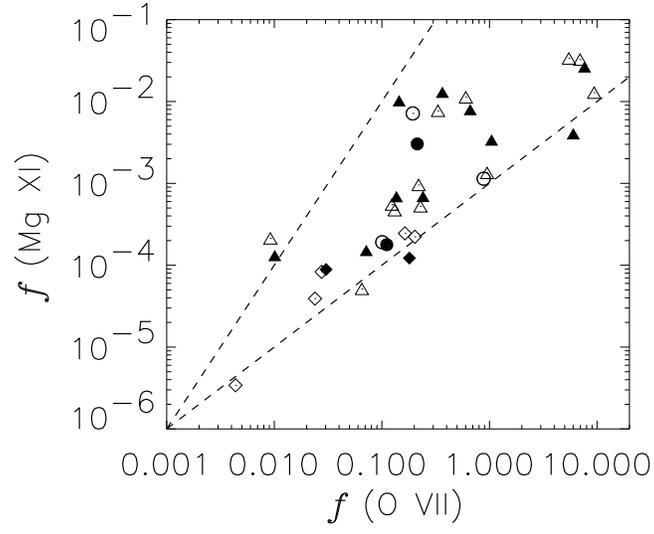,width=9.5cm}}
\caption{Mg~XI surface filling factors vs.\ O~VII surface 
	filling factors.  Symbols as in Figure~\ref{fig9}.
	The dashed lines superimposed to the data represent
	$f_{\mathrm{MgXI}}\propto f_{\mathrm{OVII}}$ and 
	$f_{\mathrm{MgXI}}\propto f^2_{\mathrm{OVII}}$.
	\label{fig19}}
\end{figure}

Finally, we examine the relative behavior of the two filling factors,
$f_{\mathrm{OVII}}$ and $f_{\mathrm{MgXI}}$ in Fig.~\ref{fig19}.  
For $f_{\mathrm{OVII}}\lesssim 0.2$, the two factors are approximately 
proportional to one another.  For $f_{\mathrm{OVII}}\gtrsim 0.2$, 
$f_{\mathrm{MgXI}}$ increases much more sharply than $f_{\mathrm{OVII}}$.  
At higher activity levels (i.e.\ higher $f$) the coronae are 
preferentially filling with hotter plasma with respect to the
cooler plasma emitting the O~VII lines.

\subsection{Summary}

The density and filling factor results suggest a scenario that has
been outlined before based on the emission measure distributions as a
function of temperature of active dwarfs as compared to those of solar
active regions by Drake et al.\ (2000; see also G\"udel 1997): going
toward activity levels higher than the Sun the trend appears to be one
of increased filling factor of solar-like active regions characterised
by essentially the same plasma density of $n_{\mathrm{e}}\sim 2\times
10^{10}$~cm$^{-3}$.  For these active stars, there is an attendant
hot, high density plasma with $n_{\mathrm{e}}\sim 10^{12}$~cm$^{-3}$ whose
contribution to the total X-ray increases toward higher activity
stars.

As the surface becomes ``saturated'', or full, of active regions, with
filling factors between 0.1 and 1, the nature of the corona begins to
change.  This surface saturation occurs at X-ray luminosities well
below the highest observed for the most active stars---by an order of
magnitude or two.  Drake et al.\ (2000) and G\"udel (1997) suggested
that interaction between adjacent active regions could give rise to
increased flaring behaviour which is manifest as the growth of the
hotter $10^7$~K corona characterised in the Mg~XI R ratio by higher
plasma densities, and with a concomitant increase in the filling
factor of this plasma.

The quite different gas densities found for ``high'' and ``low''
coronal temperatures reinforces the fact that the dominant emission at
these temperatures cannot originate from the same structures: models
seeking to explain active coronae need to account for the increase in
gas pressure with increasing temperature.

\section{Conclusions}
\label{sec:concl}

We have investigated the density of coronal plasma at different
temperatures through the analysis of the He-like triplet of O, Mg, and
Si in high spectral resolution X-ray observations obtained with {\em
Chandra} of a wide sample of active stars.  As in previous works in
the literature (e.g.\ Brickhouse\ 2002, Ness et al.\ 2003) the
unprecedented {\em Chandra} spectrometer resolution proves to be vital
for an accurate analysis: in this work we have especially emphasised
the effect of lines blending with the Mg~XI triplet lines and the need
to take these into account in order to obtain reliable measurements of
the R ratio.  Ignoring these blends results in R ratios significantly
different from values obtained when blends are accounted for.

Our measured R and G ratios provide some insights into the propriety
of recent theoretical calculations of the relative line strengths from
He-like ions.  In the case of Si~XIII, we find some evidence
suggesting that the low density limit $f/i$ ratios predicted by both
APED and Porquet et al.\ (2001) are too low by 20-40\%\ or so based on our
observed values.  Temperatures inferred from theoretical G ratios also
tend to be lower than expected.  We tentatively identify the cascade
contributions from dielectronic recombination as the likely culprit
for modelling inaccuracies.

For most of the observed sources in which O~VII yielded a useful R
ratio measurement, we find for plasma at temperatures of a few
$10^6$~K densities of $\sim2\times 10^{10}$~cm$^{-3}$.  Mg~XI lines
formed closer to $10^7$~K yielded densities $n_{\mathrm{e}} \sim
10^{12}$~cm$^{-3}$ confirming the scenario outlined in the recent
literature in which hotter ($\sim 6-10$~MK) plasma is characterized by
densities two orders of magnitude higher than the cooler plasma, and
in which the latter has both temperature and density similar to those
characterising solar coronal active regions.  No star shows good
evidence for Si~XIII R ratios significantly below the theoretical low
density limits; while this might partly be a problem associated with
the theoretical ratios, which appear too low, it seems clear that
there are no stars with electron densities significantly above
$10^{13}$~cm$^{-3}$.  This result casts some doubt on earlier studies
based on Fe XX-XXII lines seen in EUVE spectra which suggested the
presence of densities exceeding $10^{13}$~cm$^{-3}$ in active stellar
coronae.

We have found a distinct correlation between the measured Mg~XI R
ratio and X-ray luminosity, with lower R ratios, and by inference,
higher densities, found in more X-ray luminous coronae.  This trend is
also visible as a function of the X-ray ``production efficiency'',
L$_X$/L$_{bol}$.  Instead, no correlations were found between R ratio
and stellar parameters surface gravity and effective temperature, or
with Rossby number.

The surface coronal filling factors obtained from our density
estimates are significantly smaller for the hotter plasma with 
$T\sim 10^7$~K, with $f_{\mathrm{MgXI}} \sim 10^{-4} - 10^{-1}$, than
for the cooler $\sim 2$-$3\times 10^6$~K plasma characterized by 
$f_{\mathrm{OVII}}$ values in the range $10^{-3}-1$.  The remarkably 
small filling factors for hot plasma suggest the potential for 
studying X-ray rotational modulation using the hotter lines available
in high resolution X-ray spectra.  In this regard, we note that Marino
et al.\ (2003) have recently found for the first time X-ray rotational 
modulation in a supersaturated star; Orlando et al.\ (2004) also show 
that such rotational modulation may be present in the hardness ratio.

The O~VII filling factor, $f_{\mathrm{OVII}}$, is found to be directly
proportional to the X-ray surface flux, and reaches values between 0.1
and 1 at a mean surface flux of $\sim 10^7$~erg~cm$^{-2}$~s$^{-1}$---the 
same surface flux that characterises solar active regions. 
Instead, at higher temperatures the $f_{\mathrm{MgXI}}$-$F_{\mathrm{X}}$ 
relation shows a break {\em at the same surface flux level}, 
$F_{\mathrm{X}}\sim 10^7$~erg~s$^{-1}$~cm$^{-2}$, after which it 
increases more strongly with increasing surface flux.

The density and filling factor results suggest the following picture:
at plasma temperatures of up to a few $10^6$~K, stars with increasingly 
higher activity level become more and more covered with solar-like 
active regions with $n_{\mathrm{e}}\sim 2\times 10^{10}$~cm$^{-3}$.
At the same time, there is a similar growth in hotter plasma with
temperatures close to $10^7$~K, and with $n_{\mathrm{e}} \sim 
10^{12}$~cm$^{-3}$.
This hotter plasma is likely associated with continuous superimposed
flaring activity associated with the active regions, analogous to the
solar case.  For stars at the ``O~VII saturated'' level, when X-ray
surface fluxes are similar to that of active regions on the solar
surface, the whole star is essentially covered by active regions.
Further growth in X-ray emission at higher activity levels still is
in the form of increased flaring, probably arising because of the
crowding of active magnetic structures on the stellar surface.  This
activity gives rise to a growth in the surface coverage of $10^7$~K
plasma at higher densities of $n_{\mathrm{e}} \sim 10^{12}$~cm$^{-3}$.  
This picture is essentially that described earlier by G\"udel\ (1997) 
and Drake et al.\ (2000).

\begin{acknowledgements}
We acknowledge helpful discussions with Nancy Brickhouse, 
Ed DeLuca, Fabio Reale, Vinay Kashyap, Randall Smith,
Martin Laming and David Huenemoerder. 
Finally, we thank the referee, Dr. J. Linsky, for
insightful comments that enabled us to improve the manuscript 
significantly.
PT was partially supported by {\it Chandra} grants GO1-20006X
and GO1-2012X under the SAO Predoctoral Fellowship program.  
JJD was supported by NASA contract NAS8-39073 to the
{\em Chandra X-ray Center}.  GP and PT were partially 
supported by Ministero dell'Istruzione, dell'Universit\`a e 
della Ricerca and by Agenzia Spaziale Italiana.
\end{acknowledgements}

\clearpage

\begin{deluxetable}{lrcccccclcll}
\tablecolumns{12} 
\tabletypesize{\tiny}
\tablecaption{List of stellar parameters. \label{tab1}}
\tablewidth{0pt}
\tablehead{
 \colhead{Source} & \colhead{HD} & \colhead{Spectr.} 
 & \colhead{d \tablenotemark{a}} 
 & \colhead{R$_{\star}$/R$_{\odot}$} 
 & \colhead{M$_{\star}$/M$_{\odot}$ \tablenotemark{b}} 
 & \colhead{$T_{eff}$} & \colhead{B-V \tablenotemark{c}} 
 & \colhead{log(L$_{\mathrm{bol}}$) \tablenotemark{d}} 
 & \colhead{P$_{\mathrm{orb}}$ \tablenotemark{e}} 
 & \colhead{P$_{\mathrm{rot}}$ \tablenotemark{e}}
 & \colhead{$\mathcal{R}$ \tablenotemark{f}} \\
 \colhead{} & \colhead{} & \colhead{Type} & \colhead{[pc]} 
 & \colhead{} & \colhead{} & \colhead{[K]}  & \colhead{} & 
 \colhead{[erg/sec]} & \colhead{[days]} & \colhead{[days]} 
 & \colhead{}
}
\startdata
 AU~Mic   	& 197481 & M1V & 9.9 & 0.56 \tablenotemark{1} 
 		& 0.59 \tablenotemark{1} 
		& 3730 \tablenotemark{27} 
		& 1.44 \tablenotemark{37} 
		& 32.48 \tablenotemark{39} & 
		& 4.854 \tablenotemark{43} & 0.024 \\  
 Prox~Cen 	&  & M5Ve & 1.3 & 0.16 \tablenotemark{2} 
 		& 0.11 \tablenotemark{20} 
		& 2700 \tablenotemark{28} & 1.97 & 30.58 & 
		& 83.5 \tablenotemark{44} & \\ 
 EV~Lac		&  & M4.5V & 5.1 & 0.41 \tablenotemark{3} 
 		& 0.34 \tablenotemark{3} 
		& 3300 \tablenotemark{29} & 1.57 
		& 31.62 \tablenotemark{3} & 
		& 4.376 \tablenotemark{45} & \\
 AB~Dor   	& 36705 & K0V & 15 & 1. \tablenotemark{4} 
 		& 0.76 \tablenotemark{4}
 		& 5000 \tablenotemark{30} 
		& 0.80 \tablenotemark{38} 
		& 33.18 \tablenotemark{40} & 
		& 0.51479 \tablenotemark{46} & 0.0038 \\
 TW~Hya		& & K8Ve & 56 & 1. \tablenotemark{5} 
 		& 0.7 \tablenotemark{5}
 		& 4150 \tablenotemark{31} 
		& 0.7 \tablenotemark{38} & & 
		& 2.9 \tablenotemark{47} & 0.019 \\ 
	      	& 223460 & G1 III & 135 
		& 13.6 \tablenotemark{6} 
		& 2.9 \tablenotemark{21}
 		& 5110 \tablenotemark{21} 
		& 0.81 \tablenotemark{6} & 35.43 & 
		& 23.25 \tablenotemark{6} & 0.19 \\  
31~Com		& 111812 & G0 III & 94 & 8 \tablenotemark{7} 
		& 2.96 \tablenotemark{22}
 		& 5320 \tablenotemark{21} & 0.68 
		& 35.41 \tablenotemark{7} & 
		& $\geq 5$ \tablenotemark{48} & 0.23 \\ 
$\beta$~Ceti 	& 4128 & K0III & 29.4 
		& 15.1 \tablenotemark{7} 
		& 3.2 \tablenotemark{21}
 		& 4840 \tablenotemark{21} 
		& 1.2 \tablenotemark{7} & 35.77 &  
		& $\geq 255$ \tablenotemark{21} & 0.27 \\  
Canopus     	& 45348 & F0II & 95.9 & 53 \tablenotemark{8} 
		& 13 \tablenotemark{23}
 		& 7350 \tablenotemark{23} 
		& 0.15 \tablenotemark{38} 
		& 37.70 \tablenotemark{41} & 
		& $\geq 298$ \tablenotemark{23} & \\   
 $\mu$~Vel	& 93497 & G5III/... & 35.5 
		& 13 \tablenotemark{9} & 3 \tablenotemark{9}
 		& 4862 \tablenotemark{32} & 0.90 & 35.62 & 
		& $\geq 103$ \tablenotemark{32} & 0.38 \\   
 Algol		& 19356 & B8V/K1IV & 28 
		& 2.9/3.5 \tablenotemark{10} 
		& 3.7/0.8 \tablenotemark{24}
 		& 11400/5300 \tablenotemark{24} & & 34.47 
		& 2.8 \tablenotemark{10} 
		& 2.8 \tablenotemark{10} & 0.024 \\  
 ER~Vul		& 200391 & G0V/G5V & 50 
 		& 1.07/1.07 \tablenotemark{11} & 1.10/1.05 
 		& 6000/5619 \tablenotemark{33} & 0.68 
		& 33.89 \tablenotemark{42} 
		& 0.6981 & 0.6942  & 0.0077 \\ 
 44~Boo     	& 133640 & G1V/G2V & 13 
 		& 0.87/0.66 \tablenotemark{12} 
		& 0.98/0.55 \tablenotemark{12}
 		& 5300/5035 \tablenotemark{12} & 0.65 
		& 33.82 & 0.267 &  & 0.0036 \\
 TZ~CrB		& 146361 & G0V/G0V & 22 
 		& 1.1/1.1 \tablenotemark{13} & 1.12/1.14
 		& 6000/6000 \tablenotemark{13} & 0.51 
		& 33.88 \tablenotemark{42} 
		& 1.1398 & 1.1687 & 0.016 \\  
 UX~Ari		& 21242 & G5V/K0IV & 50 
 		& 1.11/5.78 \tablenotemark{14} 
		& $\geq 0.63/\geq 0.71$
		& .../4800 \tablenotemark{34} & 0.91 
		& 34.49 \tablenotemark{42} 
		& 6.4379 & 6.4379 & 0.042 \\  
 $\xi$~UMa	& 98230 & G0V/G0V & 7.7 
 		& 0.95/... \tablenotemark{15} 
		& 0.92/... \tablenotemark{25}
 		& 5650 \tablenotemark{25} & 0.59 
		& 33.35 \tablenotemark{42} & 3.9805 
		& & 0.044 \\  
 II~Peg		& 224085 & K2V/...	& 42 
 		& 3.4/... \tablenotemark{16} & 0.8/... 
 		& 4600 \tablenotemark{16} & 1.01 
		& 34.17 \tablenotemark{49} & 6.7242 
		& 6.718 & 0.056 \\  
 $\lambda$~And 	& 222107 & G8III/... & 26  
 		& 7.4/... \tablenotemark{17} 
		& 0.65/... \tablenotemark{26} 
 		& 4800 \tablenotemark{26} & 1.01 & 34.80 
		& 20.521 & 53.952 & 0.31 \\  
 TY~Pyx		& 77137 & G5IV/G5IV & 56 
 		& 1.59/1.68 \tablenotemark{18} 
		& 1.22/1.20 \tablenotemark{18} 
		& 5400/5400 \tablenotemark{18} 
		& 0.72/0.76 & 34.26 \tablenotemark{42} 
		& 3.1986 & 3.32 & 0.25 \\  
 AR~Lac		& 210334 & G2IV/K0IV & 42 
 		& 1.8/3.1 \tablenotemark{11} 
		& $\geq 1.30/\geq 1.30$
 		& .../5100 \tablenotemark{35} & 0.72 
		& 34.49 \tablenotemark{42} 
		& 1.9832 & 1.9832 & 0.018 \\   
 HR~1099	& 22468 & G5IV/K1IV & 29.0 
 		& 1.3/3.9 \tablenotemark{11} & 1.1/1.4
 		& 5400/4800 \tablenotemark{36} 
		& 0.92 \tablenotemark{11} 
		& 34.31 \tablenotemark{42} & 2.8377 
		& 2.841 & 0.0076 \\  
 IM~Peg		& 216489 & K2III-II/... & 96.8 
 		& 13/... \tablenotemark{19} 
		& 1.5/... \tablenotemark{19}
 		& 4450 \tablenotemark{19} & 1.12 
		& 35.35 \tablenotemark{42} & 24.65 
		& 24.39 & 0.036 \\  \hline  
\multicolumn{12}{l}{
\begin{tabular}{llll}
 \multicolumn{4}{l}{REFERENCES. --}   \\
 $^{1}$ Houdebine \& Doyle\ (1994)     	& $^{14}$ Duemmler \& Aarum\ (2001)
 	& $^{27}$ Linsky\ (1982)	& $^{40}$ Singh et al.\ (1999) \\
 $^{2}$ Wood et al.\ (2001) 		& $^{15}$ Sanz-Forcada et al.\ (2003)
 	& $^{28}$ Frogel et al.\ (1972) & $^{41}$ Panzera et al.\ (1999)\\
 $^{3}$ Sciortino et al.\ (1999)	& $^{16}$ Berdyugina et al.\ (1998)
 	& $^{29}$ Pettersen et al.\ (1980)& $^{42}$ Drake et al.\ (1989) \\
 $^{4}$ Maggio et al.\ (2000)		& $^{17}$ Nordgren et al.\ (1999)
 	& $^{30}$ Hussain et al.\ (1997) & $^{43}$ Pettersen et al.\ (1989) \\
 $^{5}$ Muzerolle et al.\ (2000)	& $^{18}$ Neff et al.\ (1996) 
 	& $^{31}$ Torres et al.\ (2003) & $^{44}$ Benedict et al.\ (1998) \\
 $^{6}$ Singh et al.\ (1996a)		& $^{19}$ Berdyugina et al.\ (1998) 
 	& $^{32}$ Favata et al.\ (1995) & $^{45}$ Contadakis\ (1995) \\
 $^{7}$ Ayres et al.\ (1998)		& $^{20}$ Benedict et al.\ (1999) 
 	& $^{33}$ Al-Naimiy\ (1981) 	& $^{46}$ Pakull et al.(1981) \\
 $^{8}$ Gadun\ (1994)			& $^{21}$ Gondoin\ (1999)
 	& $^{34}$ Padmakar\ (1999) 	& $^{47}$ Alencar \& Batalha\ (2002) \\
 $^{9}$ Ayres et al.\ (1999)		& $^{22}$ Pizzolato et al.\ (2000) 
 	& $^{35}$ Gehren et al.\ (1999) & $^{48}$ Simon\ (1986) \\
 $^{10}$ Ness et al.\ (2002)		& $^{23}$ Decin et al.\ (2003) 
 	& $^{36}$ Lanzafame et al.\ (2000) & $^{49}$ Marino et al.(1999) \\
 $^{11}$ Strassmeier et al.\ (1993) 	& $^{24}$ Murad \& Budding\ (1984)
 	& $^{37}$ Gliese et al.\ (1969) &  \\
 $^{12}$ Hill et al.\ (1989)		& $^{25}$ Cayrel et al.\ (1994)
 	& $^{38}$ Redfield et al.\ (2003) & \\
 $^{13}$ Gimenez et al.\ (1986) 	& $^{26}$ Donati et al.\ (1995)
 	& $^{39}$ Katsova \& Tsikoudi\ (1993) & \\
\end{tabular}
}
\enddata 
\tablenotetext{a}{from the SIMBAD database}
\tablenotetext{b}{from ref.~(11) when no other reference is indicated}
\tablenotetext{c}{from SIMBAD when no other reference is indicated}
\tablenotetext{d}{if not differently indicated, derived as 
		described by Flower\ (1996)}
\tablenotetext{e}{from ref.~(11) when no other reference is 
		indicated; for $\beta$~Cet, Canopus and 
		$\mu$~Vel the listed values are	upper limits 
		derived from $v$ sen$i$}
\tablenotetext{f}{The Rossby number, $\mathcal{R}$, is defined
		as the ratio between the observed rotation 
		period, $P_{\mathrm{rot}}$, and the convective turnover 
		time, $\tau_{\mathrm{conv}}$, derived from theoretical 
		models (Pizzolato et al.\ 2001)}
\end{deluxetable}

\begin{deluxetable}{lccccc}
\tablecolumns{6} 
\tabletypesize{\footnotesize}
\tablecaption{Parameters of the HETG observations.
		 \label{tab2}}
\tablewidth{0pt}
\tablehead{
 \colhead{Source} &  \colhead{Obs ID} & 
 \colhead{L$_{\mathrm{X}}$ \tablenotemark{a}} & 
 \colhead{L$_{\mathrm{X}}$ \tablenotemark{b}} & 
 \colhead{F$_{\mathrm{X}}$ \tablenotemark{c}} & 
 \colhead{$t_{\mathrm{exp}}$} \\
 & & [erg/sec] & [erg/sec] & [$10^5$erg/cm$^2$/sec] & [ksec] 
}
\startdata
 AU~Mic   	& 17 & 1.05$\cdot 10^{29}$ 
 		& 1.29$\cdot 10^{29}$ & 60.7 & 58.8 \\  
 Prox~Cen 	& 2388 & 3.33$\cdot 10^{26}$ 
 		&  4.56$\cdot 10^{26}$ & 2.93 & 42.4 \\ 
 EV~Lac		& 1885 & 2.59$\cdot 10^{28}$ 
 		& 3.19$\cdot 10^{28}$ & 31.1 & 100 \\
 AB~Dor   	& 16   & 6.83$\cdot 10^{29}$ 
 		& 8.26$\cdot 10^{29}$ & 136  & 52.3 \\
 TW~Hya		& 5    & 1.15$\cdot 10^{30}$ 
 		& 1.30$\cdot 10^{30}$ & 214  & 47.7 \\ 
 HD~223460 	& 1892 & 6.31$\cdot 10^{31}$ 
 		& 5.96$\cdot 10^{31}$ & 52.9 & 95.7 \\  
 31~Com 	& 1891 & 4.85$\cdot 10^{30}$ 
 		& 5.15$\cdot 10^{30}$ & 13.2 & 130.2 \\ 
 $\beta$~Ceti 	& 974 & 2.16$\cdot 10^{30}$ 
 		& 2.62$\cdot 10^{30}$ & 1.89 & 86.1 \\  
 Canopus	& 636  & 2.48$\cdot 10^{30}$ 
 		& 2.94$\cdot 10^{30}$ & 0.17 & 94.6 \\   
 $\mu$~Vel	& 1890 & 1.19$\cdot 10^{30}$ 
 		& 1.41$\cdot 10^{30}$ & 1.37 & 19.7 \\   
 $\mu$~Vel	& 3410 & 1.17$\cdot 10^{30}$ 
 		& 1.45$\cdot 10^{30}$ & 1.4  & 57 \\   
 Algol		& 604  & 8.50$\cdot 10^{30}$ 
 		& 8.88$\cdot 10^{30}$ & 119  & 51.7 \\  
 ER~Vul		& 1887 & 2.46$\cdot 10^{30}$ 
 		& 2.78$\cdot 10^{30}$ & 399  & 112.0 \\ 
 44~Boo		& 14   & 3.59$\cdot 10^{29}$ 
 		& 4.56$\cdot 10^{29}$ & 172  & 59.1 \\
 TZ~CrB		& 15   & 2.99$\cdot 10^{30}$ 
 		& 3.34$\cdot 10^{30}$ & 453  & 83.7 \\  
 UX~Ari		& 605  & 5.69$\cdot 10^{30}$ 
 		& 7.26$\cdot 10^{30}$ & 35.7 & 48.5 \\  
 $\xi$~UMa	& 1894 & 8.78$\cdot 10^{28}$ 
 		& 1.23$\cdot 10^{29}$ & 22.4 & 70.9 \\  
 II~Peg		& 1451 & 1.56$\cdot 10^{31}$ 
 		& 1.76$\cdot 10^{31}$ & 250  & 42.7 \\  
 $\lambda$~And 	& 609 & 1.34$\cdot 10^{30}$ 
 		& 1.85$\cdot 10^{30}$ & 5.55 & 81.9 \\  
 TY~Pyx		& 601  & 4.71$\cdot 10^{30}$ 
 		& 5.10$\cdot 10^{30}$ & 573  & 49.1 \\  
 AR~Lac		& 6    & 5.21$\cdot 10^{30}$ 
 		& 5.60$\cdot 10^{30}$ & 284  & 32.1 \\   
 AR~Lac		& 9    & 5.61$\cdot 10^{30}$ 
 		& 6.30$\cdot 10^{30}$ & 319  & 32.2 \\  
 HR~1099	& 62538 & 7.85$\cdot 10^{30}$ 
 		& 1.05$\cdot 10^{31}$ & 113  & 94.7 \\  
 IM~Peg		& 2527 & 2.75$\cdot 10^{31}$ 
 		& 2.79$\cdot 10^{31}$ & 27.1 & 24.6 \\ 
 IM~Peg     	& 2528 & 2.17$\cdot 10^{31}$ 
 		& 2.30$\cdot 10^{31}$ & 22.3 & 24.8 \\ 
 IM~Peg     	& 2529 & 1.86$\cdot 10^{31}$ 
 		& 1.97$\cdot 10^{31}$ & 19.2 & 24.8  
 \enddata 
\tablenotetext{a}{relative to the HEG range: 1.5-15\AA}
\tablenotetext{b}{relative to the MEG range: 2-24\AA}
\tablenotetext{c}{from L$_{\mathrm{X}}$ obtained from MEG spectra}
\tablecomments{In order to calculate the X-ray surface flux,
		$F_{\mathrm{X}}$, for binary systems, we used 
		the radius $R_{\star}$ of the component which 
		probably is the stronger X-ray source, e.g.\ the 
		K1IV component for Algol.}
\end{deluxetable}

\begin{deluxetable}{cccc}
\tablecolumns{4} 
\tabletypesize{\footnotesize}
\tablecaption{Theoretical wavelength of the He-like triplet 
		lines.	\label{tab3}}
\tablewidth{0pt}
\tablehead{
 \colhead{ion} & \multicolumn{3}{c}{wavelength (\AA)} \\
       \cline{2-4} \\[-0.2cm]
 \colhead{} & \colhead{{\em resonance (r)}} & 
 \colhead{{\em intercombination (i)}}&  
 \colhead{{\em forbidden (f)}} \\
 \colhead{} & 
 \colhead{1s$^{2}$ $^{1}$S$_0$ - 1s2p $^{1}$P$_{1}$} & 
 \colhead{1s$^{2}$ $^{1}$S$_0$ - 1s2p $^{3}$P$_{2,1}$} & 
 \colhead{1s$^{2}$ $^{1}$S$_0$ - 1s2s $^{3}$S$_{1}$}  
}
\startdata 
 Si~XIII & 6.6479  & 6.6850 / 6.6882    & 6.7403 \\
 Mg~XI	 & 9.1687  & 9.2282 / 9.2312    & 9.3143 \\
 O~VII	 & 21.6015 & 21.8010 / 21.8036  & 22.0977 
 \enddata
\end{deluxetable}

\begin{deluxetable}{lccccccccc}
\tablecolumns{10} 
\tabletypesize{\footnotesize}
\tablecaption{Neon H-like Lyman series. \label{tab4}}
\tablewidth{0pt}
\tablehead{
\colhead{transition} & \colhead{Ly$\alpha$:  2$\rightarrow$1} & 
\colhead{Ly$\beta$: 3$\rightarrow$1} & \colhead{4$\rightarrow$1} & 
\colhead{5$\rightarrow$1} & \colhead{6$\rightarrow$1} & 
\colhead{7$\rightarrow$1} & \colhead{8$\rightarrow$1} & 
\colhead{9$\rightarrow$1} & \colhead{10$\rightarrow$1}
}
\startdata
 $\lambda$ (\AA) & 12.134 &  10.239 & 9.708 & 
	 	9.481 & 9.362 & 9.291 & 9.246 & 9.215 & 9.194 \\
 {\em f \tablenotemark{a}} & 1.0000 & 0.1900 & 0.0697 & 0.0335 
 		& 0.01874 & 0.01157 & 0.0076 & 0.0053 & 0.0038  
\enddata 
\tablenotetext{a}{oscillator strength normalized to the $f$-value
		of the Ly$\alpha$ transition.}
\end{deluxetable}

\begin{deluxetable}{clcll}
\tabletypesize{\footnotesize}
\tablecaption{Mg~XI triplet lines and blending components. 
        \label{tab5}}
\tablewidth{0pt}
\tablehead{
 \colhead{$\lambda_{\mathrm{obs}}$ \tablenotemark{a}} & 
 \colhead{Ion} & \colhead{} &  \colhead{transition} & 
 \colhead{Emissivity\tablenotemark{b}}\\
 \colhead{[\AA]} &  \colhead{} & \colhead{} & \colhead{} & 
 \colhead{[ph cm$^3$s$^{-1}$]}
}
\startdata
 9.1689 & Mg~XI   & $r$ & $1s2p~^1P_1 \rightarrow 1s^2~^1S_0$ & 
 	1.10$\times 10^{-16}$ \\[0.07cm]
 9.1900 & Fe~XXI  & & 
        $1s^2 2s2p_{1/2}^24p_{3/2} \rightarrow 1s^2 2s^2 2p^2~^3P_{0}$
	& 1.14$\times 10^{-17}$\\
        & Fe~XX   & & 
        $1s^22s2p_{1/2}2p_{3/2}^24p_{3/2} \rightarrow
 	2s^22p^3~^4S_{3/2}$ & 3.29$\times 10^{-18}$ \\
        & Ne~X    & 10$\rightarrow$1 & & 1$\times 10^{-18}$ \\[0.07cm]
 9.2187 & Fe~XX   & & 
        $2s^22p^2(^3P)5d~^2F_{5/2} \rightarrow 2s^22p^3~^2D_{3/2}$ & 
	1.63$\times 10^{-18}$\\
        & Ne~X    & 9$\rightarrow$1 & & 1.4$\times 10^{-18}$ \\[0.07cm]
 9.2304 & Mg~XI   & $i$ & $1s2p~^3P_{2,1} \rightarrow 1s^2~^1S_0$ 
 	& 1.60$\times 10^{-17}$, 2.23$\times 10^{-18}$ \\[0.07cm]
 9.2467 & Ne~X    & 8$\rightarrow$1 & &  2.1$\times 10^{-18}$ \\[0.07cm]
 9.2882 & Ne~X    & 7$\rightarrow$1 & &  3.1$\times 10^{-18}$ \\
 	& Fe~XX   & & 
        $2s^22p^2(^3P)5d~^4P_{5/2} \rightarrow 2s^22p^3~^2D_{3/2}$ &
 	2.82$\times 10^{-18}$ \\
        & Fe~XXII & & 
        $1s^22s2p(^3P)4d~^4D_{3/2} \rightarrow  1s^22s2p^2~^2D_{3/2}$
 	& 1.37$\times 10^{-18}$ \\[0.07cm]
 9.3144 & Mg~XI   & $f$ & $1s2s~^3S_1 \rightarrow 1s^2~^1S_0$ & 
 	5.31$\times 10^{-17}$ 
\enddata 
\tablenotetext{a}{From the fit to the HEG coadded spectrum of
        $\beta$~Ceti, TZ~CrB, AR~Lac, and HR~1099 shown in 
        Fig.~\ref{fig7}.}
\tablenotetext{b}{Peak line emissivity from the APED database for an
        electron density $n_e=1\times 10^{10}~cm^{-3}$, and assuming 
	the chemical composition of Anders \& Grevesse (1989).
	The excitation rate for a Maxwellian distribution of electron
	velocities for higher $n$ series in hydrogenic ions is 
	essentially proportional to the $f$-value of the transition 
	(e.g., Lang 1999); since emissivities for the Ne Lyman series 
	lines do not appear to be available for upper level $n>5$, the 
	emissivities listed here are the APED emissivity of the 
	$5\rightarrow 1$ transition, scaled in proportion to the 
	transition $f$-values oscillator strengths (see Tab.~\ref{tab4}).}
\end{deluxetable}

\begin{deluxetable}{lrrrrrrrrrrrr}
\tablecolumns{13} 
\tabletypesize{\tiny}
\tablecaption{Measured Line fluxes (in 
	$10^{-6}$~photons~cm$^{-2}$~sec$^{-1}$) with 
	1$\sigma$ errors. \label{tab6}}
\tablewidth{0pt}
\tablehead{
 \colhead{Source} & \colhead{grating} & 
 \multicolumn{11}{c}{flux \tablenotemark{a}}\\
 \cline{3-13}\\[-0.15cm]
 \colhead{} & \colhead{} & 
 \multicolumn{3}{c}{Si~XIII} & \colhead{} 
 & \multicolumn{3}{c}{Mg~XI} & \colhead{} & 
 \multicolumn{3}{c}{O~VII} \\
 \cline{3-5} \cline{7-9} \cline{11-13} \\[-0.2cm]
 \colhead{} & \colhead{} & \colhead{$r$} & \colhead{$i$} 
 & \colhead{$f$} & \colhead{} & \colhead{$r$} 
 & \colhead{$i$} & \colhead{$f$} & \colhead{} 
 & \colhead{$r$} & \colhead{$i$} & \colhead{$f$} 
}
\startdata 
 AU~Mic 	& HEG & $24 \pm 5$ & $7 \pm 4$ & $19 \pm 4$ 
 		&  & - & - & - &  & - & - & -   \\
 		& MEG & $27.3 \pm 2.8$ & $6.1 \pm 1.9$ 
		& $16.5 \pm 2.1 $ &  & $19.4 \pm 1.7$ 
		& $5.0 \pm 1.8$ & $11.9 \pm 1.7$ & 
		& $180 \pm 35$ & $48 \pm 29$ & $150 \pm 40 $ \\[0.1cm]
Prox~Cen 	& HEG & - & - & - &  & - & - & - &  & - & - & - \\
 		& MEG & $7.5 \pm 2.2$ & $2.0 \pm 1.8$ 
		& $4.2 \pm 1.7$ & & $5.2 \pm 1.5$ & $2.2 \pm 1.5$ 
		& $6.5 \pm 1.5 $ &  & - & - & - \\[0.1cm]
EV~Lac 		& HEG & $31 \pm 4$ & $6.6 \pm 2.4$ 
		& $20.5 \pm 2.8$ & & $18 \pm 4$ & $6 \pm 3$ 
		& $9 \pm 3$ &  & - & - & -  \\
 		& MEG & $29.3 \pm 2.0$ & $7.8 \pm 1.4$ 
		& $17.3 \pm 1.5$ & & $18.9 \pm 0.6$ 
		& $6.3 \pm 0.7$ & $9.3 \pm 0.6$ &  
		& $207 \pm 26$ & $78 \pm 20$ & $117 \pm 25 $ \\[0.1cm]
AB~Dor 		& HEG & $56 \pm 7$ & $13 \pm 5$ 
		& $45 \pm 5$ &  & $66 \pm 9$ & $15 \pm 7$ 
		& $24 \pm 7$ &  & - & - & -   \\
 		& MEG & $49 \pm 4$ & $9.8 \pm 2.4$ & $33 \pm 3$ &  
		& $48.9 \pm 1.5$ & $13.8 \pm 1.5$ & $37.3 \pm 1.5$ 
		& & $276 \pm 45$ & $100 \pm 40$ & $176 \pm 50 $ \\[0.1cm]
TW~Hya 		& HEG & - & - & - &  & - & - & - &  & - & - & - \\
 		& MEG & $2.8  \pm 1.4$ & $1.8 \pm 1.2$ 
		& $3.0 \pm 1.2$ & & - & - & - &  & $90 \pm 30$ 
		& $73 \pm 30$ & $< 4 $ \\[0.1cm]
HD~223460 	& HEG & $32 \pm 4$ & $10 \pm 3$ & $24 \pm 3$ &  
		& - & - & - &  & - & - & -   \\
 		& MEG & $24.5 \pm 2.2$ & $9.6 \pm 1.8$ 
		& $17.2 \pm 1.9$ & & $19.4 \pm 0.9$ & $5.6 \pm 1.0$ 
		& $10.7 \pm 1.0$ &  & $36 \pm 17$ & - & - \\[0.1cm]
31~Com		& HEG & $19.0 \pm 2.6$ & $4.1 \pm 1.6$ 
		& $9.3 \pm 1.8$ &  & $13 \pm 3$ & $4.0 \pm 2.2$ 
		& $6.1 \pm 2.5$ &  & - & - & -   \\
 		& MEG & $21.1 \pm 1.5$ & $3.6 \pm 0.9$ 
		& $12.6 \pm 1.1$ &  & $15.2 \pm 0.5$ & $3.5 \pm 0.5$ 
		& $7.1 \pm 0.6$ & & $28 \pm 13$ & - & - \\[0.1cm]
$\beta$~Cet 	& HEG & $81 \pm 6$ & $21 \pm 3$ & $66 \pm 5$ &  
		& $113 \pm 7$ & $26 \pm 5$ & $78 \pm  7$ &  
		& - & - & -   \\
 		& MEG & $81 \pm 3$ & $20.6 \pm 2.1$ 
		& $49.9 \pm 2.5$ &  & $108 \pm 2.0$ 
		& $28.7 \pm 1.2$ & $64.5 \pm 1.2$ &  
		& $100 \pm 24$ & - & - \\[0.1cm]
Canopus 	& HEG & $8.6 \pm 1.7$ & $1.9 \pm 1.7$ 
		& $6.2 \pm 1.3$ &  & - & - & - &  & - & - & - \\
 		& MEG & $8.2 \pm 1.2$ & $1.5 \pm 0.8$ 
		& $6.7 \pm 1.0$ &  & $9.0 \pm 0.7$ & $2.6 \pm 0.7$ 
		& $5.0 \pm 0.7$ &  & $31 \pm 16$ & - & - \\[0.1cm]
$\mu$~Vel 	& HEG & $33 \pm 5$ & $10 \pm 3.5$ & $20 \pm 4$ 
		&  & $57 \pm 7$ & $10 \pm 5$ & $31 \pm  6$ &  
		& - & - & -   \\
 		& MEG & $30.4 \pm 2.4$ & $9.9 \pm 1.7$ 
		& $19.3 \pm 1.9$ &  & $45.0 \pm 1.1$ 
		& $7.6 \pm 1.16$ & $23.9 \pm 1.2$ &  
		& $45 \pm 26$ & - & - \\[0.1cm]
Algol 		& HEG & $133 \pm 10$ & $36 \pm 7$ & $85 \pm 8$ 
		&  & $126 \pm 13$ & $31 \pm 10$ & $78 \pm  12$ 
		&  & - & - & -   \\
 		& MEG & $125 \pm 7$ & $31 \pm 5$ & $76 \pm 4$ 
		&  & $102 \pm 2.8$ & $24.0 \pm 2.3$ 
		& $57.6 \pm 2.1$ & & $200 \pm 33$ & - & - \\[0.1cm]
ER~Vul 		& HEG & $22 \pm 3$ & $6.7 \pm 2.3$ 
		& $13.8 \pm 2.5$ &  & $33 \pm 4$ & $8 \pm 3$ 
		& $19 \pm  4$ &  & - & - & -   \\
 		& MEG & $24.1 \pm 1.7$ & $4.6 \pm 1.1$ 
		& $15.0 \pm 1.3$ &  & $22.9 \pm 0.6$ 
		& $5.59 \pm 0.6$ & $15.2 \pm 0.6 $ 
		&  & $60 \pm 19$ & - & - \\[0.1cm]
44~Boo 		& HEG & $50 \pm 7$ & $11 \pm 4$ & $38 \pm 5$ 
		&  & $95 \pm 10 $ & $21 \pm 7$ & $52 \pm  9$ 
		&  & - & - & -   \\
 		& MEG & $44 \pm 3$ & $8.9 \pm 2.4$ 
		& $28.9 \pm 2.6$ & & $63.1 \pm 2.5$ 
		& $14.8 \pm 1.2$ & $38.6 \pm 1.2 $ &  
		& $280 \pm 50$ & $100 \pm 40$ & $170 \pm 50 $ \\[0.1cm]
TZ~CrB 		& HEG & $123 \pm 7$ & $20 \pm 4$ & $72 \pm 5$ 
		&  & $176 \pm 9$ & $45 \pm 6$ & $89 \pm 8$ 
		&  & - & - & -   \\
 		& MEG & $129 \pm 4$ & $25.0 \pm 2.7$ 
		& $73 \pm 4$ & & $152 \pm 4$ & $33.0 \pm 2.8$ 
		& $78 \pm 2.0$ &  & $280 \pm 40$ & $99 \pm 27$ 
		& $230 \pm 40 $ \\[0.1cm]
UX~Ari 		& HEG & $33 \pm 7$ & $14 \pm 5$ & $24 \pm 5$ 
		&  & $31 \pm 7$ & $15 \pm 7$ & $22 \pm  7$ &  
		& - & - & -   \\
 		& MEG & $31 \pm 3$ & $7.3 \pm 2.4$ 
		& $24.0 \pm 2.7$ &  & $27.5 \pm 1.8$ 
		& $12.4 \pm 1.3$ & $17.9 \pm 1.3$ &  
		& $150 \pm 40$ & $27 \pm 27$ & $60 \pm 30 $ \\[0.1cm]
$\xi$~UMa 	& HEG & $41 \pm 5$ & $13 \pm 4$ & $27 \pm 4$ 
		&  & $76 \pm 7$ & $19 \pm 5$ & $55 \pm  7$ &  
		& - & - & -   \\
 		& MEG & $46 \pm 3$ & $8.4 \pm 1.8$ 
		& $24.7 \pm 2.1$ &  & $68.1 \pm 2.1$ 
		& $18.7 \pm 0.9$ & $42.4 \pm 1.2 $ &  
		& $360 \pm 40$ & $84 \pm 29$ & $200 \pm 40 $ \\[0.1cm]
II~Peg 		& HEG & $95 \pm 10$ & $27 \pm 7$ & $50 \pm 7$ &  
		& $68 \pm 11 $ & $32 \pm 10$ & $46 \pm  11$ &  
		& - & - & -   \\
 		& MEG & - & - & - &  & $60.3 \pm 2.2$ 
		& $20.3 \pm 1.8 $ & $34.3 \pm 2.1$ &  
		& $255 \pm 60$ & $127 \pm 50 $ 	& $210 \pm 60 $ \\[0.1cm]
$\lambda$~And 	& HEG & $61 \pm 5$ & $14 \pm 3$ & $41 \pm 4$ &  
		& $93 \pm 7$ & $15 \pm 4$ & $45 \pm  5$ &  
		& - & - & -   \\
 		& MEG & $58 \pm 3$ & $7.6 \pm 1.7$ 
		& $31.6 \pm 2.2$ &  & $84.5 \pm 0.8$ 
		& $16.8 \pm 0.7$ & $40.3 \pm 0.8 $ &  
		& $135 \pm 27$ & $36 \pm 22$ & $88 \pm 28 $ \\[0.1cm]
TY~Pyx 		& HEG & $29 \pm 6$ & $8 \pm 4$ & $22 \pm 5$ &  
		& - & - & - &  & - & - & -   \\
 		& MEG & $28 \pm 3$ & $7.7 \pm 2.2$ 
		& $19.4 \pm 2.5$ &  & $29.1 \pm 1.4$ 
		& $6.7 \pm 1.5$ & $15.8 \pm 1.4 $ 
		&  & $77 \pm 30$ & - & - \\[0.1cm]
AR~Lac 		& HEG & $61 \pm 6$ & $19 \pm 5$ & $47 \pm 5$ &  
		& $64 \pm 9 $ & $23 \pm 7$ & $44 \pm  9$ &  
		& - & - & -   \\
 		& MEG & $51 \pm 3$ & $10.9 \pm 2.3$ 
		& $30.9 \pm 2.6$ &  & $69.0 \pm 1.6$ 
		& $14.8 \pm 1.4$ & $36.8 \pm 1.5$ &  
		& $160 \pm 40$ & - & - \\[0.1cm]
HR~1099 	& HEG & $83 \pm 6.$ & $23 \pm 4$ & $64 \pm 5$ 
		&  & $108 \pm 8 $ & $26 \pm 6$ & $55 \pm  7$ &  
		& - & - & -   \\
 		& MEG & $102 \pm 4$ & $15.7 \pm 2.6$ 
		& $59 \pm 3$ &  & $97 \pm 4 $ 
		& $23 \pm 2$ & $52 \pm 2$ &  
		& $400 \pm 40 $ & $83 \pm 27$ & $240 \pm 40 $ \\[0.1cm]
IM~Peg 		& HEG & $34 \pm 6$ & $8 \pm 6$ & $22 \pm 5$ &  
		& - & - & - &  & - & - & -   \\
		& MEG & $37 \pm 3$ & $9.3 \pm 2.5$ 
		& $17.6 \pm 2.4$ &  & $25.7 \pm 1.6$ 
		& $7.1 \pm 1.7$ & $13.5 \pm 1.7$ &  
		& $30 \pm 16$ & - & - 
 \enddata
\tablenotetext{a}{r, resonance line, i, intercombination lines,	f, 
	forbidden line, corresponding to the transitions described 
	in Table~\ref{tab3}.}
\end{deluxetable}

\begin{deluxetable}{lrrrrr}
\tablecolumns{6} 
\tabletypesize{\footnotesize}
\tablecaption{Line flux measurements.
	 \label{tab7}}
\tablewidth{0pt}
\tablehead{
 \colhead{Source} & 
 \multicolumn{5}{c}{flux \tablenotemark{a}} \\
 \cline{2-6} \\
 \colhead{} & \colhead{Mg~XII} & \colhead{Ne~X} &  
 \colhead{O~VIII} & \colhead{Fe~XVIII} & 
 \colhead{Fe~XXI} \\
 \colhead{} & \colhead{Ly$\alpha$} & 
 \colhead{Ly$\alpha$} & \colhead{Ly$\alpha$} & 
 \colhead{} & \colhead{} \\
 \colhead{} & \colhead{8.419\AA} &  
 \colhead{12.132\AA} & \colhead{18.967\AA} & 
 \colhead{14.210\AA} & \colhead{12.292\AA} 
}
\startdata 
AU~Mic 		& $22 \pm 5$ & $324 \pm 13$ & $790 \pm 40$ 
		& $59 \pm 11$ & $30 \pm 6 $ \\
Prox~Cen 	& - & $60 \pm 7$ & $220 \pm 30$ & -  & -  \\
EV~Lac 		& $15.0 \pm 1.8$ & $230 \pm 8$ & $673 \pm 29$ 
		& $61 \pm 7$ & $23 \pm 3 $ \\
AB~Dor 		& $64 \pm 3$ & $683 \pm 19$ & $1470 \pm 50$ 
		& $201 \pm 15$ & $169 \pm  12 $ \\
TW~Hya 		& - & $85 \pm 8$ & $240 \pm 30$ & - & - \\
HD~223460 	& $56 \pm 3$ & $129 \pm 6$ & $168 \pm 18$ 
		& $54 \pm 8$ & $50 \pm  5 $ \\
31~Com	 	& $22.3 \pm 1.7$ & $59 \pm 4$ & $97 \pm 12$ 
		& $37 \pm 5$ & $50 \pm  4 $ \\
$\beta$~Cet 	& $100 \pm 3$ & $397 \pm 11$ & $596 \pm 29$ 
		& $403 \pm 14$ & $136 \pm  7 $ \\
Canopus 	& $8.5 \pm 1.0$ & $55 \pm 4$ & $109 \pm 15$ 
		& $33 \pm 5$ & $15.6 \pm  2.8 $ \\
$\mu$~Vel 	& $39 \pm 3$ & $143 \pm 78$ & $263 \pm 26$ 
		& $162 \pm 13$ & $49 \pm  5 $ \\
Algol 		& $207 \pm 4$ & $872 \pm 23$ & $1180 \pm 50$ 
		& $259 \pm 18$ & $258 \pm  12 $ \\
ER~Vul 		& $36 \pm 3$ & $188 \pm 7$ & $262 \pm 20$ 
		& $84 \pm 7$ & $53 \pm  4 $ \\
44~Boo 		& $60 \pm 4$ & $541 \pm 14$ & $1310 \pm 50$ 
		& $158 \pm 15$ & $75 \pm  8 $ \\
TZ~CrB 		& $188 \pm 3$ & $1059 \pm 16$ & $1830 \pm 50$ 
		& $440 \pm 19$ & $269 \pm  10 $ \\
UX~Ari 		& $48.1 \pm 2.4$ & $743 \pm 20$ & $910 \pm 50$ 
		& $76 \pm 12$ & $58 \pm  7 $ \\
$\xi$~UMa 	& $50 \pm 3$ & $371 \pm 11$ & $1160 \pm 40$ 
		& $173 \pm 12$ & $44 \pm  5 $ \\
II~Peg 		& $100 \pm 4$ & $1181 \pm 28$ & $1930 \pm 60$ 
		& $106 \pm 15$ & $105 \pm  11 $ \\
$\lambda$~And 	& $116 \pm 3$ & $553 \pm 12$ & $950 \pm 30$ 
		& $119 \pm 10$ & $95 \pm  6 $ \\
TY~Pyx 		& $60 \pm 4$ & $294 \pm 13$ & $360 \pm 30$ 
		& $78 \pm 12$ & $71 \pm  8 $ \\
AR~Lac 		& $89 \pm 4$ & $632 \pm 16$ & $820 \pm 40$ 
		& $161 \pm 15$ & $126 \pm  10 $ \\
HR~1099 	& $137 \pm 4$ & $1705 \pm 21$ & $2860 \pm 60$ 
		& $198 \pm 12$ & $207 \pm  9 $ \\
IM~Peg 		& $46 \pm 4$ & $334 \pm 11$ & $340 \pm 30$ 
		& $76 \pm 12$ & $61 \pm  7 $ 
 \enddata
\tablenotetext{a}{fluxes, in 
		$10^{-6}$~photons~cm$^{-2}$~sec$^{-1}$,
		from MEG spectra, with 1$\sigma$ errors.}
\end{deluxetable}

\begin{deluxetable}{lcccc}
\tablecolumns{5} 
\tabletypesize{\footnotesize}
\tablecaption{Densities. 
	\label{tab8}}
\tablewidth{0pt}
\tablehead{
 \colhead{Source\tablenotemark{a}} 
 & \multicolumn{4}{c}{$n_{\mathrm{e}}$ (cm$^{-3}$)} \\
 \cline{2-5} \\[-0.15cm]
 \colhead{} & \multicolumn{2}{c}{Mg~XI} 
 & \colhead{} & \colhead{O~VII} \\
 \cline{2-3} \cline{5-5} \\[-0.15cm]
 \colhead{} & \colhead{HEG} & \colhead{MEG} 
 & \colhead{} & \colhead{MEG}
}
\startdata 
 AU~Mic		& - & $<5.62 \cdot 10^{12}$ &  
 		& $<5.62 \cdot 10^{11}$ \\[0.05cm]
 Prox~Cen 	& - & $<5.6 \cdot 10^{12}$ &  & - \\[0.05cm]
 EV~Lac		& $5.6^{+26}_{-4} \cdot 10^{12}$
 		& $5.6^{+4}_{-0.6} \cdot 10^{12}$
 		&  & $5.6^{+2.5}_{-1.4} \cdot 10^{10}$ \\[0.05cm]
 AB~Dor 	& $<3.2 \cdot 10^{13}$ & $<1.8 \cdot 10^{12}$
 		& & $5.6^{+7}_{-1.4} \cdot 10^{10}$ \\[0.05cm]
 TW~Hya 	& - & - &  & $>7.8 \cdot 10^{11}$ \\[0.05cm] \hline
 HD~223460 	& - & $3.2^{+2.5}_{-1.4} \cdot 10^{12}$ 
 		& & - \\[0.05cm]
 31~Com 	& $5.6^{+26}_{-5} \cdot 10^{12}$
 		& $3.2^{+2.5}_{-1.4} \cdot 10^{12}$ 
		& &  - \\[0.05cm]
 $\beta$~Cet 	& $<1.8 \cdot 10^{12}$
		& $1.8^{+0.6}_{- 0.6} \cdot 10^{12}$ 
		& &  - \\[0.05cm]
 Canopus 	& - & $3.2^{+7}_{-2.2} \cdot 10^{12}$ 
 		& &  - \\[0.05cm]
 $\mu$~Vel 	& $<5.6 \cdot 10^{12}$ 
 		& $<1.0 \cdot 10^{12}$ &  & - \\[0.05cm] \hline
 Algol 		& $<5.6 \cdot 10^{12}$
 		& $1.8^{+0.6}_{-1.2} \cdot 10^{12}$ 
		& & - \\[0.05cm]
 ER~Vul 	& $<5.6 \cdot 10^{12}$ 
 		& $<1.8 \cdot 10^{12}$ & &  - \\[0.05cm]
 44~Boo 	& $<5.6 \cdot 10^{12}$ 
 		& $1.0^{+0.8}_{-0.8} \cdot 10^{12}$
 		& & $1.0^{+7}_{-0.4} \cdot 10^{10}$ \\[0.05cm]
 TZ~CrB 	& $3.2^{+2.5}_{-1.4} \cdot 10^{12}$
 		& $1.8^{+0.6}_{-0.6} \cdot 10^{12}$
 		& & $1.8^{+1.4}_{-1.2} \cdot 10^{10}$ \\[0.05cm]
 UX~Ari 	& $5.6^{+12}_{-4} \cdot 10^{12}$
 		& $10^{+0.6}_{-4} \cdot 10^{12}$
 		& &  - \\[0.05cm]
 $\xi$~UMa 	& $1.8^{+30}_{-6} \cdot 10^{11}$
 		& $3.2^{+0.6}_{-1.4} \cdot 10^{12}$
 		& & $1.8^{+1.4}_{-1.2} \cdot 10^{10}$ \\[0.05cm]
 II~Peg 	& $5.6^{+12}_{-2.5} \cdot 10^{12}$
 		& $5.6^{+0.6}_{-2.5} \cdot 10^{12}$
 		& & $3.2^{+7}_{-1.4} \cdot 10^{10}$ \\[0.05cm]
 $\lambda$~And 	& $<3.2 \cdot 10^{12}$
 		& $1.8^{+0.6}_{-0.8} \cdot 10^{12}$ & 
		& $<1.0 \cdot 10^{11}$ \\[0.05cm]
 TY~Pyx 	& - & $1.8^{+1.4}_{-1.6} \cdot 10^{12}$ 
 		& &  - \\[0.05cm]
 AR~Lac 	& $3.2^{+7}_{-2.2} \cdot 10^{12}$ 
 		& $<1.8 \cdot 10^{12}$ &  & - \\[0.05cm]
 HR~1099 	& $3.2^{+2.5}_{-2.2} \cdot 10^{12}$
 		& $1.8^{+0.6}_{-0.6} \cdot 10^{12}$
 		& & $1.0^{+2.2}_{-0.8} \cdot 10^{10}$ \\[0.05cm]
 IM~Peg 	& - & $3.2^{+2.5}_{-1.4} \cdot 10^{12}$ 
 		& &  - 
\enddata
\tablenotetext{a}{Horizontal lines divide the different classes 
	of source into single dwarfs (top), single giants (middle), 
	and active binaries (bottom).}
\end{deluxetable}

\begin{deluxetable}{lccl|cccllc}
\tablecolumns{10} 
\tabletypesize{\footnotesize}
\tablecaption{Comparison of derived densities with previous work: 
	{\it single stars}. \label{tab9}}
\tablewidth{0pt}
\tablehead{
 \colhead{} & \multicolumn{3}{c}{This work} & 
 \colhead{} & \multicolumn{5}{c}{Previous works} \\
 \cline{2-4} \cline{6-10}\\[-0.2cm]
 \colhead{Source\tablenotemark{a}} & \colhead{$\log$T\tablenotemark{b}} & 
 \colhead{$\log n_{\mathrm{e}}$} & \colhead{Ion\tablenotemark{c}} & 
 \colhead{} & \colhead{$\log$T\tablenotemark{b}} & 
 \colhead{$\log n_{\mathrm{e}}$} & \colhead{Ion\tablenotemark{c}} &
 \colhead{data\tablenotemark{d}} & \colhead{Ref.} \\
 \colhead{} & \colhead{[MK]} & \colhead{[cm$^{-3}$]} & \colhead{} & 
 \colhead{} & \colhead{[MK]} & \colhead{[cm$^{-3}$]} & \colhead{} & 
 \colhead{} & \colhead{}
}
\startdata 
 AU~Mic		& 6.8 & $< 12.75$ & MgXI &
 		& 7 & $\sim 13$ & FeXXI & D & 1 \\
 		& 6.3 & $< 11.75$ & OVII & & 6.3 
		& $10.0-10.3$ (flaring) & OVII & C & 2 \\
 		& & & & & 6.3 & $\lesssim 9$ (quiescent)
		& OVII & C & 2 \\
 		&  &  & & & 5.3 & $10.7-11$ & OV, OIV, SIV 
		& D,E,F & 3,4 \\[0.1cm] 
 Prox~Cen 	& 6.8 & $< 11.75$ & MgXI &    
		& 6.3 & $\sim 11.6$ (flaring) & OVII & C & 5 \\
 	 	&  &  & & & 6.3 & $\sim 10.3$ (quiescent) 
		& OVII & C & 5 \\[0.1cm]
 AB~Dor 	& 6.8 & $< 12.25$ & MgXI & 
 		& 7 & $12-13$ & FeXXI-XXII & A,C,D & 6,7 \\
 		& 6.3 & $10.7^{+0.4}_{-0.1}$ & OVII & 
		& 6.3 & $\sim 10.8$ & OVII & A,C & 8 \\[0.1cm]
 TW~Hya 	& 6.3 & $> 11.9$  & OVII &
 		& 6.3 & $\gtrsim 12$ & OVII 
		& A,C & 9,10 \\[0.05cm] \hline \vspace{0.05cm}
 HD~223460 	& 6.8 & $12.5\pm 0.25$  & MgXI &
 		& 6.3-7 & -- & OVII, NeIX, SiXII & C & 11 \\[0.1cm]
 31~Com 	& 6.8 & $12.5\pm 0.25$  & MgXI &
		& 6.3 & $9.8-11.3$ & OVII & C & 12 \\
 	 	&  &  & & & 5.2 & $10.0-10.6$ & OIV & F & 13 \\[0.1cm]
 $\beta$~Cet 	& 6.8 & $12.25^{+0.1}_{-0.2}$ & MgXI & 
 		& 7 & $11.9-13.1$ & FeXIX-XXII & D & 6 \\
 	 	&  &  & & & 5.2 & $9.7-10.3$ & OIV 
		& F & 13 \\[0.1cm]
 $\mu$~Vel 	& 6.8 & $< 12.25$ &  MgXI &
 		&  & -- & & D & 14 
\enddata
\tablenotetext{a}{Horizontal lines divide the different classes 
    of source into single dwarfs (top) and single giants
    (bottom).}
\tablenotetext{b}{Temperature of maximum formation of the ion
    emitting the density sensitive lines used for the density
    diagnostics.}
\tablenotetext{c}{Ion emitting the density sensitive lines used
    for the diagnostics.}
\tablenotetext{d}{A: {\em Chandra}-HETGS; B: {\em Chandra}-LETGS;
    C: {\em XMM-Newton}-RGS; D: {\em EUVE}; E: {\em FUSE}; F: {\em HST}.}
\tablerefs{(1) Monsignori Fossi et al.\ (1996);
    (2) Magee et al.\ (2003); (3) DelZanna, Landini \& Mason (2002);
    (4) Redfield et al.\ (2002); (5) G{\" u}del et al.\ (2002);
    (6) Sanz-Forcada et al.\ (2002); (7) Sanz-Forcada et al.\ (2003a);
    (8) Sanz-Forcada et al.\ (2003b); (9) Kastner et al.\ (2002);
    (10) Stelzer \& Schmitt\ (2004); (11) Gondoin\ (2003); 
    (12) Scelsi et al.\ (2004); (13) Ayres et al.\ (1998); 
    (14) Ayres et al.\ (1999).}
\end{deluxetable}

\begin{deluxetable}{lccl|cccllc}
\tablecolumns{10} 
\tabletypesize{\footnotesize}
\tablecaption{Comparison of derived densities with previous work: 
	{\it active binaries}. \label{tab10}}
\tablewidth{0pt}
\tablehead{
 \colhead{} & \multicolumn{3}{c}{This work} & 
 \colhead{} & \multicolumn{5}{c}{Previous works} \\
 \cline{2-4} \cline{6-10}\\[-0.2cm]
 \colhead{Source} & \colhead{$\log$T\tablenotemark{a}} & 
 \colhead{$\log n_{\mathrm{e}}$} & \colhead{Ion\tablenotemark{b}} & 
 \colhead{} & \colhead{$\log$T\tablenotemark{a}} & 
 \colhead{$\log n_{\mathrm{e}}$} & \colhead{Ion\tablenotemark{b}} &
 \colhead{data\tablenotemark{c}} & \colhead{Ref.} \\
 \colhead{} & \colhead{[MK]} & \colhead{[cm$^{-3}$]} & \colhead{} & 
 \colhead{} & \colhead{[MK]} & \colhead{[cm$^{-3}$]} & \colhead{} & 
 \colhead{} & \colhead{}
}
\startdata 
 Algol 		& 6.8 & $12.3^{+0.1}_{-0.5}$ & MgXI &
		& 6.1-6.3 & $\sim 10.5$ & NVI, OVII & B & 1,2 \\[0.1cm]
 ER~Vul 	& 6.8 & $< 12.3$ & MgXI & & 7 
 		& $12.2-12.8$ & FeXIX-XXI & D & 3 \\[0.1cm]
 44~Boo 	& 6.8 & $12^{+0.3}_{-0.7}$ & MgXI &
 		& 7 & $12-14$ & FeXIX-XXI & D & 4 \\
 	 	& 6.3 & $10^{+0.8}_{-0.2}$  & OVII & & & & & \\[0.1cm]
 TZ~CrB 	& 6.8 & $12.3^{+0.1}_{-0.1}$ & MgXI &
 		& $6.6-7$ & $12.0-12.5$ & NeIX, MgXI, SiXII & A & 5 \\
 	 	& 6.3 & $10.3^{+0.2}_{-0.5}$  & OVII &
		& 6.3 & $10.0-10.8$ & OVII & A & 5 \\[0.1cm]
 UX~Ari 	& 6.8 & $13^{+0.03}_{-0.3}$ & MgXI &
		& 7 & $12.1-12.9$ & FeXIX-XXI & D & 6 \\
 		& & & & & 6.6 & $11.1-11.7$ & NeIX & B & 1 \\
  		& & & & & 6.1 & $9.8-10.8$ & NVI & B & 1 \\[0.1cm]
 $\xi$~UMa      & 6.8 & $12.5^{+0.1}_{-0.2}$ & MgXI & 
 		& 7 & $11.8-12.3$ & FeXIX-XXI & D & 3 \\
	        & 6.3 & $10.3^{+0.2}_{-0.5}$ &  OVII & & & & & \\[0.1cm]
 II~Peg         & 6.8 & $12.7^{+0.1}_{-0.3}$ & MgXI & 
 		& 7 & $12.4-13.4$ & FeXIX-XXI & D & 6 \\
	        & 6.3 & $10.5^{+0.5}_{-0.3}$ & OVII & 
		& 6.3 & $10.6-11.6$ & OVII & A & 7 \\
		& & & & & 6.8 & $12.8-13.8$ & MgXI & A & 7\\
		& & & & & 6.6 & $11-12$ & NeIX & A & 7 \\[0.1cm]
 $\lambda$~And  & 6.8 & $12.3^{+0.1}_{-0.3}$ & MgXI & 
 		& 7 & $11.7-12.4$ & FeXIX-XXI & D & 6 \\[0.1cm]
 AR~Lac 	& 6.8 & $12.5 \pm 0.5$ &  MgXI &
 		& 6.8 & $\lesssim 12.8$ & MgXI & A & 8 \\
 	 	& &  &  & & 6.3 & $9-12$ & OVII & A & 8 \\[0.1cm]
 HR~1099 	& 6.8 & $12.3^{+0.1}_{-0.1}$ & MgXI &
		& 6.8 & $11-12$ & MgXI & B & 1 \\
 	 	& 6.3 & $10.0^{+0.5}_{-0.7}$  & OVII &
 		& 6.6 & $< 11$ & NeIX & B & 1 \\
  		& & & & & 6.3 & $10.1-10.6$ & OVII & A,B & 9,1 \\
		& & & & & 6.3-7 & $\lesssim 11$ 
		& MgXI, SiXIII & A & 9 \\ 
 	 	& &  & & & 7 & $\lesssim 12-13$ & FeXXI &  D & 9 
\enddata
\tablenotetext{a}{Temperature of maximum formation of the ion
    emitting the density sensitive lines used for the density
    diagnostics.}
\tablenotetext{b}{Ion emitting the density sensitive lines used
    for the diagnostics.}
\tablenotetext{c}{A: {\em Chandra}-HETGS; B: {\em Chandra}-LETGS;
    C: {\em XMM-Newton}-RGS; D: {\em EUVE}.}
\tablerefs{(1) Ness et al.\ (2002a); (2) Ness et al.\ (2002b); 
    (3) Sanz-Forcada et al.\ (2003a); (4) Brickhouse \& Dupree\ (1998); 
    (5) Osten et al.\ (2003); (6) Sanz-Forcada et al.\ (2002); 
    (7) Huenemoerder et al.\ (2001); (8) Huenemoerder et al.\ (2003); 
    (9) Ayres et al.\ (2001).}
\end{deluxetable}

\clearpage

\clearpage

\clearpage

\begin{thebibliography}{}
\bibitem[Alencar \& Batalha (2002)]{alenba02} Alencar, 
	S.~H.~P. \& Batalha, C.\ 2002, \apj, 571, 378 
\bibitem[Al-Naimiy (1981)]{aln81} Al-Naimiy, H.~M.~K.\ 1981, 
	\aaps, 43, 85
\bibitem[Anders \& Grevesse (1989)]{ag89} Anders, E.
	\& Grevesse, N.\ 1989, \gca, 53, 197
\bibitem[Argiroffi et al. (2003)]{argi03} Argiroffi, C.,
	Maggio, A., Peres, G.\ 2003, \aap, 404, 1033
\bibitem[Audard (2003)]{Audard03} Audard, M.\ 2003, Advances 
	in Space Research, 32, 927
\bibitem[Audard et al. (2001)]{Audard01} Audard, M., Behar, 
	E., G{\" u}del, M., Raassen, A.~J.~J., Porquet, D., 
	Mewe, R., Foley, C.~R. \& Bromage, G.~E.\ 2001, 
	\aap, 365, L329
\bibitem[Ayres et al. (2001)]{ayres01} Ayres, T.R, Brown, 
	A., Osten, R.A., Huenemoerder, D.~P., Drake, J.~J.,
	Brickhouse, N.~S. \& Linsky, J.~L.\ 2001, \apj, 549,
	554
\bibitem[Ayres et al. (1999)]{ayres99} Ayres, T.R, Osten, 
	R.A. \& Brown, A.\ 1999, \apj, 526, 445
\bibitem[Ayres et al. (1998)]{ayres98} Ayres, T.R., Simon, 
	T., Stern, R.A., Drake, S.A., Wood, B.E. \& Brown, 
	A.\ 1998, \apj, 496, 428
\bibitem[Benedict et al. (1998)]{bened98} Benedict, 
	G.~F.\ et al.\ 1998, \aj, 116, 429
\bibitem[Benedict et al. (1999)]{bened99} Benedict, G.~F.\ 
	et al.\ 1999, \aj, 118, 1086
\bibitem[Berdyugina et al. (1999)]{berd99} Berdyugina, S.V., 
	Ilyin, I. \& Tuominen, I.\ 1999, \aap 347, 932
\bibitem[Berdyugina et al. (1998)]{berd98} Berdyugina, S.V., 
	Jankov, S., Ilyin, I., Tuominen, I. \& Fekel, F.C.\ 
	1998, \aap, 334, 863
\bibitem[Bowyer et al. (2000)]{bowyer00} Bowyer, S., Drake, 
	J.~J. \& Vennes, S.\ 2000, \araa, 38, 231
\bibitem[Brickhouse (2002)]{brick02} Brickhouse, N.~S.\ 
	2002, in Stellar Coronae in the Chandra and 
	XMM-NEWTON Era,	eds. Favata, F. and Drake, J.J., 
	ASP Conf. Ser. 277, 13 
\bibitem[Brickhouse et al. (2001)]{BDY01} Brickhouse, N.S., 
	Dupree, A.K. \& Young, P.R.\ 2001, \apj, 562, L75
\bibitem[Brickhouse \& Dupree (1998)]{brick98} Brickhouse, 
	N.~S. \& Dupree, A.~K.\ 1998, \apj, 502, 918 
\bibitem[Brinkman et al. (2000)]{brink00} Brinkman, A.~C.
	et al.\ 2000, \apjl, 530, L111
\bibitem[Canizares et al. (2000)]{HETG00} Canizares, 
	C.~R.\ et al.\ 2000, \apjl, 539, L41 
\bibitem[Cayrel et al. (1994)]{cay94} Cayrel de 
	Strobel, G., Cayrel, R., Friel, E., Zahn, J.-P., 
	\& Bentolila, C.\ 1994, \aap, 291, 505 
\bibitem[Contadakis (1995)]{cont95} Contadakis, M.~E.\ 
	1995, \aap, 300, 819 
\bibitem[Decin et al. (2003)]{decin03} Decin, L., 
	Vandenbussche, B., Waelkens, K., Eriksson, C., 
	Gustafsson, B., Plez, B. \& Sauval, A.~J.\ 2003,
	\aap, 400, 695 
\bibitem[Del Zanna et al. (2002)]{delz02} Del Zanna, G.,
	Landini, M. \& Mason, H.E.\ 2002, \aap, 385, 968
\bibitem[Donati et al. (1995)]{donati95}Donati, J.-F., 
	Henry, G.~W., \& Hall, D.~S.\ 1995, \aap, 293, 107 
\bibitem[Drake et al. (2000)]{drake00} Drake, J.~J., Peres, G.,
	Orlando, S., Laming, J.~M. \& Maggio, A.\ 2000, \apj, 
	545, 1074
\bibitem[Drake(2001)]{2001xras.conf...53D} Drake, J.~J.\ 2001, 
	ASP Conf.~Ser.~234: X-ray Astronomy 2000, 53 
\bibitem[Drake et al. (2001)]{drake01} Drake, J.~J., Brickhouse, 
	N.~S., Kashyap, V., Laming, J.~M., Huenemoerder, D.~P., 
	Smith, R. \& Wargelin, B.~J.\ 2001, \apjl, 548, L81
\bibitem[Drake (2003)]{drake03} Drake, J.~J.\ 2003, Advances in 
	Space Research, 32, 945
\bibitem[Drake et al. (1989)]{drake89} Drake, S.~A., 
	Simon, T., \& Linsky, J.~L.\ 1989, \apjs, 71, 905 
\bibitem[Duemmler \& Aarum (2001)]{dumm01} Duemmler, R. 
	\& Aarum, V.\ 2001, \aap, 370, 974
\bibitem[Favata et al. (1995)]{favata95} Favata, F., 
	Barbera, M., Micela, G., \& Sciortino, S.\ 1995, 
	\aap, 295, 147 
\bibitem[Favata et al. (2000)]{favata00} Favata, F., Reale, 
	F., Micela, G., Sciortino, S., Maggio, A. \& 
	Matsumoto, H.\ 2000, \aap, 353, 987 
\bibitem[Flower(1996)]{flower96} Flower, P.~J.\ 1996, \apj, 
	469, 355 
\bibitem[Frogel et al. (1972)]{frog72} Frogel, J.~A., 
	Kleinmann, D.~E., Kunkel, W., Ney, E.~P., \& 
	Strecker, D.~W.\ 1972, \pasp, 84, 581 
\bibitem[Gabriel \& Jordan (1969)]{GJ69} Gabriel, A.H. \& 
	Jordan, C.\ 1969, MNRAS, 145, 241
\bibitem[Gadun (1994)]{gadun94} Gadun, A.S.\ 1994, AN, 
	315, 413
\bibitem[Gehren et al. (1999)]{gehr99} Gehren, T., Ottmann, 
	R., \& Reetz, J.\ 1999, \aap, 344, 221 
\bibitem[Gimenez et al. (1986)]{gime86} Gimenez, A., 
	Fernandez-Figueroa, M.J., de Castro, E., Ballester, 
	J.L. \&	Reglero, V.\ 1986, \aj, 92, 131
\bibitem[Gliese et al. (1969)]{gliese69} Gliese, W.\ 1969, 
	Veroeffentlichungen des Astronomischen 
	Rechen-Instituts Heidelberg, 22, 1
\bibitem[Gondoin (2003)]{gondn03} Gondoin, P.\ 2003, \aap, 
	409, 263 
\bibitem[Gondoin (1999)]{gondn99} Gondoin, P.\ 1999, \aap, 
	352, 217 
\bibitem[G{\" u}del et al. (2002)]{Guedel02} G{\" u}del, 
	M., Audard, M., Skinner, S.~L. \& Horvath, M.~I.\ 
	2002, \apjl, 580, L73
\bibitem[G{\" u}del et al. (2001a)]{Guedel01a} G{\" u}del, 
	M., Audard, M., Briggs, K, Haberl, F., Magee, H., 
	Maggio, A., Mewe, R., Pallavicini, R. \& Pye, J.\ 
	2001a, \aap, 365, L336
\bibitem[G{\" u}del et al. (2001b)]{Guedel01b} G{\" u}del, 
	M., Audard, M., Magee, H., Franciosini, E., Grosso, 
	N., Cordova, F.~A., Pallavicini, R. \& Mewe, R.\ 
	2001b, \aap, 365, L344
\bibitem[G{\" u}del (1997)]{Guedel97} G{\" u}del, M.\ 1997, 
	\apjl, 480, L121
\bibitem[Hill et al. (1989)]{hill98} Hill, G., Fisher, W.A. 
	\& Holmgren, D.\ 1989, \aap, 211, 81
\bibitem[Houdebine \& Doyle (1994)]{hd94} Houdebine, E.~R. 
	\& Doyle, J.~G.\ 1994, \aap, 289, 185 
\bibitem[Huenemoerder et al. (2001)]{HCS01} Huenemoerder,
	D.P., Canizares, C.R. \& Schulz, N.S.\ 2001, \apj, 
	559, 1135
\bibitem[Hussain (1997)]{hus97} Hussain, G.~A.~J., Unruh, 
	Y.~C. \& Collier Cameron, A.\ 1997, \mnras, 288, 343
\bibitem[Kashyap \& Drake (2000)]{KD00} Kashyap, V. \&
	Drake, J.J.\ 2000, Bull. Astron. Soc. India, 28, 475
\bibitem[Kastner et al. (2002)]{Kast02} Kastner, J.~H., 
	Huenemoerder, D.~P., Schulz, N.~S., Canizares, C.~R.
	\& Weintraub, D.~A.\ 2002, \apj, 567, 434
\bibitem[Kato \& Nakazaki (1989)]{kato89} Kato, T. \& Nakazaki, 
	S.\ 1989, ADNDT, 42, 313
\bibitem[Katsova \& Tsikoudi (1993)]{ks93} Katsova, M.~M. 
	\& Tsikoudi, V.\ 1993, \apjl, 402, L9
\bibitem[Laming (1998)]{LAM98} Laming, J.M.\ 1998, ASP Conf. 
	Ser. 154, The Tenth Cambridge Workshop on Cool Stars, 
	Stellar Systems and the Sun, Edited by R. A. Donahue 
	and J. A. Bookbinder, p.447
\bibitem[Lang(1999)]{lang99} Lang, K.~R.\ 1999, 
	Astrophysical formulae  Vol.\ 1/ K.R.~Lang.~New York : 
	Springer, 1999.~(Astronomy and astrophysics library,
	ISSN0941-7834), p.107
\bibitem[Lanzafame et al. (2000)]{Lanz00} Lanzafame, A.~C., 
	Bus{\` a}, I., \& Rodon{\` o}, M.\ 2000, \aap, 362, 
	683 
\bibitem[Liedahl et al.\ (1995)]{hullac} Liedahl, D.A., Osterheld, 
	A.L. \& Goldstein, W.H.\ 1995, \apjl, 438, L115 
\bibitem[Linsky et al. (1982)]{Lin82} Linsky, J.~L., 
	Bornmann, P.~L., Carpenter, K.~G., Hege, E.~K., 
	Wing, R.~F., Giampapa, M.~S., \& Worden, S.~P.\ 
	1982, \apj, 260, 670 
\bibitem[Magee et al. (2003)]{magee03} Magee, H.~R.~M.,
	G{\" u}del, M., Audard, M. \& Mewe, R.\ 2003, 
	Advances in Space Research, 32, 1149	 
\bibitem[Maggio et al. (2000)]{maggio00} Maggio, A., 
	Pallavicini, R., Reale, F. \& Tagliaferri, G.\ 
	2000, \aap, 356, 627 
\bibitem[Mariska (1992)]{mariska92} Mariska, J.~T.\ 1992, 
	in "The solar transition region", Cambridge 
	Astrophysics Series, vol 22, Cambridge University 
	Press
\bibitem[Marino et al. (2003)]{marino03} Marino, A., Micela, 
	G., Peres, G. \& Sciortino, S.\ 2003, \aap, 407L, 63
\bibitem[Marino et al. (1999)]{marino99} Marino, G., 
	Rodon{\' o}, M., Leto, G. \& Cutispoto, G.\ 1999, 
	\aap, 352, 189
\bibitem[Mewe et al.(2003)]{mewe03} Mewe, R., Porquet, D., 
	Raassen, A.J.J., Kaastra, J.S., Dubau, J. \& Ness, 
	J-U.\ 2003, Proceedings of the 12th Cambridge Workshop 
	on Cool Stars, Stellar Systems, \& the Sun,
	http://origins.Colorado.EDU/cs12/proceedings/poster/mewe.ps
\bibitem[Monsignori Fossi et al. (1996)]{mf96} Monsignori Fossi, 
	B.~C., Landini, M., Del Zanna, G., \& Bowyer, S.\  
	1996, \apj, 466, 427 
\bibitem[Murad \& Budding (1984)]{muba84} Murad, I.~M. \& 
	Budding, E.\ 1984, \apss, 98, 163 
\bibitem[Muzerolle et al. (2000)]{muz00} Muzerolle, J., 
	Calvet, N., Briceño, C., Hartmann, L. \& 
	Hillenbrand, L.\ 2000, \apj, 535L, 47
\bibitem[Nordgren et al. (1999)]{nord99} Nordgren, T.E. et
	al.\ 1999, \aj, 118, 3032
\bibitem[Neff et al. (1996)]{neff96} Neff, J.E., Pagano, I., 
	Rodon{\` o}, M., Brown, A., Dempsey, R.C., Fox, D.C.
	\& Linsky, J.L.\ 1996, \aap, 310, 173
\bibitem[Ness et al. (2003)]{ness03} Ness, J.U., Brickhouse, 
	N.~S., Drake, J.~J. \& Huenemoerder, D.~P.\ 2003, 
	\apj, 598,1277
\bibitem[Ness et al. (2002a)]{ness02a} Ness, J.U., Schmitt, 
	J.H.M.M., Burwitz, V., Mewe, R., Raassen, A.~J.~J., 
	van der Meer, R.~L.~J. \& Predehl, P., Brinkman, 
	A.~C.\ 2002a, \aap, 394, 911
\bibitem[Ness et al. (2002b)]{ness02b} Ness, J.U., Schmitt, 
	J.H.M.M., Burwitz, V., Mewe, R. \& Predehl, P.\ 
	2002b, \aap, 387, 1032
\bibitem[Ness et al. (2001)]{ness01} Ness, J.U., Mewe, R., 
	Schmitt, J.~H.~M.~M., Raassen, A.~J.~J., Porquet, 
	D., Kaastra, J.~S., van der Meer, R.~L.~J., Burwitz, 
	V. \& Predehl, P.\ 2001, \aap, 367, 282
\bibitem[Orlando, Peres, \& Reale (2000)]{orlando00} 
	Orlando, S., Peres, G., \& Reale, F.\ 2000, \apj, 
	528, 524 
\bibitem[Orlando et al. (2004)]{orlando04} Orlando, S., 
	Peres, G. \& Reale\ 2004, \aap, submitted
\bibitem[Osten et al. (2003)]{osten03} Osten, R.~A., Ayres, 
	T.~R., Brown, A., Linsky, J.~L. \& Krishnamurthi, 
	A.\ 2003, \apj, 582, 1073	
\bibitem[Padmakar (1999)]{pad99} Padmakar \& Pandey, S.~K.\ 
	1999, \aaps, 138, 203 
\bibitem[Pakull et al. (1981)]{pak81} Pakull, M.~W.\ 1981, 
	\aap, 104, 33 
\bibitem[Panzera et al. (1999)]{panz99} Panzera, M.~R., 
	Tagliaferri, G., Pasinetti, L., \& Antonello, E.\ 
	1999, \aap, 348, 161
\bibitem[Peres, Orlando \& Reale (2004)]{peres04} Peres, 
	G., Orlando, S., \& Reale, F.\ 2003, \apj, submitted
\bibitem[Pettersen et al. (1980)]{petter80} Pettersen, 
	B.~R.\ 1980, \aj, 85, 871 
\bibitem[Pettersen et al. (1989)]{petter89} Pettersen, 
	B.~R. \& Hawley, S.~L.\ 1989, \aap, 217, 187 
\bibitem[Pizzolato et al. (2000)]{pizz00} Pizzolato, N., 
	Maggio, A., \& Sciortino, S.\ 2000, \aap, 361, 614 
\bibitem[Plucinsky et al. (2002)]{Pluc02} Plucinsky, P.~P.,
	Edgar, R.~J., Virani, S.~N., Townsley, L.~K. \&
	Broos, P.~S.\ 2002, ASP Conf. Ser. 262: The High 
	Energy Universe at Sharp Focus: Chandra Science, 391
\bibitem[Porquet \& Dubau (2000)]{PD00} Porquet, D. \& 
	Dubau, J.\ 2000, A\&AS, 143, 495
\bibitem[Porquet et al. (2001)]{Porquet01} Porquet, D.,
	Mewe, R., Dubau, J., Raassen, A.~J.~J. \& Kaastra, 
	J.~S.\ 2001, \aap, 376, 1113
\bibitem[Pradhan \& Shull (1981)]{pradhan81} Pradhan, A.~K.
	\& Shull, J.~M.\ 1981, \apj, 249, 821
\bibitem[Raassen et al. (2002)]{Raassen02} Raassen et al.\ 
	2002, \aap, 389, 228
\bibitem[Redfield et al. (2003)]{red03} Redfield, S., 
	Ayres, T.~R., Linsky, J.~L., Ake, T.~B., Dupree, 
	A.~K., Robinson, R.~D., \& Young, P.~R.\ 2003, 
	\apj, 585, 993 
\bibitem[Redfield et al. (2002)]{red02} Redfield, S., 
	Linsky, J.~L., Ake, T.~B., Ayres, T.~R.,  Dupree, 
	A.~K., Robinson, R.~D., Wood, B.~E. \& Young, 
	P.~R.\ 2002, \apj, 581, 626	
\bibitem[Rosner, Tucker \& Vaiana (1978)]{RTV} Rosner,
	R., Tucker, W.~H. \& Vaiana, G.~S.\ 1978, \apj, 
	220, 643
\bibitem[Rosner \& Vaiana (1977)]{RV77} Rosner, R. \& 
	Vaiana, G.~S.\ 1977, \apj, 216, 141
\bibitem[Saar (1996)]{saar96} Saar, S.~H.\ 1996, IAU Symp. 
	176: Stellar Surface Structure, 237
\bibitem[Sampson et al. (1983)]{samp83} Sampson, D.H., Goett, S.J., 
	Clark, R.E.H.\ 1983, ADNDT, 29, 467
\bibitem[Sanz-Forcada et al. (2003a)]{sanz03a} 
	Sanz-Forcada, J., Brickhouse, N.S \& Dupree, A.K.\ 
	2003a, \apjs, 145, 147
\bibitem[Sanz-Forcada et al. (2003b)]{sanz03b} 
	Sanz-Forcada, J., Maggio, A. \& Micela, G.\ 2003b, 
	\aap, 408, 1087
\bibitem[Sanz-Forcada et al. (2002)]{sanz02} Sanz-Forcada, 
	J., Brickhouse, N.S \& Dupree, A.K.\ 2002, \apj, 
	570, 799
\bibitem[Sanz-Forcada et al. (2001)]{sanz01} Sanz-Forcada, 
	J., Brickhouse, N.S \& Dupree, A.K.\ 2001, \apj, 
	554, 1079
\bibitem[Scelsi et al. (2004)]{scelsi04} Scelsi, L., Maggio, 
	A., Peres, G. \& Gondoin, P.\ 2004, \aap, 413, 643
\bibitem[Sciortino et al. (1999)]{sciorti99} Sciortino, S., 
	Maggio, A., Favata, F. \& Orlando, S.\ 1999, \aap, 
	342, 502
\bibitem[Simon (1986)]{simon86} Simon, T.\ 1986, \aj, 91, 1233 
\bibitem[Singh et al. (1996a)]{singh96a} Singh, K.P., Drake, 
	S.A., White, N.E. \& Simon, T.\ 1996a, \aj, 112, 221
\bibitem[Singh et al. (1996b)]{singh96b} Singh, K.P., White, 
	N.E. \& Drake, S.A.\ 1996b, \apj, 456, 766
\bibitem[Singh et al. (1999)]{singh99} Singh, K.~P., Drake, 
	S.~A., Gotthelf, E.~V., \& White, N.~E.\ 1999, \apj, 
	512, 874 
\bibitem[Smith et al. (2001)]{smith2001} Smith, R.~K., 
	Brickhouse, N.~S., Liedahl, D.~A. \& Raymond, 
	J.~C.\ 2001, \apjl, 556, L91
\bibitem[Stelzer et al. (2002)]{stelzer02} Stelzer, B. et
	al.\ 2002, \aap, 392, 585
\bibitem[Stelzer \& Schmitt (2004)]{stelzer04} Stelzer, B. 
	\& Schmitt, J.~H.~M.~M.\ 2004, \aap, in press
	(astro-ph/0402108)
\bibitem[Strassmeier et al. (1993)]{strass93}  Strassmeier, 
	K.G., Hall, D.S., Fekel, F.C. \& Scheck, M.\ 1993,
	\aaps, 100, 173
\bibitem[Testa et al. (2004)]{testa04} Testa, P., Drake, 
	J.J., Peres, G. \& DeLuca, E.E.\ 2004, ApJL, 
	submitted
\bibitem[Torres et al. (2003)]{torr03} Torres, G., Guenther, 
	E.~W., Marschall, L.~A., Neuh{\" a}user, R., Latham,
	D.~W., \& Stefanik, R.~P.\ 2003, \aj, 125, 825
\bibitem[Wood et al. (2001)]{wood01} Wood, B.E., Linsky, 
	J.L., Müller, H-R. \& Zank, G.P.\ 2001, \apj, 547, 
	L49
\bibitem[Withbroe \& Noyes (1977)]{withb77} Withbroe, G.~L.
	\& Noyes, R.~W.\ 1977, \araa, 15, 363
\bibitem[Zhang \& Sampson (1987)]{zhang87} Zhang, H. \& Sampson, 
	D.H.\ 1987, \apjs, 63, 487
\end{thebibliography}
\end{document}